\documentclass[a4paper,10pt]{article}
\pdfoutput=1

\usepackage{graphics,epsfig}
\usepackage{xspace,colortbl}
\usepackage{amsfonts,amsmath,amssymb} 
\usepackage{cite} 
\usepackage{hyperref} 

\newcommand{\bea}{\begin{eqnarray}}
\newcommand{\eea}{\end{eqnarray}}
\newcommand{\beq}{\begin{equation}}
\newcommand{\eeq}{\end{equation}}

\def\slash#1{\mbox{$\not\!\! #1$}}

\def\simge{\mathrel{\rlap{\raise 0.511ex \hbox{$>$}}{\lower 0.511ex
 \hbox{$\sim$}}}}
\def\simle{\mathrel{\rlap{\raise 0.511ex \hbox{$<$}}{\lower 0.511ex
 \hbox{$\sim$}}}}
\def\slash#1{\setbox0=\hbox{$#1$}\dimen0=\wd0 \setbox1=\hbox{/} \dimen1=\wd1
 \ifdim\dimen0>\dimen1 \rlap{\hbox to \dimen0{\hfil/\hfil}} #1
 \else \rlap{\hbox to \dimen1{\hfil$#1$\hfil}} / \fi}

\def\rmi{g}
\def\rmif{c}
\def\rmiff{d}
\def\rmii{a}
\def\infntv{b}
\def\rmiii{e}
\def\infntre{f}

\newcommand{\sla}[1]%
        {\kern .25em\raise.18ex\hbox{$/$}\kern-.6em #1}

\newcommand{\gou}{\raisebox{-0.4\totalheight}{\includegraphics[scale=.4]{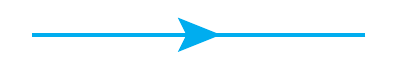}}}
\newcommand{\god}{\raisebox{-0.4\totalheight}{\includegraphics[scale=.4]{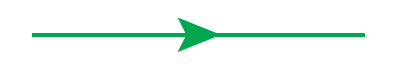}}}
\newcommand{\gol}{\raisebox{-0.4\totalheight}{\includegraphics[scale=.4]{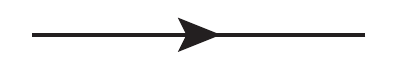}}}
\newcommand{\goi}{\raisebox{-0.4\totalheight}{\includegraphics[scale=.4]{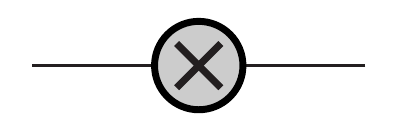}}}
        
\newcommand{\gdud}{\raisebox{-0.4\totalheight}{\includegraphics[scale=.3]{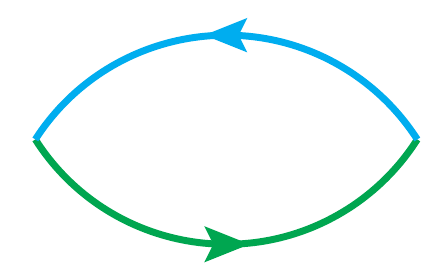}}}
\newcommand{\gdil}{\raisebox{-0.4\totalheight}{\includegraphics[scale=.3]{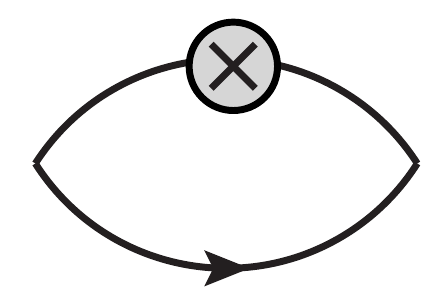}}}
\newcommand{\gdll}{\raisebox{-0.4\totalheight}{\includegraphics[scale=.3]{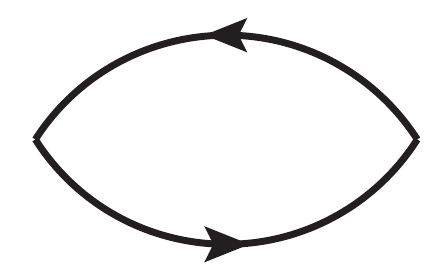}}}
\newcommand{\gdsi}{\raisebox{-0.4\totalheight}{\includegraphics[scale=.3]{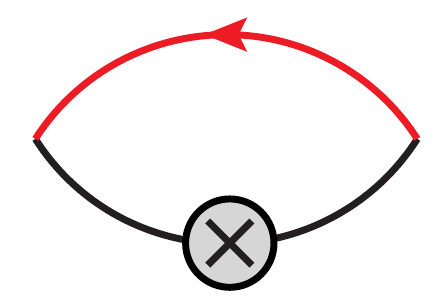}}}

\newcommand{\gdsu}{\raisebox{-0.4\totalheight}{\includegraphics[scale=.3]{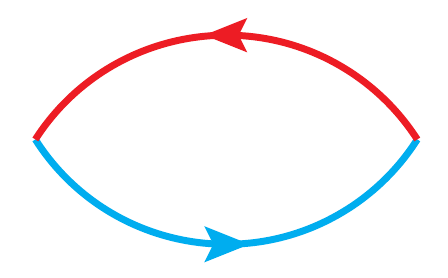}}}
\newcommand{\gduu}{\raisebox{-0.4\totalheight}{\includegraphics[scale=.3]{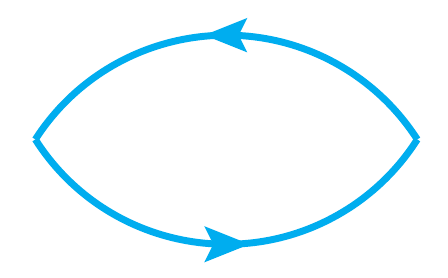}}}
\newcommand{\gddd}{\raisebox{-0.4\totalheight}{\includegraphics[scale=.3]{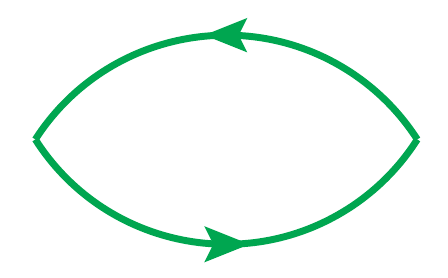}}}
\newcommand{\gdli}{\raisebox{-0.4\totalheight}{\includegraphics[scale=0.3]{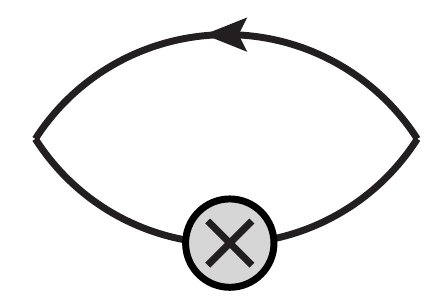}}}
\newcommand{\gdsd}{\raisebox{-0.4\totalheight}{\includegraphics[scale=.3]{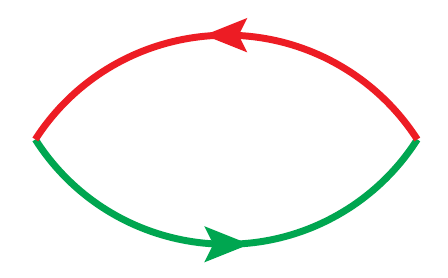}}}
\newcommand{\gdsl}{\raisebox{-0.4\totalheight}{\includegraphics[scale=.3]{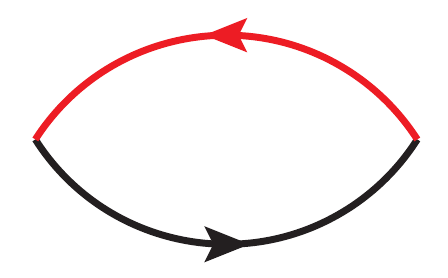}}}

\newcommand{\gdslselfl}{\raisebox{-0.6\totalheight}{\includegraphics[scale=.3]{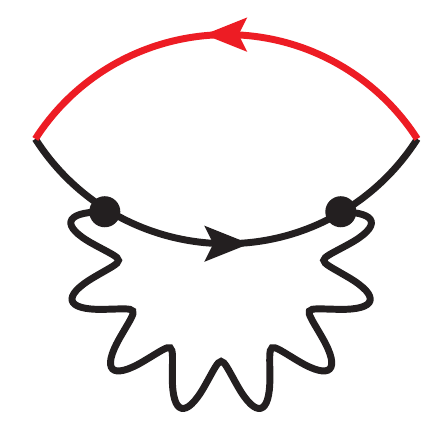}}}
\newcommand{\gdslexch}{\raisebox{-0.4\totalheight}{\includegraphics[scale=.3]{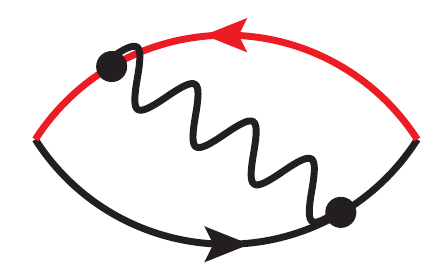}}}

\newcommand{\gdii}{\raisebox{-0.4\totalheight}{\includegraphics[scale=0.3]{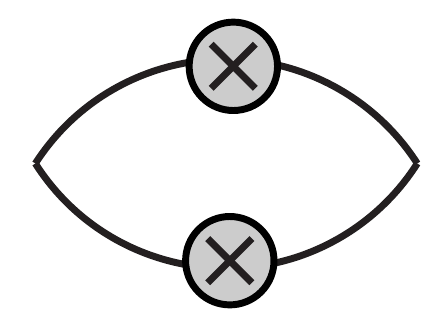}}}
\newcommand{\discgdii}{\raisebox{-0.4\totalheight}{\includegraphics[scale=0.4]{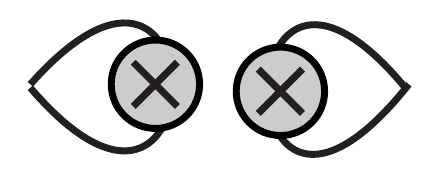}}}

\newcommand{\protond}{\raisebox{-0.4\totalheight}{\includegraphics[scale=.16]{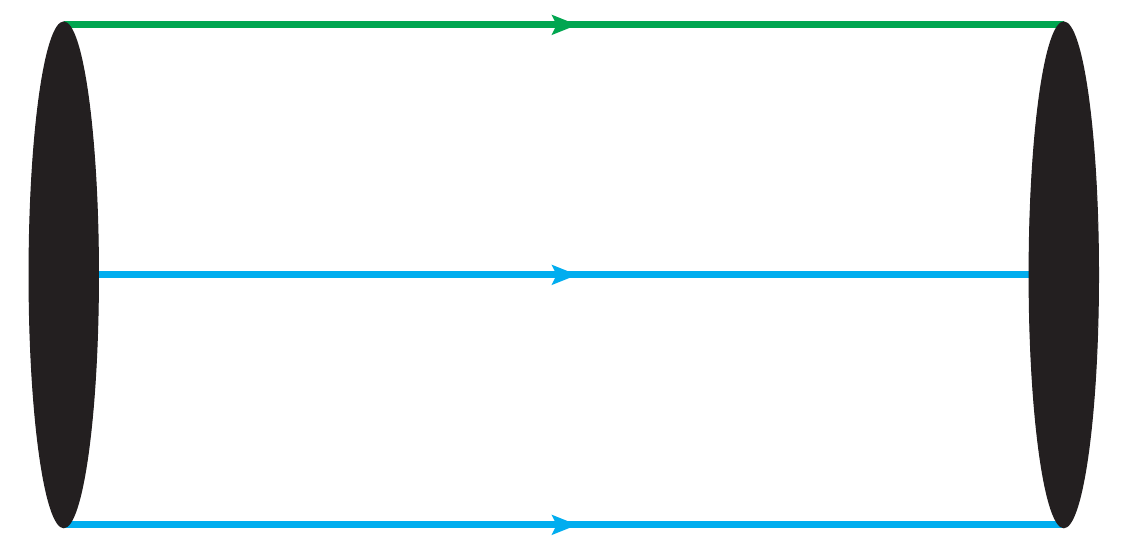}}}
\newcommand{\neutrond}{\raisebox{-0.4\totalheight}{\includegraphics[scale=.16]{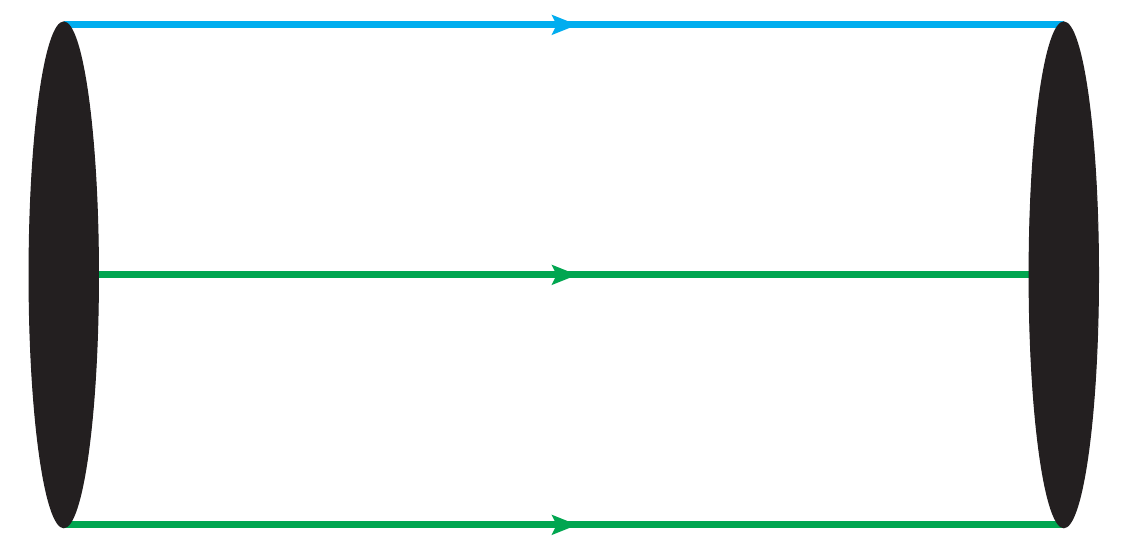}}}
\newcommand{\protondlll}{\raisebox{-0.4\totalheight}{\includegraphics[scale=.16]{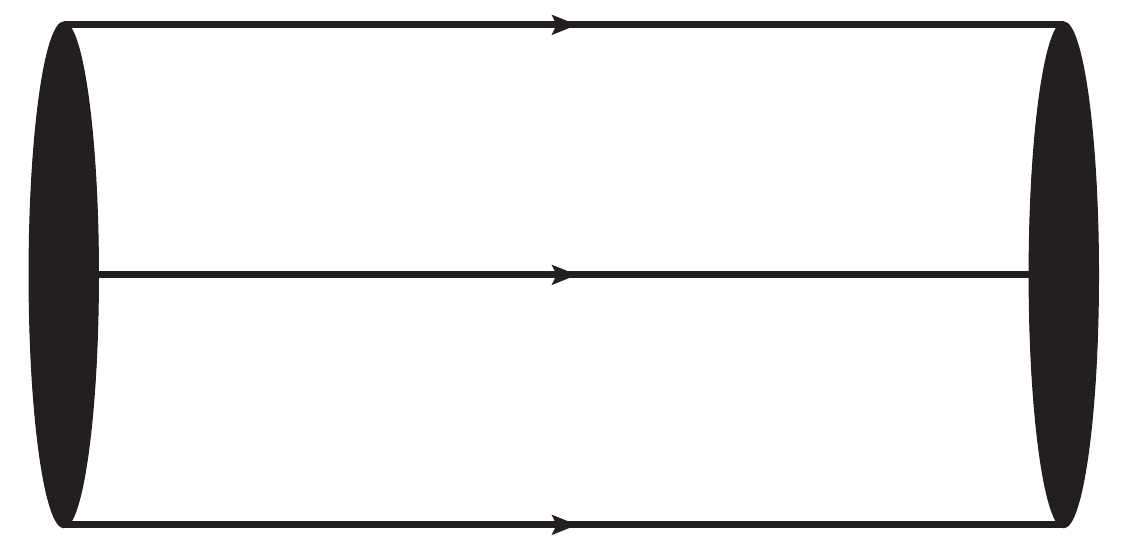}}}
\newcommand{\protondill}{\raisebox{-0.4\totalheight}{\includegraphics[scale=.16]{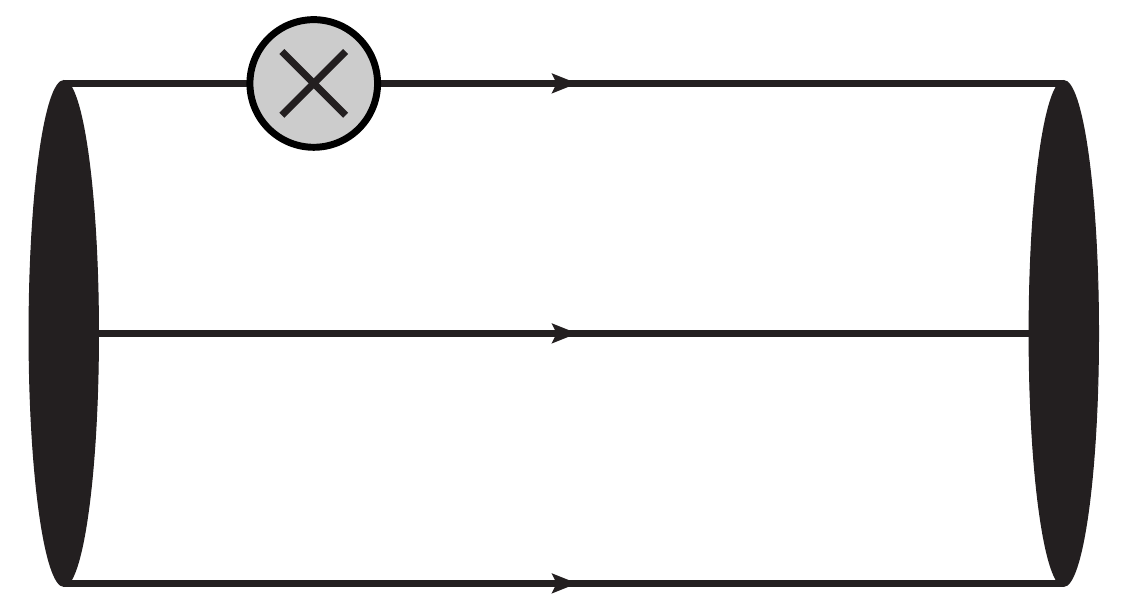}}}
\newcommand{\protondlil}{\raisebox{-0.4\totalheight}{\includegraphics[scale=.16]{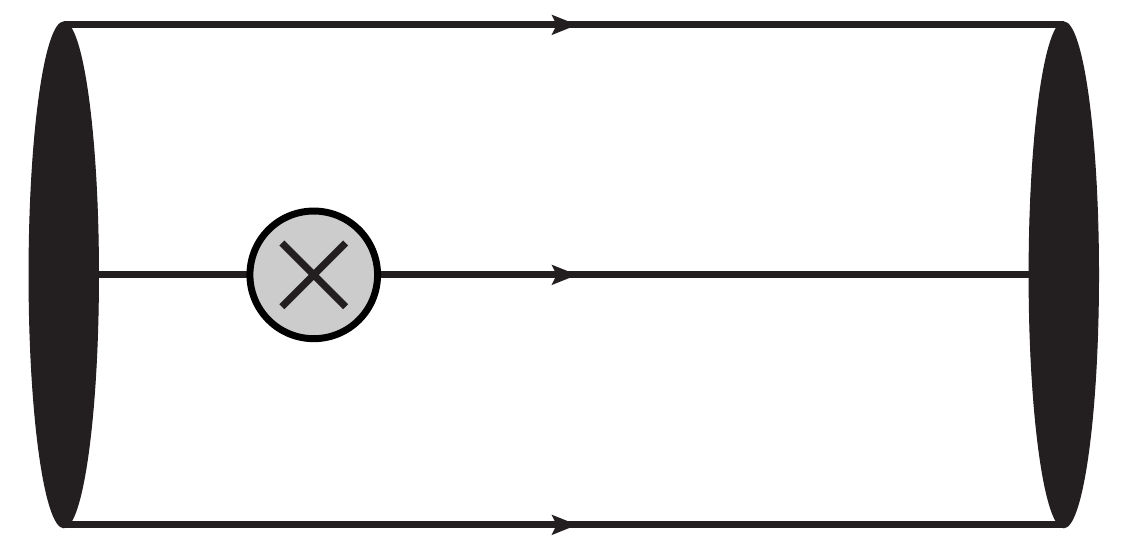}}}
\newcommand{\protondlli}{\raisebox{-0.4\totalheight}{\includegraphics[scale=.16]{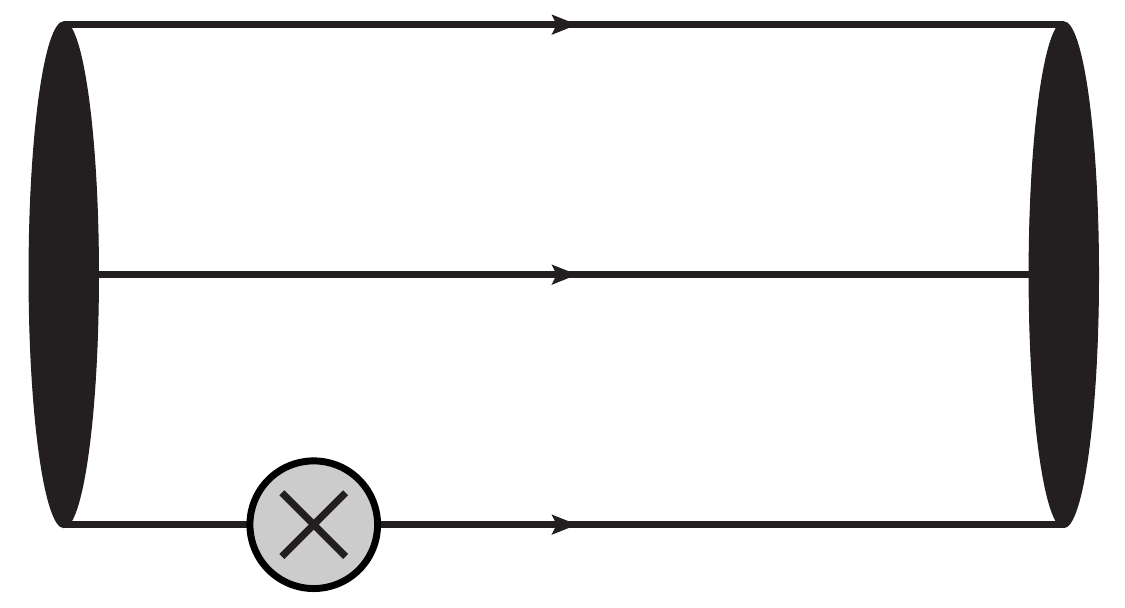}}}

\newcommand{\protonc}{\raisebox{-0.4\totalheight}{\includegraphics[scale=.16]{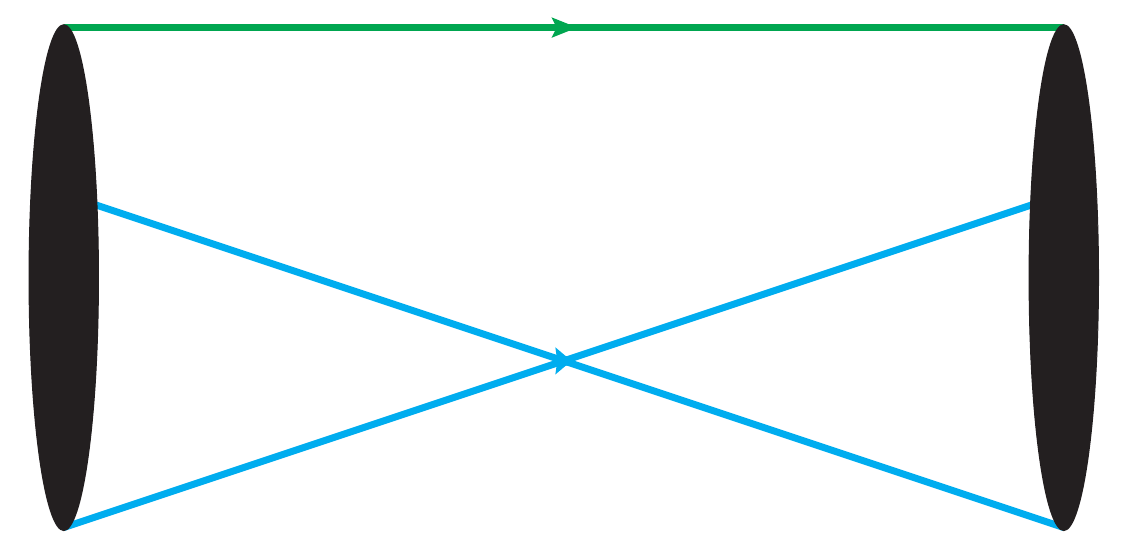}}}
\newcommand{\neutronc}{\raisebox{-0.4\totalheight}{\includegraphics[scale=.16]{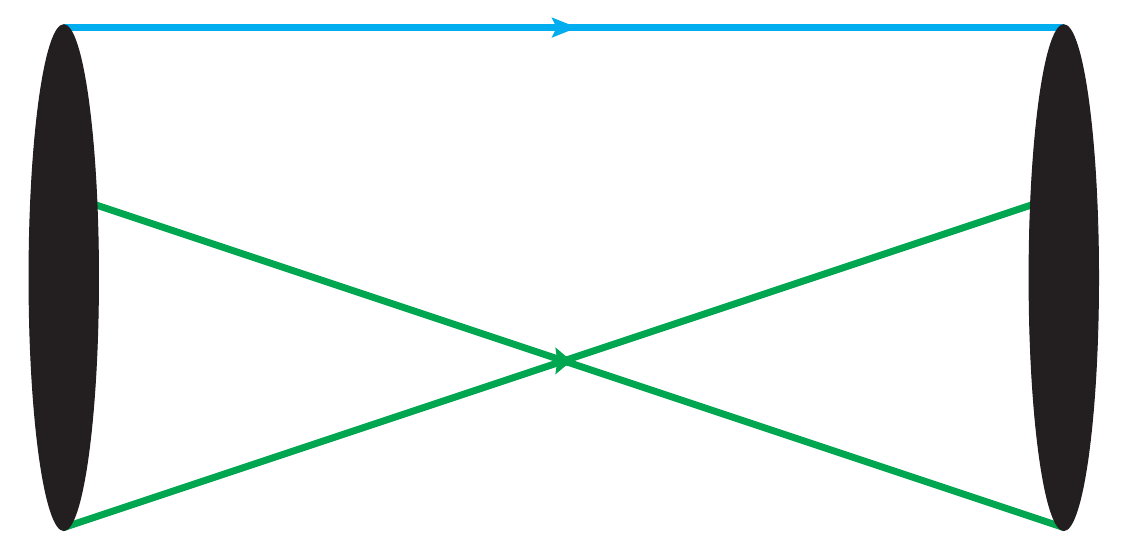}}}
\newcommand{\protonclll}{\raisebox{-0.4\totalheight}{\includegraphics[scale=.16]{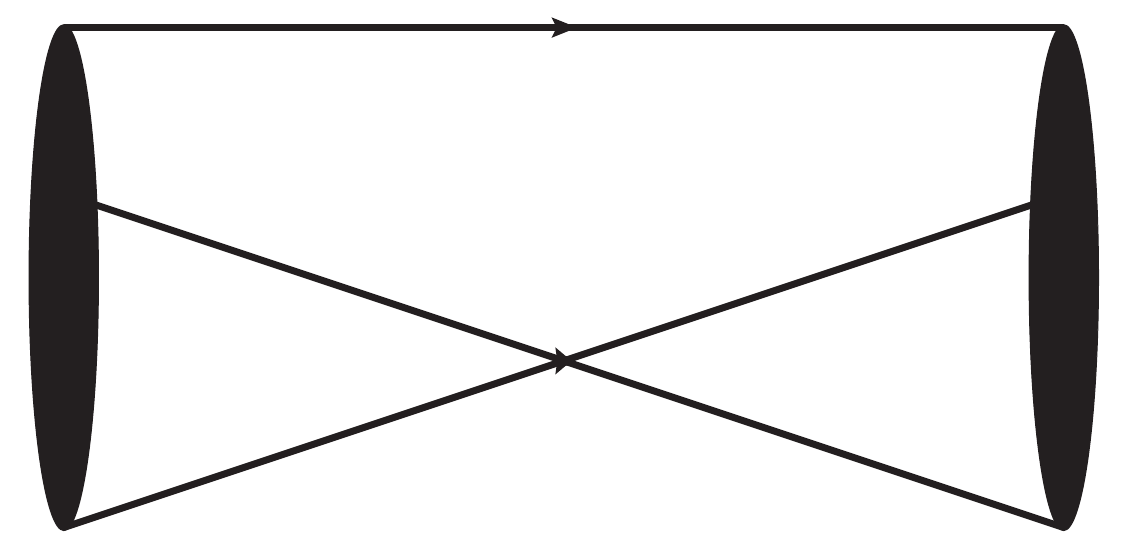}}}
\newcommand{\protoncill}{\raisebox{-0.4\totalheight}{\includegraphics[scale=.16]{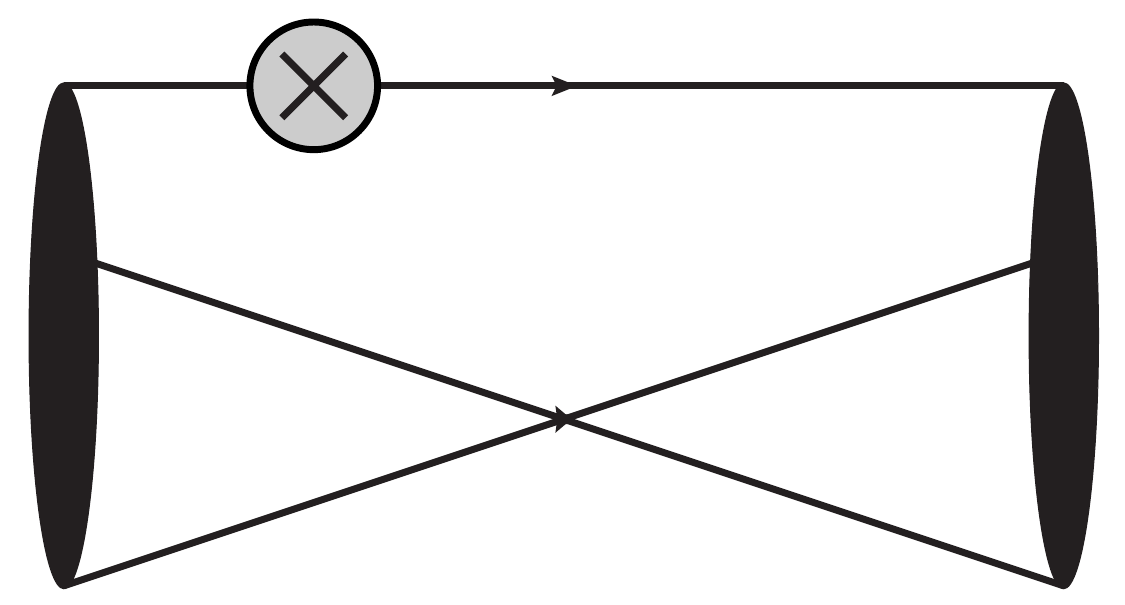}}}
\newcommand{\protonclil}{\raisebox{-0.4\totalheight}{\includegraphics[scale=.16]{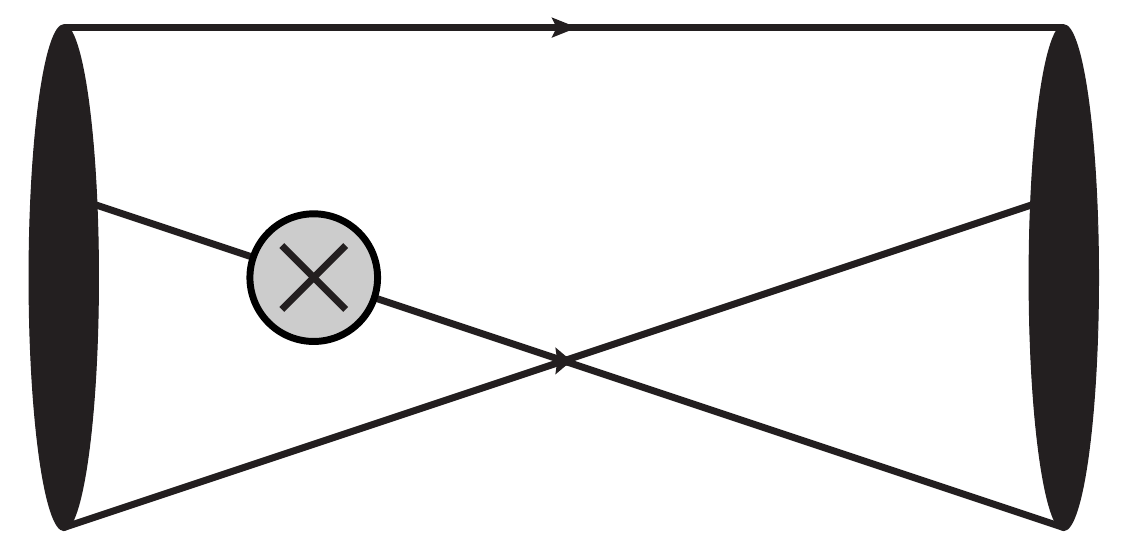}}}
\newcommand{\protonclli}{\raisebox{-0.4\totalheight}{\includegraphics[scale=.16]{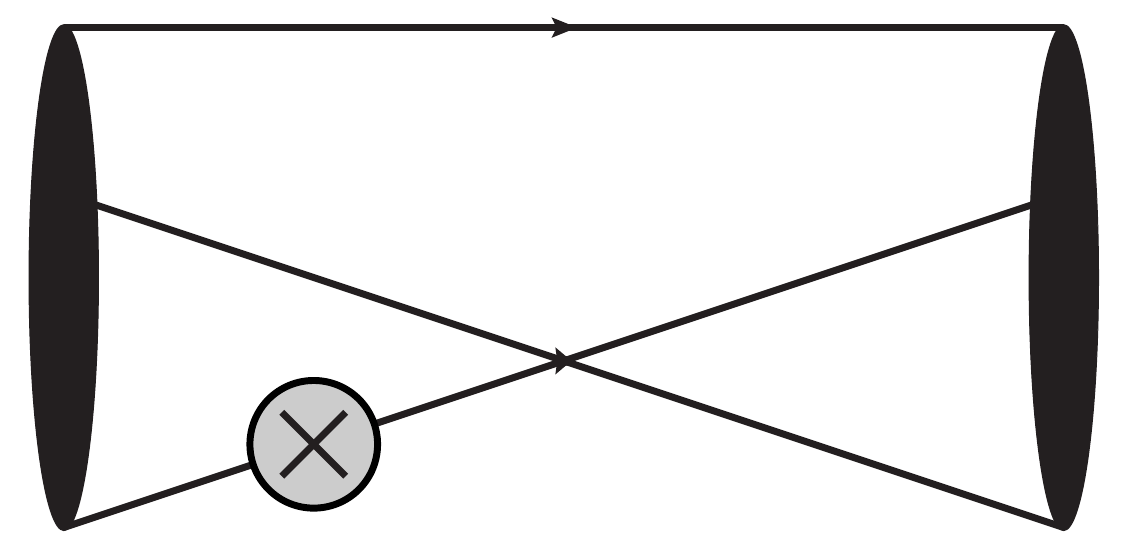}}}

\newcommand{\discgtsud}{\raisebox{-0.4\totalheight}{\includegraphics[scale=.3]{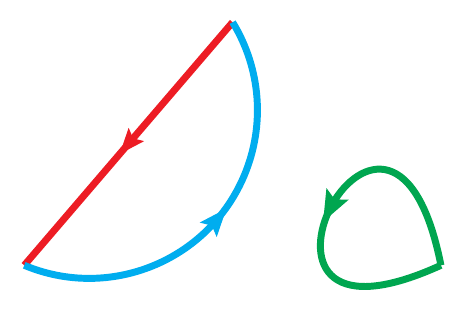}}}        
\newcommand{\discgtsuu}{\raisebox{-0.4\totalheight}{\includegraphics[scale=.3]{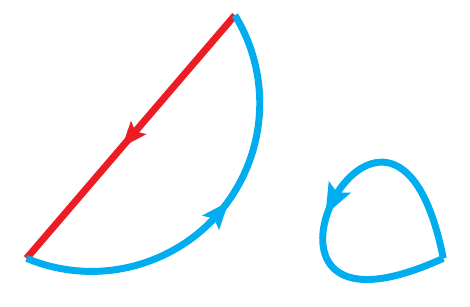}}}        
\newcommand{\discgtsll}{\raisebox{-0.4\totalheight}{\includegraphics[scale=.3]{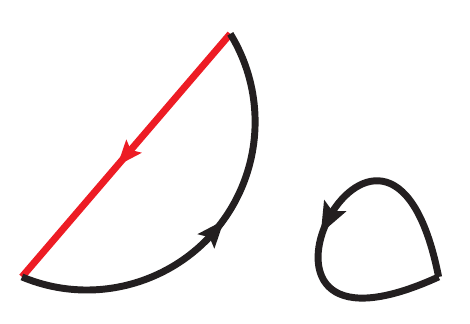}}}        
\newcommand{\discgtsil}{\raisebox{-0.4\totalheight}{\includegraphics[scale=.3]{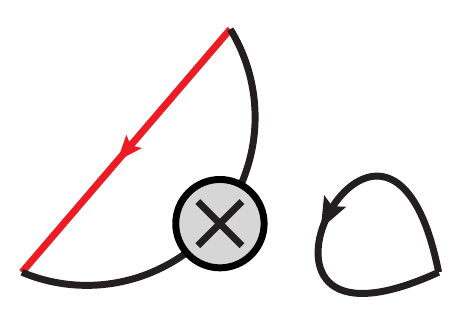}}}        
\newcommand{\discgtils}{\raisebox{-0.4\totalheight}{\includegraphics[scale=.3]{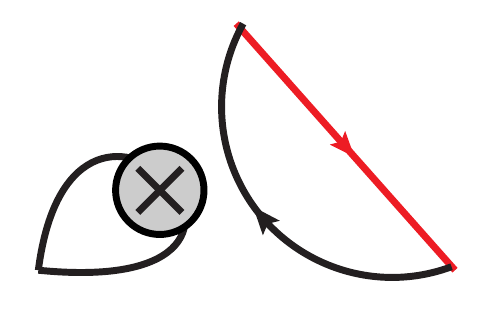}}}        
\newcommand{\discgtsli}{\raisebox{-0.4\totalheight}{\includegraphics[scale=.3]{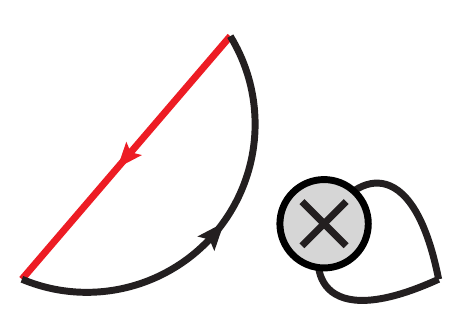}}}        
\newcommand{\ins}{\raisebox{-0.4\totalheight}{\includegraphics[scale=.3]{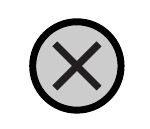}}}

\newcommand{\gtsli}{\raisebox{-0.3\totalheight}{\includegraphics[scale=.3]{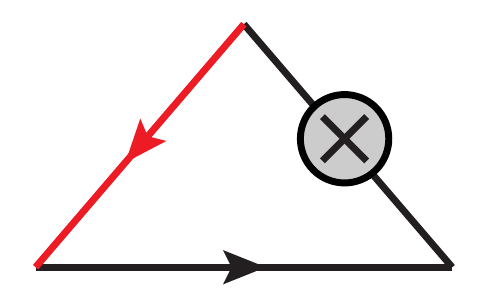}}}
\newcommand{\gtils}{\raisebox{-0.3\totalheight}{\includegraphics[scale=.3]{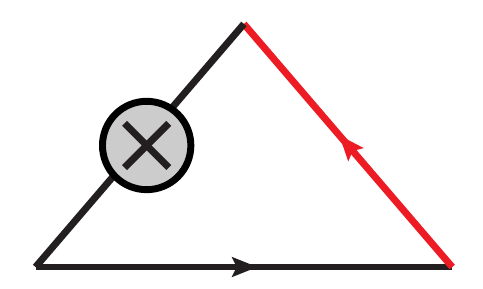}}}

\newcommand{\gtcus}{\raisebox{-0.3\totalheight}{\includegraphics[scale=.3]{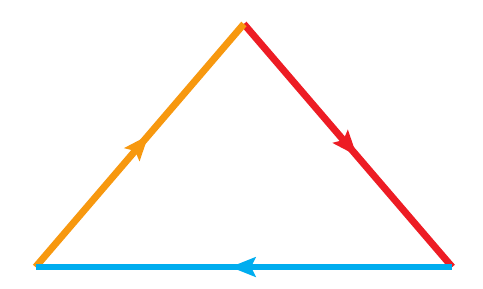}}}
\newcommand{\gtcds}{\raisebox{-0.3\totalheight}{\includegraphics[scale=.3]{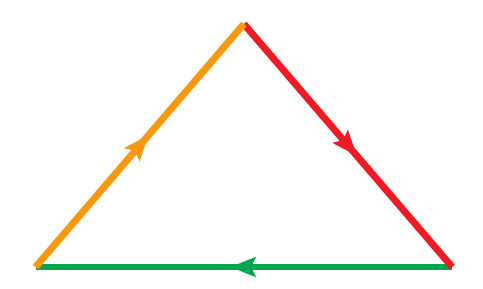}}}
\newcommand{\gtcls}{\raisebox{-0.3\totalheight}{\includegraphics[scale=.3]{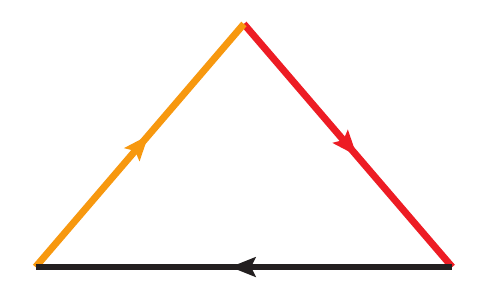}}}
\newcommand{\gtcis}{\raisebox{-0.3\totalheight}{\includegraphics[scale=.3]{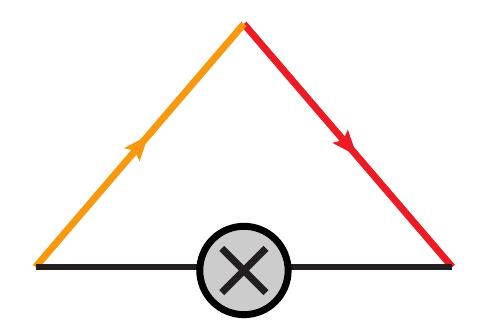}}}

\newcommand{\gtsdc}{\raisebox{-0.3\totalheight}{\includegraphics[scale=.3]{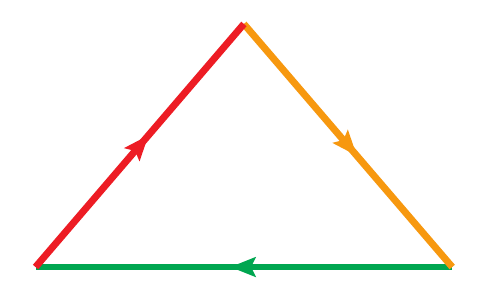}}}
\newcommand{\gtslc}{\raisebox{-0.3\totalheight}{\includegraphics[scale=.3]{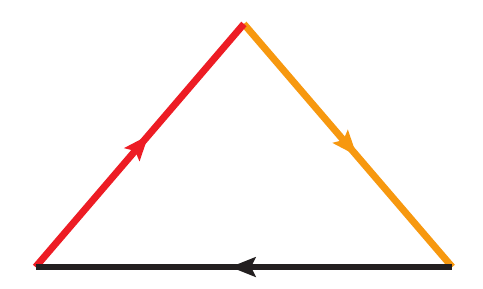}}}
\newcommand{\gtsic}{\raisebox{-0.3\totalheight}{\includegraphics[scale=.3]{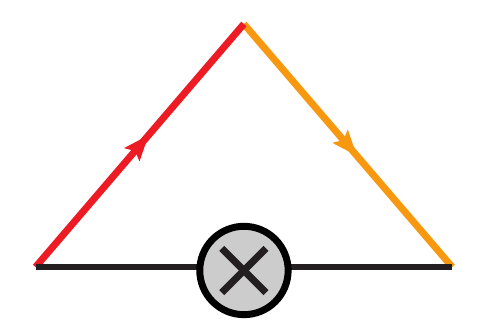}}}

\newcommand{\gtcdc}{\raisebox{-0.3\totalheight}{\includegraphics[scale=.3]{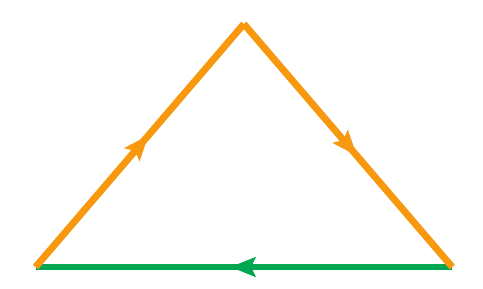}}}
\newcommand{\gtclc}{\raisebox{-0.3\totalheight}{\includegraphics[scale=.3]{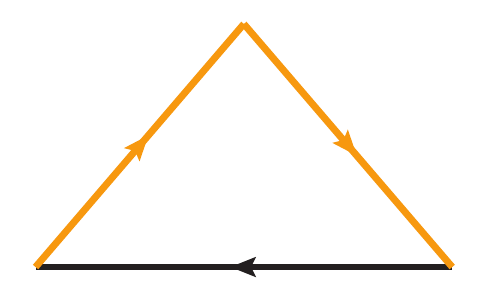}}}
\newcommand{\gtcic}{\raisebox{-0.3\totalheight}{\includegraphics[scale=.3]{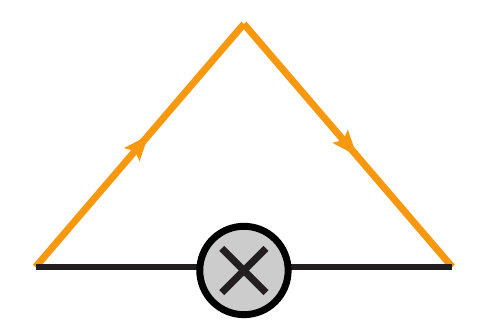}}}

\newcommand{\gtsuu}{\raisebox{-0.3\totalheight}{\includegraphics[scale=.3]{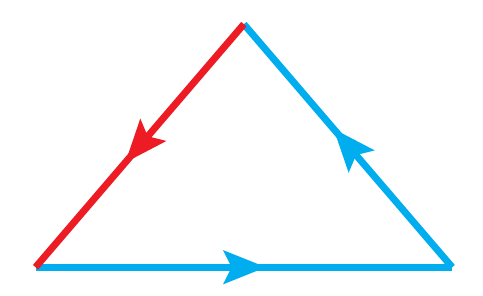}}}
\newcommand{\gtsil}{\raisebox{-0.3\totalheight}{\includegraphics[scale=.3]{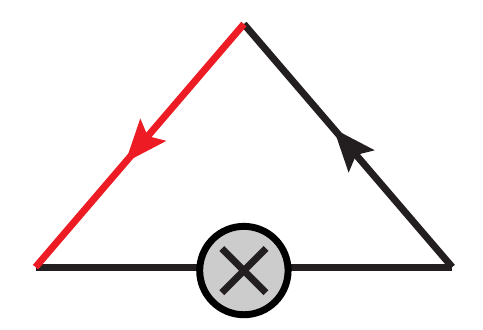}}}
\newcommand{\gtlis}{\raisebox{-0.3\totalheight}{\includegraphics[scale=.3]{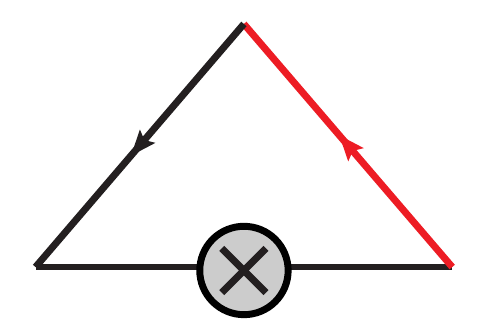}}}
\newcommand{\gtsll}{\raisebox{-0.3\totalheight}{\includegraphics[scale=.3]{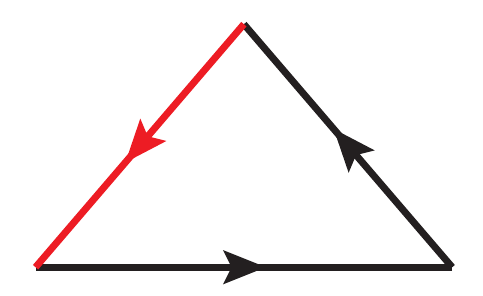}}}
\newcommand{\gtlls}{\raisebox{-0.3\totalheight}{\includegraphics[scale=.3]{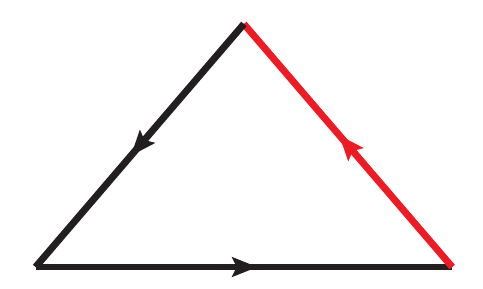}}}

\newcommand{\gtlll}{\raisebox{-0.3\totalheight}{\includegraphics[scale=.3]{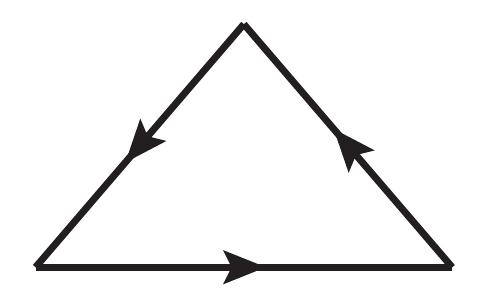}}}

\newcommand{\gtsdu}{\raisebox{-0.3\totalheight}{\includegraphics[scale=.3]{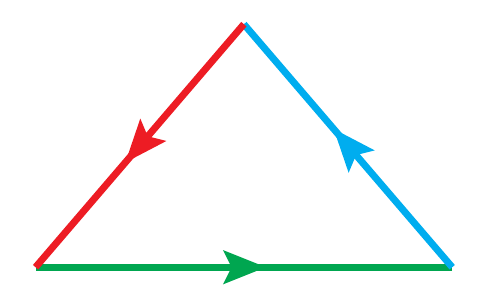}}}
\newcommand{\gtsls}{\raisebox{-0.3\totalheight}{\includegraphics[scale=.3]{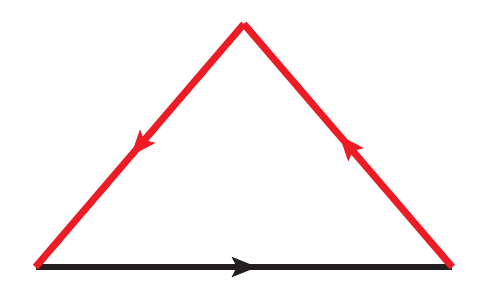}}}
\newcommand{\gtsis}{\raisebox{-0.3\totalheight}{\includegraphics[scale=.3]{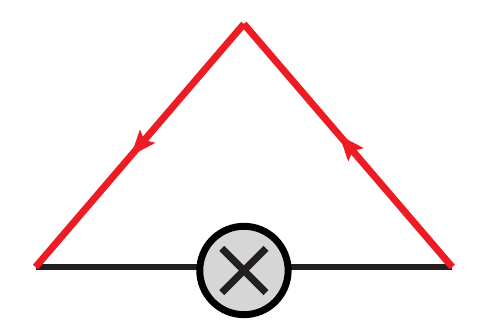}}}

\newcommand{\gtsds}{\raisebox{-0.3\totalheight}{\includegraphics[scale=.3]{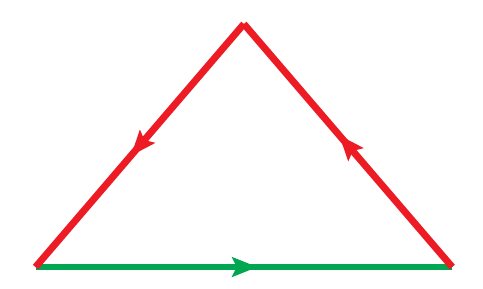}}}

\newcommand{\bear}[1]{\begin{equation}\begin{array}{#1}}
\newcommand{\eear}{ \end{array}\end{equation}}
\newcommand{\barr}[1]{\begin{array}{#1}}
\newcommand{\earr}{\end{array}}
 
\newcommand{\half}{\ensuremath{\frac{1}{2} }}

\newcommand{\ket}[1]{\ensuremath{| {#1} \rangle }}
\newcommand{\bra}[1]{\ensuremath{\langle {#1} |}}

\newcommand{\braket}[2]{\ensuremath{\langle {#1} | {#2} \rangle}}

\usepackage[title,titletoc]{appendix}

\begin{document}

\begin{titlepage}

\vspace{-1cm} 
\begin{flushright}\footnotesize
RM3-TH/11-14\\
RM1/1472 28/10/2011\\
ROM2F/2011/17\\
SISSA  58/2011/EP
\end{flushright}
\vspace{1cm}

\begin{center}
{\LARGE\sc Isospin breaking effects due to the up-down\\
\vskip 0.3cm 
mass difference in Lattice QCD} \\
\end{center}

\vspace{2cm}
\baselineskip 20pt plus 2pt minus 2pt

\begin{center}
{\sc \Large RM123 Collaboration}\\
\vspace{0.5cm}
{\it
G.M.~de Divitiis$^{(\rmii,\infntv)}$, 
P.~Dimopoulos$^{(\rmif,\rmiff)}$, 
R.~Frezzotti$^{(\rmii,\infntv)}$,
V.~Lubicz$^{(\rmiii,\infntre)}$,\\
G.~Martinelli$^{(\rmi,\rmiff)}$,  
R.~Petronzio$^{(\rmii,\infntv)}$, 
G.C.~Rossi$^{(\rmii,\infntv)}$, 
F.~Sanfilippo$^{(\rmif,\rmiff)}$,\\
S.~Simula$^{(\infntre)}$, 
N.~Tantalo$^{(\rmii,\infntv)}$, 
C.~Tarantino$^{(\rmiii,\infntre)}$}
\end{center}

\begin{center}
\begin{footnotesize}
\noindent \vspace{0.2cm}

$^{(\rmii)}$ Dip. di Fisica, Universit{\`a} di Roma ``Tor Vergata", 
Via della Ricerca Scientifica 1, I-00133 Rome, Italy

\vspace{-6pt}
$^{(\infntv)}$ INFN, Sez. di Roma ``Tor Vergata",
Via della Ricerca Scientifica 1, I-00133 Rome, Italy

\vspace{-6pt}
$^{(\rmif)}$ Dip. di Fisica, Universit\`a di Roma ``La Sapienza'',
   P.le A.~Moro 5, I-00185 Rome, Italy

\vspace{-6pt}
$^{(\rmiff)}$  INFN, Sezione di Roma, P.le A.~Moro 5, I-00185 Rome, Italy

\vspace{-6pt}
$^{(\rmiii)}$ Dip. di Fisica, Universit{\`a} Roma Tre, Via della Vasca Navale
84, I-00146 Rome, Italy

\vspace{-6pt}
$^{(\infntre)}$ INFN, Sez. di Roma III, Via della Vasca Navale 84, I-00146 Rome,
Italy

\vspace{-6pt}
$^{(\rmi)}$ SISSA - Via Bonomea 265 - 34136, Trieste - Italy 

\end{footnotesize}
\end{center}

\vspace{1.5cm}

\begin{abstract}
We present a new method to  evaluate  with high precision leading isospin breaking effects due to the small mass difference between the up and down quarks using lattice QCD. Our proposal is applicable in principle to any hadronic observable which can be computed on the lattice. It is based on the expansion of the path-integral in powers of the small parameter $m_d - m_u$. In this paper, we apply this method to compute the leading isospin breaking effects for several physical quantities of interest: the kaon meson masses, the kaon decay constant, the form factors of semileptonic $K_{\ell 3}$ decays and the neutron-proton mass splitting.
\end{abstract}

\end{titlepage}

\setlength\abovedisplayskip{22pt plus 3pt minus 7pt}
\setlength\belowdisplayskip{22pt plus 3pt minus 7pt}

\section{Introduction}
\label{sec:intro}
Isospin symmetry $SU(2)_V$, is an almost exact property of strong interactions as described by the  QCD Lagrangian. This happens because the difference between the up and down quark masses is much smaller than the QCD scale, $(m_d- m_u)/\Lambda_{QCD} \ll 1$, and it remains true also when electromagnetic interactions are switched on, because isospin breaking effects due to the different quark electric charges ($e_u \neq e_d$) are suppressed by the electromagnetic coupling constant, $\alpha_{em} \sim 1/137$. 
For these reasons most of theoretical predictions of several physical quantities assume isospin symmetry, i.e. the masses of the up and down quarks are taken equal and electromagnetic effects are neglected.  

Nowadays, with the increasing precision of the experimental determinations of many physical quantities, and in some cases with the improvement of the theoretical predictions, the control over isospin breaking effects is becoming phenomenologically relevant. This is the case, for example, of the form factors parametrizing $K_{\ell 3}$ decays.  Isospin breaking effects are important also for hadron spectroscopy,  for the meson decay constants, for the $\pi$-$\pi$ scattering length, for the quark condensate and for many other quantities.  

In the past, isospin breaking effects due to the light quarks mass difference (in the following referred to as QIB effects for \emph{QCD isospin breaking effects}) have been accommodated within the chiral perturbation theory ($\chi$pt) framework (see~\cite{Gasser:1984gg,Gasser:1984pr,Amoros:2001cp,Bijnens:2007xa,Kastner:2008ch,Cirigliano:2011tm} for a largely incomplete list of references on the subject), while several attempts to compute electromagnetic effects for the hadron spectroscopy in lattice QCD have been presented~\cite{Duncan:1996xy,Basak:2008na,Blum:2010ym,Portelli:2010yn}.  It is very difficult to take into account in numerical simulations QCD isospin breaking (see refs.~\cite{Tiburzi:2005na,Beane:2006fk,McNeile:2009mx,WalkerLoud:2009nf,WalkerLoud:2010qq} for a selection of previous lattice works on the subject) because the effect is in general rather small and comparable with the errors in the determination of, say, the  meson masses or decay constants. Furthermore, in order to perform unitary dynamical simulations of two light quarks of different mass the single quark determinant must be positive and this happens only in the case of lattice discretizations of the fermion action that are very expensive from the numerical point of view.

In this paper we present a new method to compute  the leading QIB effects with high precision. The method is based on the expansion of the lattice path-integral in powers of the small parameter $m_d - m_u$ and is applicable in principle to any hadronic observable which can be computed on the lattice. As a first application, and to show that the method works, we  have applied it to the computation of the leading QIB effects for several physical quantities of interest: the kaon meson masses, the kaon decay constants, the form factors of semileptonic $K_{\ell 3}$ decays and the neutron-proton mass splitting. In the future we plan to apply the method to other physical quantities, to include QED corrections and to try also the calculation of next-to-leading corrections such as the $\pi^+$-$\pi^0$ mass difference.

The  main physical results of this work are
\bea
\left[m_d-m_u\right]^{QCD}(\overline{MS},2 GeV) &=& 2.35(8)(24)\ \mbox{MeV} 
\quad \times \quad 
\frac{\left[M_{K^0}^2-M_{K^+}^2\right]^{QCD}}{6.05\times 10^3\ \mbox{MeV}^2} \, ,
\nonumber \\
\nonumber \\
\left[\frac{F_{K^+}/F_{\pi^+}}{F_{K}/F_{\pi}}-1\right]^{QCD} &=& -0.0039(3)(2)
\quad \times \quad 
\frac{\left[M_{K^0}^2-M_{K^+}^2\right]^{QCD}}{6.05\times 10^3\ \mbox{MeV}^2} \, ,
\nonumber \\
\nonumber \\
\left[M_n-M_p\right]^{QCD} &=& 2.8(6)(3)\ \mbox{MeV}
\quad \times \quad 
\frac{\left[M_{K^0}^2-M_{K^+}^2\right]^{QCD}}{6.05\times 10^3\ \mbox{MeV}^2} \, ,
\label{eq:mainresults}
\eea
and the QIB corrections to semileptonic decay rates discussed in sec.~\ref{sec:semileptonic}.
In previous equations the lattice error (the one in the first parenthesis) has been obtained with a rather modest statistics,  $\sim 150$ gauge field configurations, and can be reduced in the future. Most of the systematic error comes from the ambiguity in the definition of the electromagnetic corrections (see uncertainties in the second parenthesis). 
In this study we have extracted the $m_d-m_u$ mass difference in QCD by using as an external input the QIB effect in the kaon masses, $[M_{K^0}^2-M_{K^+}^2]^{QCD}$. This quantity requires in turn a definition of 
the electromagnetic isospin breaking corrections that we have taken from other calculations, as explained below in the paper, and that will be treated more properly in the future.
The result for the neutron-proton mass difference has been obtained at fixed lattice spacing and will be improved in a separate publication. 

Next-to-leading QIB corrections, i.e. second or higher orders in the $(m_u-m_d)$ expansion, have not been calculated in the present paper. The associated effects are estimated to be  negligible at the current level of (both theoretical and experimental) precision on flavour physics observables. 
Indeed, on the basis of dimensional arguments, higher orders corrections are expected to be suppressed by additional powers of the small expansion parameter $(m_d-m_u)/\Lambda_{QCD}$. The actual numerical size of the leading order contributions presented in eq.~(\ref{eq:mainresults}) is consistent with this power counting expectation. Clearly, we cannot exclude the existence of specific observable for which higher order QIB effects may be larger than expected. Nevertheless if necessary, the method suggested in the present paper can be extended to evaluate higher order contributions as discussed, for example, in sec.~\ref{sec:observables} for the $\pi^+$-$\pi^0$ mass splitting.

In the case of $M_K^2$ and $F_K$, we quote also the first order derivatives with respect to $m_d - m_u$, 
\bea
\frac{\left[M_{K^0}^2-M_{K^+}^2\right]^{QCD}}{[m_d-m_u]^{QCD}(\overline{MS},2GeV)}
&=& 2.57(8)\times 10^{3}\ \mbox{MeV}\, ,
\nonumber \\
\nonumber \\
\frac{\left[F_{K^0}/F_{K^+}-1\right]^{QCD}}{[m_d-m_u]^{QCD}(\overline{MS},2GeV)}
&=& 3.3(3)\times 10^{-3}\ \mbox{MeV}^{-1}\, ,
\eea
so that  one can use his/her preferred value of the up-down mass difference to get the physical QIB effect.

The paper is organized as follows: in sec.~\ref{sec:method} we introduce the method and set our notation. In sec.~\ref{sec:observables} we derive the correlation functions needed to extract isospin breaking effects whose calculation is discussed in detail in the remaining sections. In particular, in sec.~\ref{sec:kaontwopoint} we study kaon two point correlation functions and in sec.~\ref{sec:electromagnetic} we discuss electromagnetic isospin breaking effects and extract $m_d-m_u$ and $F_{K^0}-F_{K^+}$. In sec.~\ref{sec:nucleons} we calculate the neutron-proton mass difference while in sec.~\ref{sec:semileptonic} we discuss the calculation of isospin breaking effects for the $K_{\ell 3}$ form factors. Our conclusions and outlooks are given in sec.~\ref{sec:outlooks}. Some technical details are discussed in the appendices.

\section{Description of the method}
\label{sec:method}
In this section we present the basic ingredients of our method, which is simply based on a perturbative expansion in the small parameter $(m_d- m_u)/\Lambda_{QCD}$. Let us start by considering the evaluation of a generic euclidean correlation function $\langle {\cal O}\rangle $ used to extract information about physical quantities as masses, decay constants, form factors etc.,
\bea 
\langle {\cal O}\rangle = 
\frac{\int{ D\phi \ {\cal O} \, e^{-S}} }{\int{ D\phi \ e^{-S} }}\, , 
\label{eq:fi}
\eea 
where $D\phi$ represents synthetically the full functional integration measure of the theory. By neglecting for the moment electromagnetic corrections and possible isospin breaking terms that may arise because of lattice artifacts with particular choices of the lattice fermion action, we can write the Lagrangian density as a term which is $SU(2)_V$ symmetric plus a  term which violates  the isospin symmetry.
\bea 
{\cal L}  &=& {\cal L}_{kin} + {\cal L}_{m}  
\nonumber \\
\nonumber \\
&=& {\cal L}_{kin}  + \frac{m_u+m_d}{2}(\bar u u + \bar d d) -  
\frac{m_d-m_u}{2} (\bar u u - \bar d d)   
\nonumber \\
\nonumber \\
&=& {\cal L}_{kin}  + m_{ud}\, \bar q q -  
\Delta m_{ud}\, \bar q \tau^3 q   
\nonumber \\
\nonumber \\
&=& {\cal L}_{0} - \Delta m_{ud} \, \hat {\cal L}\, , 
\label{eq:lag} 
\eea 
where $q^T= (u,d)$, $m_{ud}=(m_d+m_u)/2$ and $\Delta m_{ud}=(m_d-m_u)/2$.
By expanding  at first order the exponential of the action, $S=\sum_x {\cal L}(x)$, with respect to $\Delta m_{ud}$ we obtain
\bea 
\langle {\cal O}\rangle 
&\simeq& 
\frac{\int{ D\phi \ {\cal O}\, (1+  \Delta m_{ud} \, \hat S)\, e^{-S_0} }}
{\int{ D\phi \ \, (1+  \Delta m_{ud} \, \hat S) \, e^{-S_0} }}
= 
\frac{\langle {\cal O}\rangle_0 +  \Delta m_{ud} \, \langle {\cal O}\hat S\rangle_0 }
{1+ \Delta m_{ud} \, \langle \hat S\rangle_0 }\, 
\nonumber \\
\nonumber \\
&=& 
\langle {\cal O}\rangle_0 +  \Delta m_{ud} \, \langle {\cal O}\hat S\rangle_0 \, ,
\label{eq:basicofthemethod}
\eea
where $\langle \cdot \rangle_0$ represent the vacuum expectation value in the isospin symmetric theory and $\hat S$ is the isospin breaking term,
\bea
\hat S=\sum_x{[\bar q \tau_3 q](x)}=\sum_x{[\bar u u - \bar d d](x)} \, .
\eea
The correction in the denominator vanishes, $\langle \hat S\rangle_0=0$, because of isospin symmetry. Concerning the Wick contractions of the correlation functions $ \langle {\cal O}\hat S\rangle_0$, isospin symmetry makes also to vanish some fermionic disconnected contributions of the form
\bea 
\Delta m_{ud} \, \langle\ \left[ \mbox{fermionic Wick contractions of }{\cal O} \right]\times \mbox{tr}[\hat S]\ \rangle_0 = 0 \, ,
\label{eq:nodet}
\eea
since these are proportional to the trace of the flavour matrix $\tau_3$. 
We can now describe a general recipe to be used in order to compute leading QIB effects on the lattice:
\begin{itemize}
\item consider a given correlation function in the full theory, i.e. with $m_u\neq m_d$, and for each gauge configuration draw all the fermionic Wick contractions;
\item expand the up and down quark propagators with respect to $\Delta m_{ud}$ according to
\bea
G_u(x_1,x_2) &=& G_\ell(x_1,x_2) + \Delta m_{ud} \ \sum_y{G_\ell(x_1,y)\ G_\ell(y,x_2)}+\cdots\, ,
\nonumber \\
G_d(x_1,x_2) &=& G_\ell(x_1,x_2) - \Delta m_{ud} \ \sum_y{G_\ell(x_1,y)\ G_\ell(y,x_2)}+\cdots\, ;
\label{eq:propsexpformulae}
\eea
\item retain the terms linear in $\Delta m_{ud}$ and compute the corresponding diagrams (or fermionic Wick contractions).
\end{itemize}
In the following sections we shall discuss in detail how to extract physical information from the resulting correlation functions. To this end we need to set the notation we are going to use in drawing diagrams. Eqs.~(\ref{eq:propsexpformulae}) can be represented diagrammatically as follows
\bea
&&\overset{u} {\gou} = \gol + \goi  + \cdots \, ,
\nonumber \\
&&\overset{d} {\god} = \gol - \goi  + \cdots \, ,
\eea
where here and in the following the up quark line in the full theory is drawn in light blue color while the down quark line in green. All the black lines refer to $G_\ell$, the propagator with the symmetric mass $m_{ud}$ in the isospin symmetric theory whose square, entering eqs.~(\ref{eq:propsexpformulae}), can be easily calculated on the lattice by using $G_\ell$ itself as the source vector of a new inversion.

The insertion of the scalar density is represented by a cross according to
\bea
_y\gol_x &=& G_\ell(x-y) = \langle \ell(x)\bar \ell(y) \rangle \, ,
\nonumber \\
\nonumber \\
\ins 
& = & \Delta m_{ud} \,  \sum_z{\bar \ell(z) \ell (z)}
\ = \ \Delta m_{ud}^L \,  \sum_z{\left[\bar \ell(z) \ell (z)\right]^L}
\, ,
\eea
with $\ell$ either $u$ or $d$. Here and in the following the superscript $L$ stays for \emph{bare lattice quantity}. In particular we have
\bea
\Delta m_{ud}=Z_{\Delta m}\ \Delta m_{ud}^L
\eea
where $Z_{\Delta m}$ is scale and scheme dependent while the combination $\Delta m_{ud} \,  \sum_z{\bar \ell(z) \ell (z)}$ is renormalization group invariant. According to eq.~(\ref{eq:basicofthemethod}), a generic correlator $\langle \mathcal{O} \hat S \rangle_0$ with a single insertion of the isospin breaking term  can be obtained from $\langle \mathcal{O} \rangle$, a correlator in the full theory, as the derivative of the latter with respect to $\Delta m_{ud}$ evaluated at $\Delta m_{ud}=0$. It follows that, by working within a mass independent renormalization scheme, $\langle \mathcal{O} \hat S \rangle_0$ is finite provided that $\mathcal{O}$ and $\hat S$ have been separately renormalized. By iterating the previous argument it can be easily understood that the connected parts of correlators with multiple insertions of the renormalized operator $\hat S$ are finite.

In this paper we have applied the method discussed above by using the so called Twisted Mass lattice discretization of the QCD action. This choice has advantages and drawbacks. The big advantage is automatic $O(a)$ improvement. The drawback is the breaking of isospin symmetry at finite lattice spacing even with $\Delta m_{ud}=0$.
The associated $O(a^2)$ cutoff effects are eliminated by performing continuum extrapolations. 
Since our method is general and can be applied with any lattice regularization of the quark action (e.g. Wilson, Overlap, etc.), in the main body of the paper we illustrate our results without entering into the specific details of the Twisted Mass fermion action that we discuss in Appendix~\ref{sec:tm}.

We close this section by explaining the notation used in the following to express and calculate variations of correlation functions and matrix elements. To this end it is useful to introduce the following operators acting as absolute variations
\bea
&&\Delta_f \mathcal{O}= \mathcal{O}(f)-\mathcal{O}(i)\, ,
\nonumber \\
\nonumber \\
&&\Delta_b \mathcal{O}= \mathcal{O}(i)-\mathcal{O}(b)\, ,
\nonumber \\
\nonumber \\
&&\Delta \mathcal{O}= \frac{\Delta_f\mathcal{O}+\Delta_b\mathcal{O}}{2}= \frac{\mathcal{O}(f)-\mathcal{O}(b)}{2}\, ,
\eea
and the corresponding relative variations
\bea
\delta_f \mathcal{O}= \frac{\Delta_f\mathcal{O}}{\mathcal{O}(i)}\, ,
\qquad \qquad
\delta_b \mathcal{O}= \frac{\Delta_b\mathcal{O}}{\mathcal{O}(i)}\,,
\qquad \qquad
\delta \mathcal{O}= \frac{\Delta\mathcal{O}}{\mathcal{O}(i)} \, .
\eea
Here $\mathcal{O}(f)$ and $\mathcal{O}(b)$ are quantities calculated with the light quark propagators at first order in $\Delta m_{ud}$,  while $\mathcal{O}(i)$ is the corresponding quantity calculated in the unperturbed isospin symmetric theory. The labels $f$ (forward), $b$ (backward) and $i$ (iso-symmetric) are generic. More precisely, $f$ will stand for
\bea
f&=&\left\{\
d, \
K^0,\ 
n,\ 
D^+K^0,\ 
K^0\pi^-\ 
\right\}
\eea 
when we shall discuss in turn quark masses, kaon masses and decay constants, the neutron and proton masses, the semileptonic decay $D\rightarrow K\ell\nu$ and the semileptonic decay
$K\rightarrow \pi\ell\nu$. Correspondingly $b$ and $i$ will represent
\bea
b&=&\left\{\
u, \
K^\pm,\ 
p,\ 
D^0K^+,\ 
K^+\pi^0\ 
\right\} \, ,
\nonumber \\
\nonumber \\
i&=&\left\{\
\ell, \
K,\ 
N,\ 
DK,\ 
K\pi\ 
\right\}\, ,
\eea 
where $N$ stands for nucleon.

\section{Correlation functions at first order}
\label{sec:observables}
In this section we shall derive the correlation functions that need to be calculated in order to extract the leading QIB corrections to meson masses and decay constants, nucleon masses, and to the form factors parametrizing semileptonic meson decays. In particular, we shall consider the following two point correlation functions 
\bea
&&
C_{\pi^+\pi^-}(t,\vec{p})=\sum_{\vec x}
e^{-i\vec{p}\cdot \vec{x}}
\langle\, \bar u \gamma_5 d(x)\ \bar d \gamma_5 u(0)\, \rangle
\ ,
\nonumber \\
\nonumber \\
&&
C_{\pi^0\pi^0}(t,\vec{p})=
\frac{1}{2}\sum_{\vec x}
e^{-i\vec{p}\cdot \vec{x}}
\langle\, 
(\bar u \gamma_5 u -\bar d \gamma_5 d)(x)\ 
(\bar u \gamma_5 u -\bar d \gamma_5 d)(0)
\, \rangle \, ,
\nonumber \\
\nonumber \\
&&
C_{K^+K^-}(t,\vec{p})=
\sum_{\vec x}
e^{-i\vec{p}\cdot \vec{x}}
\langle\, \bar u \gamma_5 s(x)\ \bar s \gamma_5 u(0)\, \rangle
\ ,
\nonumber \\
\nonumber \\
&&
C_{K^0K^0}(t,\vec{p})=
\sum_{\vec x}
e^{-i\vec{p}\cdot \vec{x}}
\langle\, 
\bar d \gamma_5 s(x)\ 
\bar s \gamma_5 d(0)
\, \rangle \, ,
\nonumber \\
\nonumber \\
&&
C_{pp}^\pm(t,\vec{p})=
\sum_{\vec x}
e^{-i\vec{p}\cdot \vec{x}}
\langle\, 
\left [\epsilon_{abc}(\bar u_a C\gamma_5 \bar d^T_b ) 
\bar u_c  \frac{1\pm \gamma^0}{2}\right](x)\, 
\left [\epsilon_{def}\frac{1\pm \gamma^0}{2} u_d
(u^T_e C\gamma_5 d_f ) \right](0)
\, \rangle \, ,
\nonumber \\
\nonumber \\
&&
C_{nn}^\pm(t,\vec{p})=
\sum_{\vec x}
e^{-i\vec{p}\cdot \vec{x}}
\langle\, 
\left [\epsilon_{abc}(\bar d_a C\gamma_5 \bar u^T_b ) 
\bar d_c  \frac{1\pm \gamma^0}{2}\right](x)\, 
\left [\epsilon_{def}\frac{1\pm \gamma^0}{2} d_d
(d^T_e C\gamma_5 u_f ) \right](0)
\, \rangle \, ,
\label{eq:twopoint}
\eea
and the following three point correlation functions
\bea
&&
C_{D^0K^+}^\mu(t;\vec{p}_D,\vec{p}_K)=\sum_{\vec x, \vec y}
e^{-i\vec{p}_K\cdot \vec{x}}e^{-i\vec{p}_D\cdot (\vec{x}-\vec{y})}
\langle\,
\bar u \gamma^5 c(\vec y, T/2)\
\bar c \gamma^\mu s(\vec x, t)\
\bar s \gamma^5 u(0)
\, \rangle \, ,
\nonumber \\
\nonumber \\
&&
C_{D^+K^0}^\mu(t;\vec{p}_D,\vec{p}_K)=\sum_{\vec x, \vec y}
e^{-i\vec{p}_K\cdot \vec{x}}e^{-i\vec{p}_D\cdot (\vec{x}-\vec{y})}
\langle\,
\bar d \gamma^5 c(\vec y, T/2)\
\bar c \gamma^\mu s(\vec x, t)\
\bar s \gamma^5 d(0)
\, \rangle \, ,
\nonumber \\
\nonumber \\
&&
C_{K^0\pi^-}^\mu(t;\vec{p}_K,\vec{p}_\pi)=\sum_{\vec x, \vec y}
e^{-i\vec{p}_\pi\cdot \vec{x}}e^{-i\vec{p}_K\cdot (\vec{x}-\vec{y})}
\langle\,
\bar d \gamma^5 s(\vec y, T/2)\
\bar s \gamma^\mu u(\vec x, t)\
\bar u \gamma^5 d(0)
\, \rangle \, ,
\nonumber \\
\nonumber \\
&&
C_{K^+\pi^0}^\mu(t;\vec{p}_K,\vec{p}_\pi)=\sum_{\vec x, \vec y}
e^{-i\vec{p}_\pi\cdot \vec{x}}e^{-i\vec{p}_K\cdot (\vec{x}-\vec{y})}
\langle\,
\bar u \gamma^5 s(\vec y, T/2)\
\bar s \gamma^\mu u(\vec x, t)\
(\bar u \gamma^5 u - \bar d \gamma^5 d)(0)
\, \rangle \, .
\eea
In previous expressions $C$ denotes the charge conjugation matrix while $\epsilon_{abc}$ the totally antisymmetric tensor in color space.

A first trivial observation comes from eq.~(\ref{eq:nodet}) telling us that all the quantities that do not involve a light valence quark propagator do not get corrected at first order in $\Delta m_{ud}$. This is the case for example of heavy-heavy and heavy-strange meson masses and decay constants, etc. 
Pion masses and decay constants too do not get corrected at first order. This can be shown diagrammatically for the charged pions two point function
\bea
C_{\pi^+\pi^-}(t)&=&-\overset{u} {\underset{d} {\gdud}} =-\gdll - \gdil + \gdli + \cdots 
\nonumber \\
\nonumber \\
&=& -\gdll
+{\cal O}(\Delta m_{ud})^2\, ,
\eea
and for the connected diagrams entering neutral pion two point function $C_{\pi^0\pi^0}(t)$, 
\bea
&&\overset{u} {\underset{u} {\gduu}} =\gdll + \gdil + \gdli + \cdots =   \gdll + 2 \gdil + {\cal O}(\Delta m_{ud})^2\, ,
\nonumber  \\
&&\overset{d} {\underset{d} {\gddd}} = \gdll - \gdil - \gdli + \cdots =  \gdll - 2 \gdil  +{\cal O}(\Delta m_{ud})^2\, ,
\nonumber \\
\nonumber \\
\nonumber \\
&&C_{\pi^0\pi^0}(t)=-\half \left[\overset{u} {\underset{u} {\gduu}}+  \overset{d} {\underset{d} {\gddd}}\right]
= -\gdll +{\cal O}(\Delta m_{ud})^2\,  .
\eea
It is easy to show that the first order corrections cancel also for the disconnected diagrams contributing to $C_{\pi^0\pi^0}(t)$ in the full theory, a known result that can be understood in terms of isospin quantum numbers. The Wigner-Eckart reduced matrix element of the operator $\hat S$  between pion states is indeed zero for $G$-parity,
\bea
\langle\pi\| \hat  S \|\pi\rangle  = \langle 1,I_3\| 1,0 \| 1,I_3\rangle  = 0\, .
\eea
This is certainly not the case at second order where the relevant ${\cal O}(\Delta m_{ud})^2$ diagrams are
\bea
C_{\pi^0\pi^0}(t)-C_{\pi^+\pi^-}(t)=-2\left[
\gdii-\discgdii
\right]+{\cal O}(\Delta m_{ud})^3\, .
\label{eq:pionssecondorder}
\eea
 
For flavoured mesons first order corrections to masses and decay constants are instead different from zero. Here we discuss the case of strange particles but the discussion proceeds unchanged if the strange is replaced with a charm or a bottom quark. The QIB correction to the two point correlation functions of the strange mesons are
\bea
C_{K^+K^-}(t) &=& -\overset{s} {\underset{u} {\gdsu}}=   -\gdsl -   \gdsi + {\cal O}(\Delta m_{ud})^2\, ,
\nonumber \\
\nonumber \\
C_{K^0K^0}(t) &=& -\overset{s} {\underset{d} {\gdsd}}= -  \gdsl +   \gdsi + {\cal O}(\Delta m_{ud})^2  \, .   
\label{eq:kpcorr}
\eea
In the diagrams above and in the following the strange quark line is red.
Note that the correction to the neutral mesons is equal in magnitude to that to the charged particles,  
$\Delta_f C_{KK}(t)=\Delta_b C_{KK}(t)= \Delta C_{KK}(t)$.

We now consider first order corrections in the case of nucleon masses. The neutron--proton mass difference can be extracted at first order in $\Delta m_{ud}$ by the diagrammatic analysis of $C_{nn}^\pm(t)$ and $C_{pp}^\pm(t)$,
\bea
&&C_{nn}^\pm(t) = - \neutrond+\neutronc\, ,
\nonumber \\
\nonumber \\
\nonumber \\
&&C_{pp}^\pm(t) = - \protond+\protonc\, ,
\nonumber \\
\nonumber \\
\nonumber \\
\nonumber \\
&&\neutrond-\protond=2\left[
\protondill-\protondlil-\protondlli
\right] + {\cal O}(\Delta m_{ud})^2\, ,
\nonumber \\
\nonumber \\
\nonumber \\
\nonumber \\
&&\neutronc-\protonc=2\left[
\protoncill-\protonclil-\protonclli
\right] + {\cal O}(\Delta m_{ud})^2\, .
\nonumber \\
\nonumber \\
\label{eq:nucleondia}
\eea
As usual this is obtained by expanding all the light quark propagators appearing into the correlation functions. Also in this case we find 
$\Delta_f C_{NN}(t)=\Delta_b C_{NN}(t)=\Delta C_{NN}(t)$.

Concerning the form factors parametrizing semileptonic decays, we start by considering a charmed meson decaying into a strange meson. The discussion would proceed along the same lines in the cases of $B\rightarrow D$ transitions. The charm quark line is drawn in yellow. We get
\bea
C^\mu_{D^0K^+}(t)&=& -\underset{u}{ \sideset{^c}{^s}{\operatorname{ \gtcus} }}= 
-\gtcls  -\gtcis + {\cal O}(\Delta m_{ud})^2 \, ,
\nonumber \\ 
\nonumber \\ 
C^\mu_{D^+K^0}(t)&=&-\underset{d}{ \sideset{^c}{^s}{\operatorname{ \gtcds} }}= 
-\gtcls  +\gtcis +{\cal O}(\Delta m_{ud})^2 \, .
\label{eq:dtokcorr}
\eea
As for the correlation functions analyzed above
the correction is equal in magnitude between the two processes, because the weak flavour changing current does not contain a light quark field.
From the previous two equations  one can extract $f_+^{D^0K^+}(q^2)-f_+^{D^+K^0}(q^2)$.  At present the experimental and theoretical uncertainties on this quantity are such that QIB effects can be safely neglected but we have chosen to discuss these processes first because of their simplicity. Indeed, the  phenomenologically relevant $K\rightarrow\pi$ case involves disconnected diagrams\footnote{This is not in contradiction with eq.~(\ref{eq:nodet}) since disconnected diagrams arise by making the fermionic Wick contractions of the observable $\cal{ O}$ and not from $\mbox{Tr}[\hat S]$.} that complicate the analysis.

The expansion of $C^\mu_{K^0\pi^-}(t)$ is given by
\bea
C^\mu_{K^0\pi^-}(t)=-\underset{d}{ \sideset{^s}{^u}{\operatorname{ \gtsdu} }}= 
-\gtsll  +   \gtsil -   \gtsli+{\cal O}(\Delta m_{ud})^2\, .
\label{eq:k0pim}
\eea
The correction in the $K^+ \rightarrow \pi^0 l^+ \nu$ case is obtained from the correlation function $C^\mu_{K^+\pi^0}(t)$ in the full theory, 
whose disconnected diagrams survive at first order in $\Delta m_{ud}$,
\bea
C^\mu_{K^+\pi^0}(t)&=&
-\underset{u}{ \sideset{^s}{^u}{\operatorname{ \gtsuu} }}\ + \
\underset{u}{ \sideset{^s}{^u}{\operatorname{ \discgtsuu} }}\ -\
\underset{u}{ \sideset{^s}{^d}{\operatorname{ \discgtsud} }}
\nonumber \\
\nonumber \\
&=&
-\gtsll +\discgtsll-\discgtsll
\nonumber \\
&&-\gtsil-\gtsli+\discgtsli
\nonumber \\
&&+\discgtsil-\discgtsil+\discgtsli
\nonumber \\
\nonumber \\
\nonumber \\
&=&-\gtsll-\gtsil-\gtsli+2\discgtsli 
+ {\cal O}(\Delta m_{ud})^2\, .
\label{eq:kppi0}
\eea
The correction to $K^+ \rightarrow \pi^0 l^+ \nu$ is not just equal to the $K^0 \rightarrow \pi^- l^+ \nu$ one.
From eqs.~(\ref{eq:kppi0}) and (\ref{eq:k0pim}),
one can extract $f_+^{K^0\pi^-}(q^2)-f_+^{K^-\pi^0}(q^2)$.
At $q^2=0$ this quantity has been estimated by using the measured form factors and chiral perturbation theory~\cite{Antonelli:2010yf},
\bea
\left[ \frac{f_+^{K^0\pi^-}(0)-f_+^{K^+\pi^0}(0) }{f_+^{K^0\pi^-}(0)}\right]^{\chi pt}
= -0.029 \pm 0.004 \, , 
\label{eq:expdeltafkpi}
\eea
and found of the same size  of the deviation from unity of the form factor at $q^2=0$, being $f_+^{K\pi}(0)=f_0^{K\pi}(0)=1$ the value at the $SU(3)_V$ symmetric point ($M_K=M_\pi$).

\section{Kaon masses and decay constants}
\label{sec:kaontwopoint}
In this section we discuss in detail the strategy used to derive the isospin corrections to the kaon masses and decay constants. To this end we start by considering the Euclidean correlation functions of eqs.~(\ref{eq:kpcorr}) both in the full  theory and in the isospin symmetric one.
The spectral decomposition of $C_{K^0K^0}$ (the analysis of $C_{K^+K^-}$ proceeds along similar lines) is
\bea 
C_{K^0K^0}(\vec p, t) &=& 
\sum_{\vec x} 
e^{-i \vec p \cdot \vec x } 
\langle \bar d \gamma_5 s(\vec x,t)\, \bar s \gamma_5 d(0)\rangle \, 
=
\sum_n{ 
\frac{\bra{0} \bar d \gamma_5 s(0) \ket{n^{\Delta}}\, 
\bra{n^{\Delta}} \bar s \gamma_5 d(0) \ket{0}}{2E_n^\Delta}\ 
e^{-E_n^{\Delta}t}}
\nonumber \\
\nonumber \\
&=& \frac{G_{K^0}^2}{2E_{K^0}}\ e^{-E_{K^0}t}+\cdots\, ,
\label{eq:spectral}
\eea
where the dots represent sub leading exponentials and
where $\ket{n^{\Delta}}$ and $E_n^{\Delta}$ are the states and the eigenvalues of the perturbed theory corresponding respectively to $\ket{n}$ and $E_n$ in the isospin symmetric unperturbed theory. These are related at first order in perturbation theory with respect to $\Delta m_{ud}$ according to
\bea
&&\ket{K^0} \equiv \ket{K^\Delta}= \ket{K} + \ket{\Delta K} + {\cal O}(\Delta m_{ud})^2\, ,
\nonumber \\
\nonumber \\
&&E_{K^0} \equiv E_K^\Delta= E_K+\Delta E_K + {\cal O}(\Delta m_{ud})^2\, .
\label{eq:ptformuale}
\eea
Explicit expressions for $\Delta E_K$ and $\ket{\Delta K}$ are derived in Appendix~\ref{sec:proof}. 
By substituting eqs.~(\ref{eq:ptformuale}) into eq.~(\ref{eq:spectral}) and by recalling the diagrammatic analysis of eqs.~(\ref{eq:kpcorr}) we obtain
\bea
C_{KK}(\vec p, t)= -\gdsl &=& \frac{G_K^2}{2E_K}\ e^{-E_Kt}+\cdots \, ,
\nonumber
\\
\Delta C_{KK}(\vec p, t)= \gdsi
&=& 
\frac{G_K^2}{2E_K}\ e^{-E_Kt}
\left[
\frac{\Delta (G_K^2/2E_K)}{G_K^2/2E_k}-t\Delta E_K
\right]
+\cdots \, ,
\label{eq:ckkone}
\eea
where $\Delta G_K=\bra{0} \bar s \gamma_5 \ell(0) \ket{\Delta K}$.
Note that the insertion of the QIB term $\hat {\mathcal L}$ constitutes a flavour diagonal perturbation and that, consequently, the kaons are the lightest states contributing both to $C_{KK}(\vec p, t)$ and to $\Delta C_{KK}(\vec p, t)$. 
The analysis would be considerably more complicated in the case of a perturbation (typically insertions of the weak hamiltonian) opening a decay channel for the kaons because the physical information would be hidden into sub-leading exponential terms.

In our case, by studying the ratio of the two correlators of eqs.~(\ref{eq:ckkone}),
\bea
\delta C_{KK}(\vec p, t) 
=\frac{\Delta C_{KK}(\vec p, t)}{C_{KK}(\vec p, t)}
=-\frac{\gdsi}{\gdsl}
=
\delta\left(\frac{G_K^2}{2E_K}\right)-t\Delta E_K
+\cdots \, ,
\label{eq:ckkratiocontinuum}
\eea
it is possible to extract the leading QIB corrections to kaon energies and decay constants. Indeed $\Delta E_K$ appears directly in the previous equation as the ``slope" with respect to $t$ whereas   $\delta F_K$ can be extracted from the ``intercept" according to
\bea
F_K&=&(m_s+m_{ud})\frac{G_K}{M_K^2} \, ,
\nonumber \\
\nonumber \\
\delta F_K &=& \frac{\Delta m_{ud}}{m_s+m_{ud}} + \delta G_K -2 \delta M_K \, .
\eea  
On a lattice of finite time extent $T$ with quark fields satisfying anti-periodic boundary conditions along the time direction and given our choice of the kaon source and sink operators, the pseudoscalar densities, eq.~(\ref{eq:ckkratiocontinuum}) has to be modified according to
\bea
\delta C_{KK}(\vec p, t)
=\delta\left(\frac{G_K^2 e^{-E_KT/2}}{2E_K}\right)+\Delta E_K (t-T/2)\tanh\left[E_K(t-T/2) \right]
+\cdots \, .
\label{eq:ckkratiolattice}
\eea

\begin{figure}[!t]
\begin{center}
\includegraphics[width=0.45\textwidth]{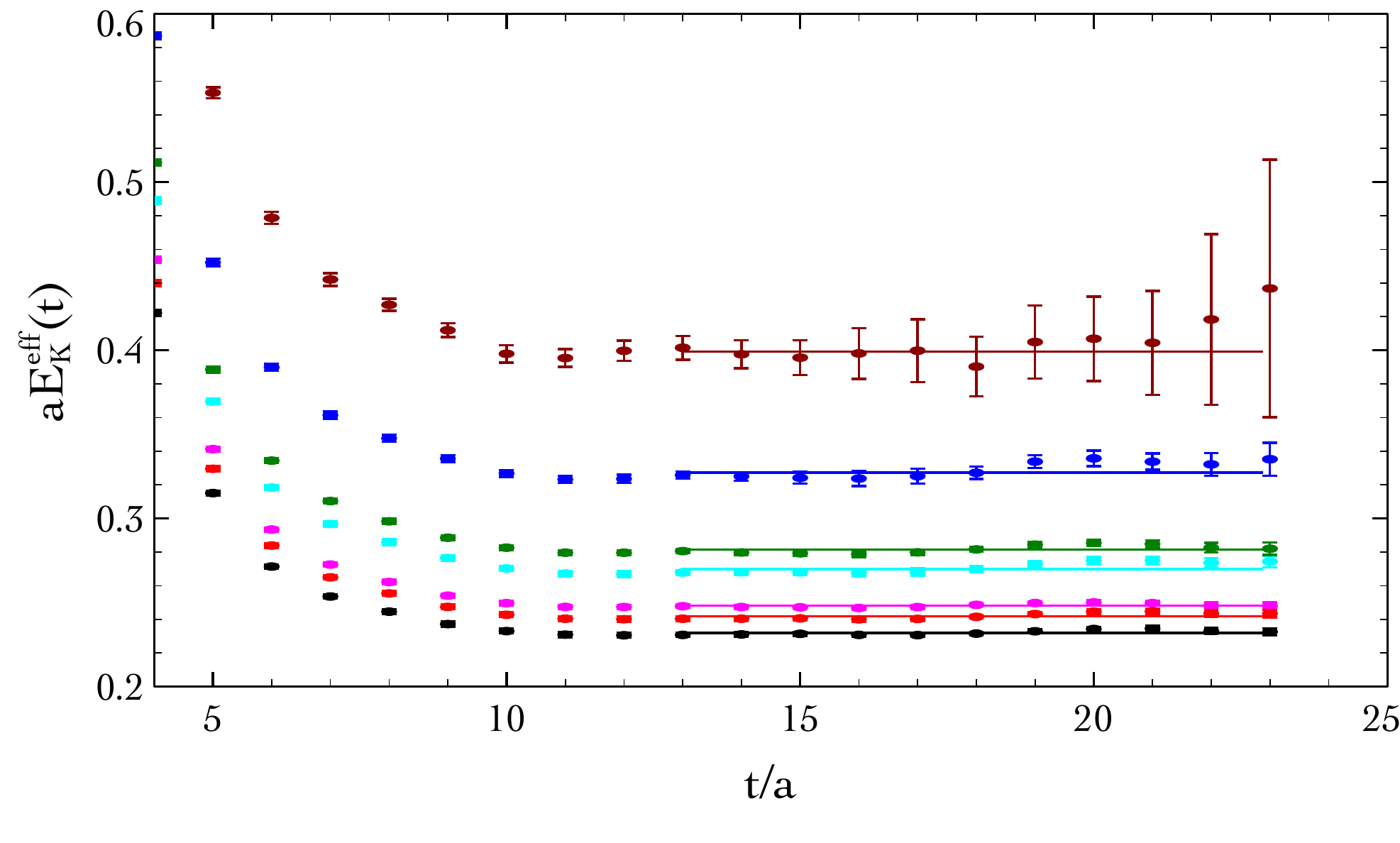}\hfill
\includegraphics[width=0.45\textwidth]{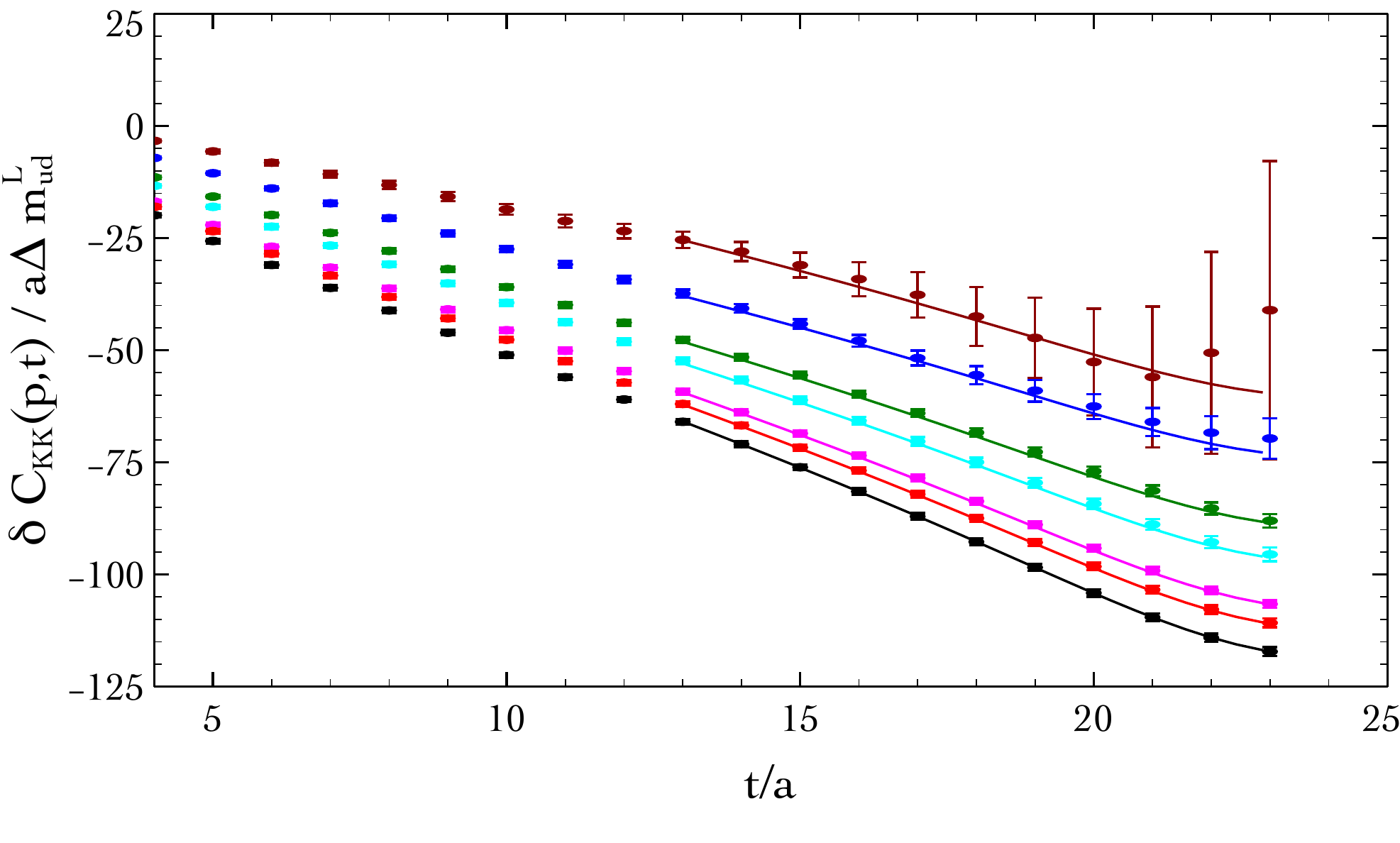}
\caption{\label{fig:Kslopes} \footnotesize
{\it Left panel}: extraction of meson energies from the effective mass of $C_{KK}(\vec p,t)$.
{\it Right panel}: fits of $\delta C_{KK}(\vec p,t) / a\Delta m_{ud}^L$ according to eq.~(\ref{eq:ckkratiolattice}): as it can be seen numerical data follow theoretical expectations. The data correspond to $\beta=3.9$, $am_{ud}^L=0.0064$,
$am_{s}^L=0.0177$ (see Appendix~\ref{sec:tm}).
}
\end{center}
\end{figure}
\begin{figure}[!t]
\begin{center}
\includegraphics[width=0.45\textwidth]{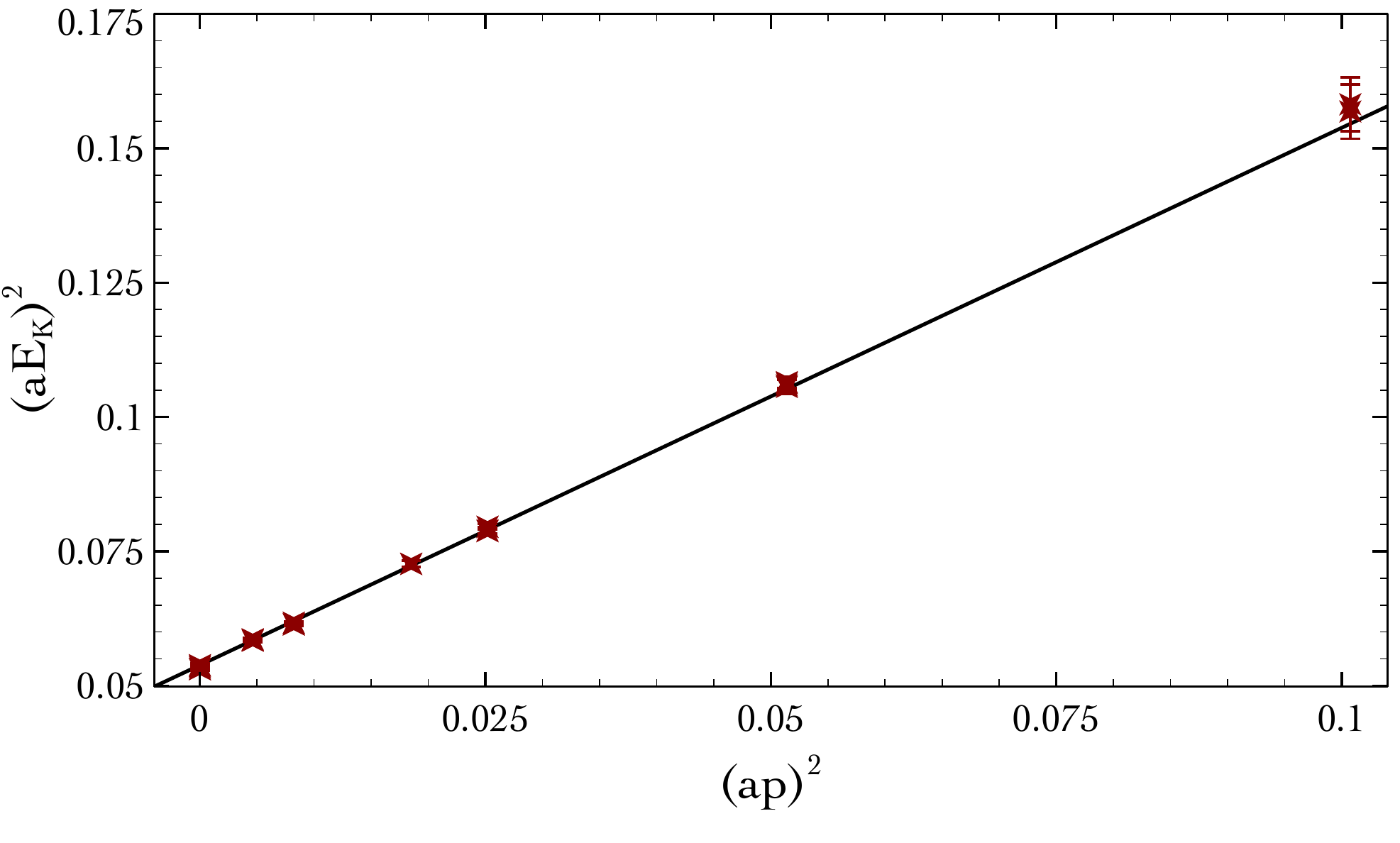}\hfill
\includegraphics[width=0.45\textwidth]{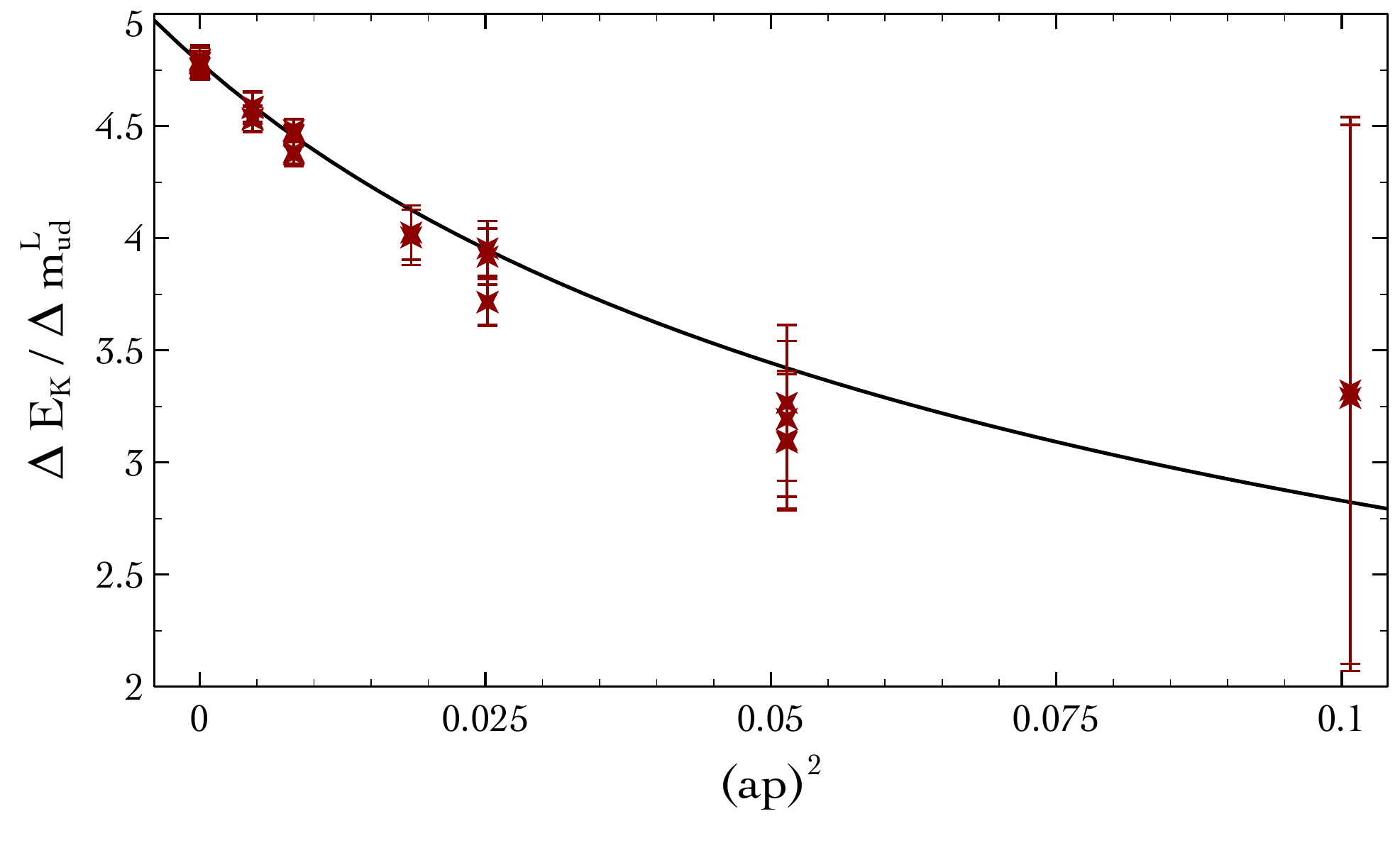} \\
\vskip 0.2cm
\includegraphics[width=0.45\textwidth]{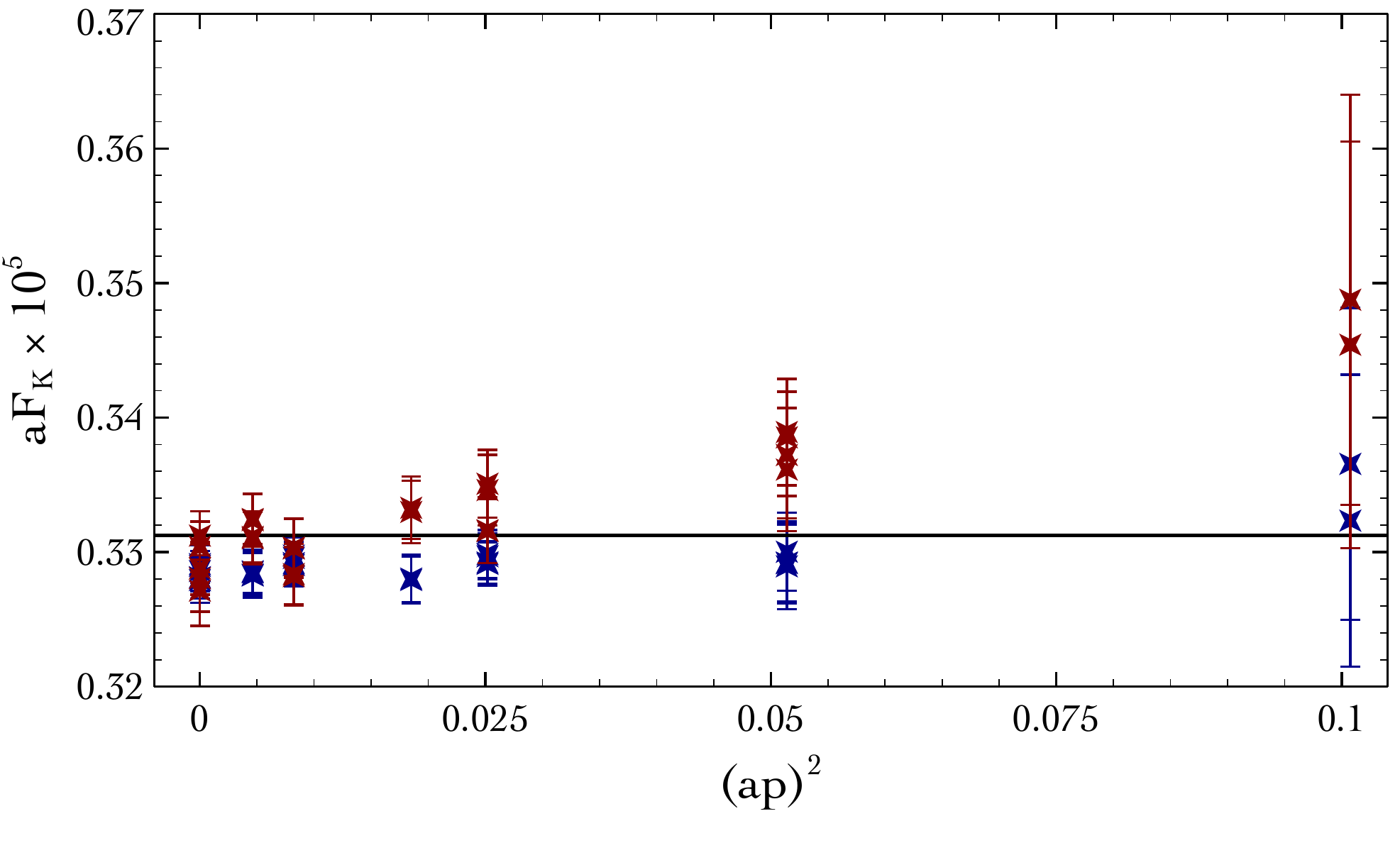}\hfill
\includegraphics[width=0.45\textwidth]{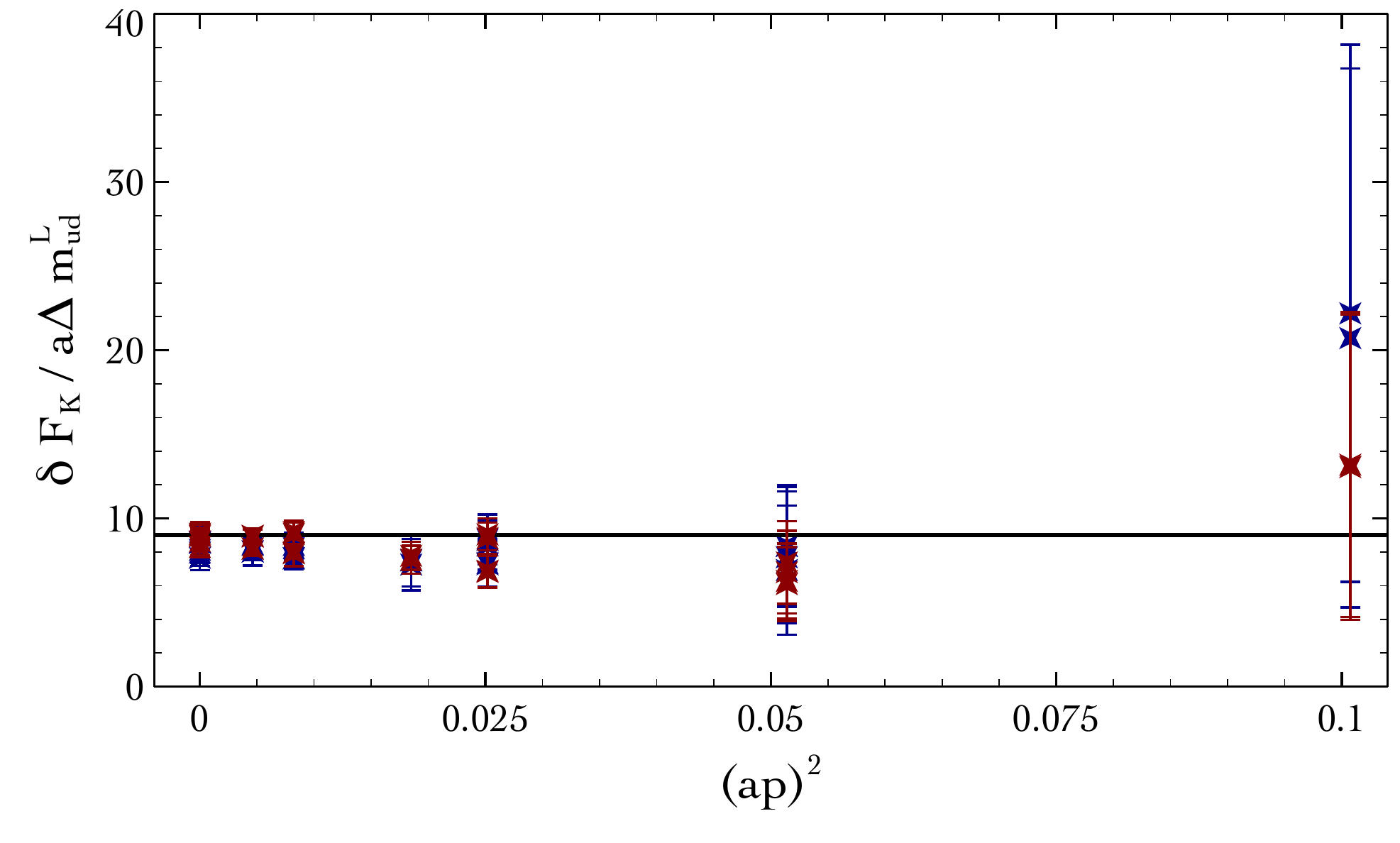}
\caption{\label{fig:Kdisp} \footnotesize
{\it Top-left panel}: the effective energies extracted from $C_{KK}(\vec p,t)$ are
compared with the expected continuum-like dispersion relation $E_K^2(p)=p^2+M_K^2$ (no fit).
{\it Top-right panel}: the corrections to the effective energies are compared with the theoretical expectation $\Delta E_K(p)=M_K\Delta M_K/E_K(p)$ (no fit).
{\it Bottom-left panel}: $F_K$ extracted from $C_{KK}(\vec p,t)$ (dark red points) and from $C_{KK}^{A^0}(\vec p,t)$ (dark blue points, see eq.~(\ref{eq:pseudoaxial})) for different values of $\vec p$; the solid line correspond to the value extracted at $\vec p =0$.
{\it Bottom-right panel}: $\delta F_K / a\Delta m_{ud}^L$ extracted from $\delta C_{KK}(\vec p,t)$ (dark red points) and from $\delta C_{KK}^{A^0}(\vec p,t)$ (dark blue points) for different values of $\vec p$; the solid line correspond to the value extracted at $\vec p =0$.
The data correspond to $\beta=3.9$, $am_{ud}^L=0.0064$,
$am_{s}^L=0.0177$ (see Appendix~\ref{sec:tm}).}
\end{center}
\end{figure}
As can be seen from Figure~\ref{fig:Kslopes}, $ \delta C_{KK}(\vec p, t)$ is determined with high precision, given the strong statistical correlation  existing between the numerator and the denominator of the ratio in eq.~(\ref{eq:ckkratiocontinuum}).  A consistency check of our procedure consists in verifying the dispersion relation $E_K^2(p) = p^2 + M_K^2$ and in comparing the variation $\Delta E_K(p)$ against its expected behaviour $\Delta E_K(p) = M_K \, \Delta M_K/E_K(p)$. Excellent agreement is found between numerical data and the theoretical curves shown in Figure~\ref{fig:Kdisp} both for $E_K^2(p)$, top-left panel, and $\Delta E_K^2(p)$, top-right panel.
In the bottom panels of Figure~\ref{fig:Kdisp} we also show that two different definitions of $\delta F_K$ (blue and red points) extracted from correlators at several $\vec p$-values give consistent results. The second definition of $F_K$ and of $\delta F_K$  has been obtained by considering the correlation function between the pseudoscalar density and the axial vector current
\bea
C_{KK}^{A^0}(t)&=&
\sum_{\vec x}\langle \bar \ell \gamma^0 \gamma_5 s(x)\  \bar s \gamma_5 \ell(0)\rangle
\ = \
\frac{F_K G_K}{2}\ e^{-E_Kt}+\cdots
\label{eq:pseudoaxial}
\eea 
and its correction at first order in $\Delta m_{ud}$.
  
We shall now discuss the  chiral and continuum extrapolations. Concerning chiral extrapolation,  it is useful to consider the correction to the meson mass square because this is a finite quantity in the chiral limit. The chiral formulae for $\Delta M_K^2$ and $\delta F_K$ have been obtained long ago,  in ref.~\cite{Gasser:1984gg}, within the unitary $SU(3)_L\times SU(3)_R$ effective theory ($N_f=2+1$),
\bea
  \frac{\Delta M_K^2}{\Delta m_{ud}}
  &=&B_0\left\{
  1+\frac{2}{3}\mu_\eta + M_K^2 \frac{\mu_\eta-\mu_\pi}{M_K^2-M_\pi^2}+ \right.
  \nonumber \\
  \nonumber \\
  &&\qquad\qquad\left.  
  +(m_s+m_{ud})\frac{16B_0}{F_0^2}(2L_8^r-L_5^r)+
  (m_s+2m_{ud})\frac{16B_0}{F_0^2}(2L_6^r-L_4^r)
  \right\} \,  ,
  \nonumber \\
  \nonumber \\
  \nonumber \\
  \frac{\delta F_K}{\Delta m_{ud}}
  &=&B_0\left\{
  \frac{4L_5^r}{F_0^2}
  -\frac{1}{64\pi^2F_0^2}-\frac{\mu_K}{2M_K^2}
  -\frac{\mu_\eta-\mu_\pi}{M_\eta^2-M_\pi^2}
  \right\}  \, ,
\label{eq:chiralformulae}
\eea
where $B_0$, $F_0$ and $L_i^r$ are low energy constants while
\bea
  &&M_\pi^2=2B_0m_{ud} \, ,
  \nonumber \\
  \nonumber \\
  &&M_K^2=B_0(m_{ud}+m_s) \, ,
  \nonumber\\
  \nonumber \\
  &&M_\eta^2=2B_0(m_{ud}+2m_s)/3 \, ,
  \nonumber \\
  \nonumber \\
  &&\mu_P=\frac{M_P^2}{32\pi^2F_0^2}\ln(M_P^2/\mu^2)\, ,
  \qquad P=\{\pi,\eta,K \}
  \, .
\eea  
\begin{figure}[!t]
\begin{center}
\includegraphics[width=0.49\textwidth]{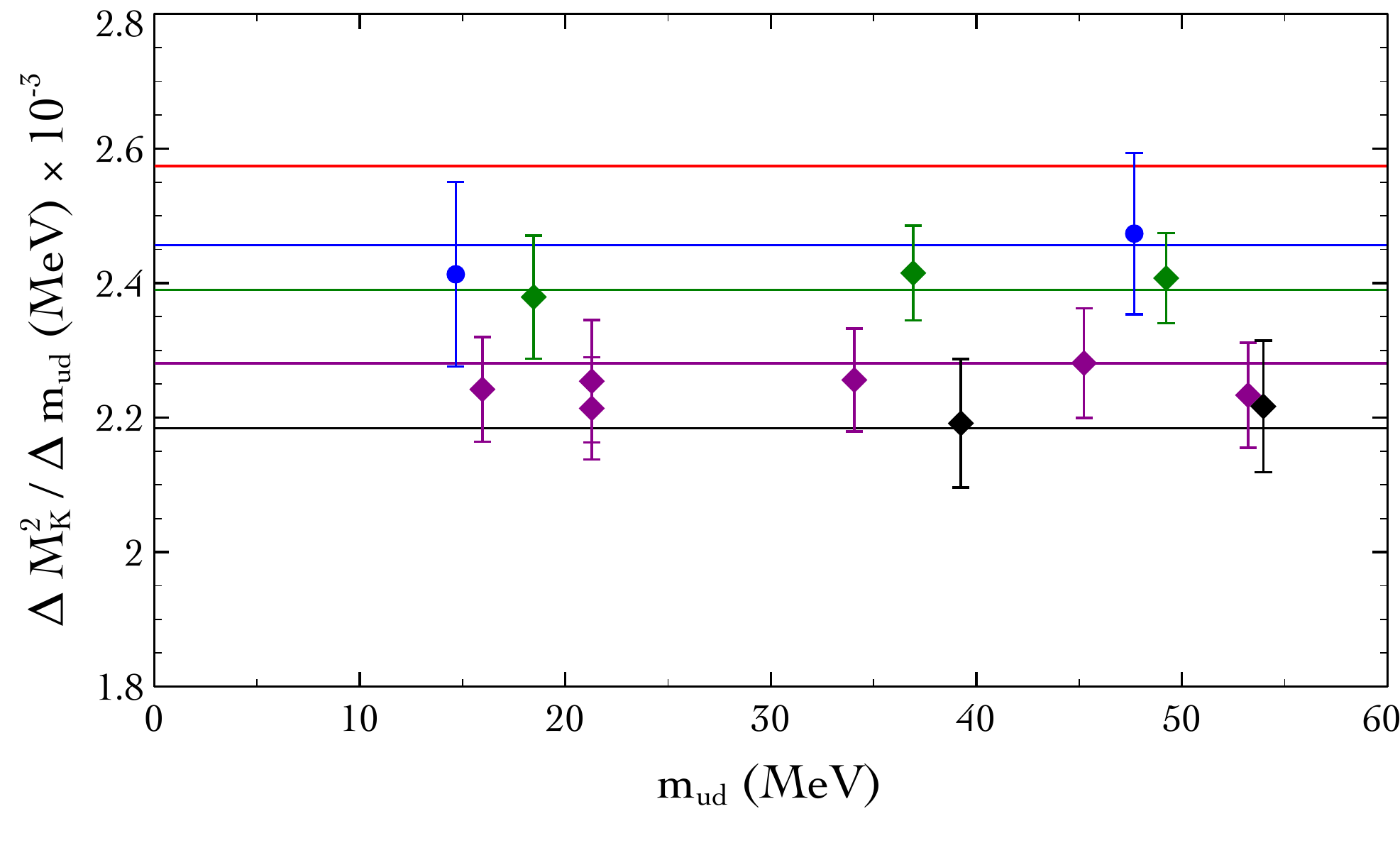}\hfill
\includegraphics[width=0.49\textwidth]{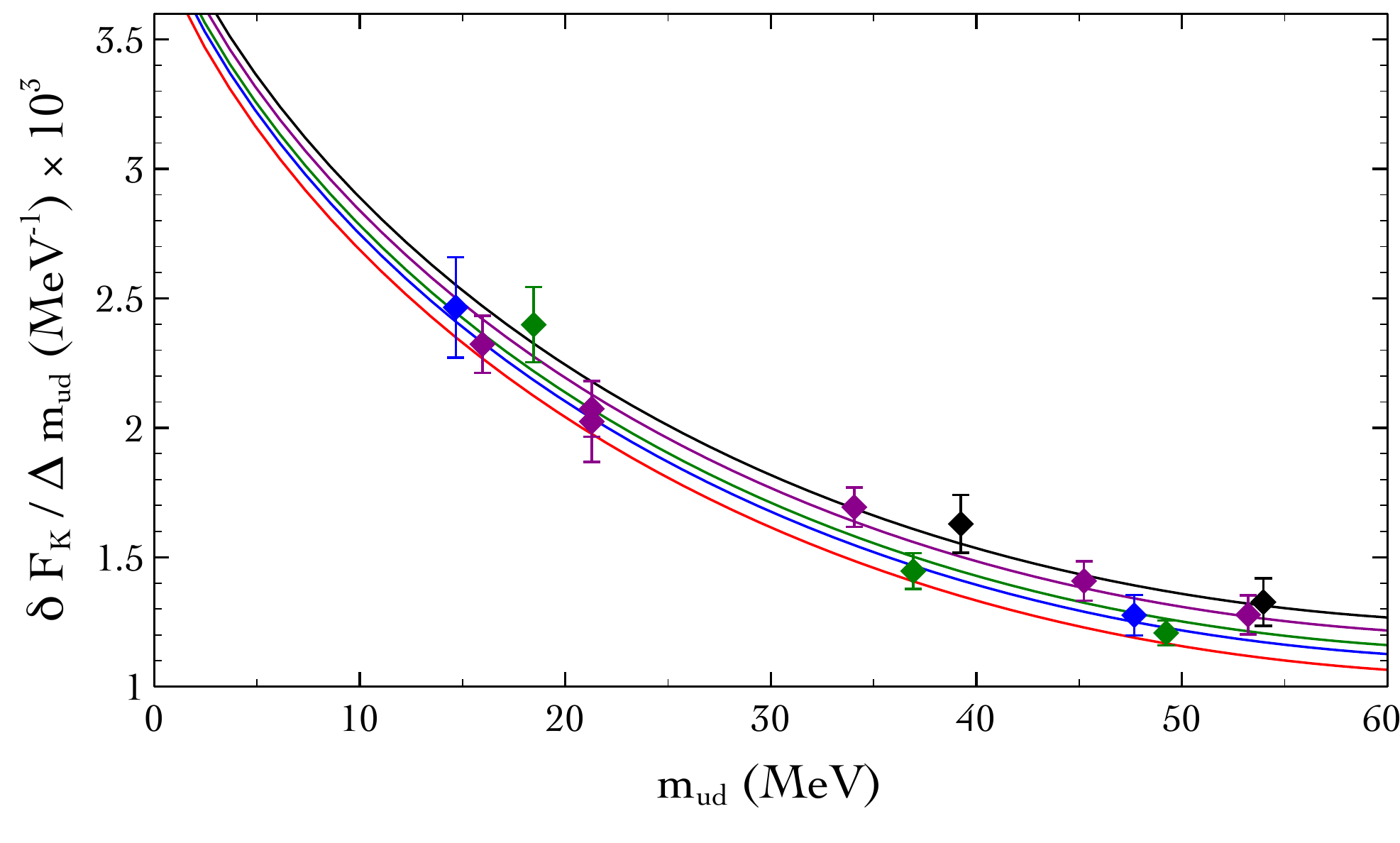}
\caption{\label{fig:Kchiral} \footnotesize
{\it Left panel}: combined chiral and continuum extrapolations of $\Delta M_K^2/\Delta m_{ud}$. 
{\it Right panel}: combined chiral and continuum extrapolations of $\delta F_K/\Delta m_{ud}$. Black points correspond to the coarser lattice spacing, $a=0.098$~fm, dark magenta points correspond to $a=0.085$~fm, green points to $a=0.067$~fm and blue points to $a=0.054$~fm. Red lines are the results of the continuum extrapolations.}
\end{center}
\end{figure}

In view of the poor convergence properties of $SU(3)_L\times SU(3)_R$ effective theory and also to cope with the fact that our results have been obtained by quenching the strange quark ($N_f=2$), we have chosen to fit our data with the formulae resulting by expanding the r.h.s. of eqs.~(\ref{eq:chiralformulae}) with respect to $m_{ud}/m_s$
(see ref.~\cite{Allton:2008pn} for a detailed and quantitative discussion of this point). This procedure is justified when the average light quark mass is sufficiently small compared to both $\Lambda_{QCD}$ and the (valence) strange quark mass, as it appears to be the case by looking at our data shown in Figure~\ref{fig:Kchiral}.

More precisely, in the left panel of Figure~\ref{fig:Kchiral} we show the combined chiral and continuum extrapolation of $\Delta M_K^2/\Delta m_{ud}$. Our results, obtained at four different lattice spacings for several values of the average light quark mass (see Appendix~\ref{sec:tm}), do not show a visible dependence with respect to $m_{ud}$ that can be quantified within the quoted errors. These errors combine in quadrature the statistical error and the systematic one coming from the uncertainties on the lattice spacing and from the renormalization constants (see Appendix~\ref{sec:tm}). We have consequently fit the numerical data to the following functional form
\bea
\left[\frac{\Delta M_K^2}{\Delta m_{ud}}\right](m_{ud},a)=
\left[\frac{\Delta M_K^2}{\Delta m_{ud}}\right]^{QCD}+C_M a^2 \, .
\label{eq:fitmk}
\eea
where $[\Delta M_K^2/\Delta m_{ud}]^{QCD}$ is a constant thus representing the physical value of this quantity in the continuum limit and at the physical light quark mass. We have also performed a fit of the numerical data by adding to the previous expression a linear and a (chiral) logarithmic term. In this case we have checked that the fitted coefficient of the chiral logarithm is compatible within its (large) error with the value $-B_0^2/(8\pi^2F_0^2)$ resulting from the expansion of eqs.~(\ref{eq:chiralformulae}) with respect to $m_{ud}/m_s$. The two fits give compatible results for $[\Delta M_K^2/\Delta m_{ud}]^{QCD}$ within the associated errors and we consider the difference between the central values as an estimate of the chiral extrapolation systematics, an error of about $3\%$ that we have added in quadrature to the lattice uncertainty.

In the case of $\delta F_K/\Delta m_{ud}$, right panel of Figure~\ref{fig:Kchiral}, the dependence upon $m_{ud}$ is significant within quoted errors (again combining in quadrature statistical and systematical ones) and we have included in the fitting function the leading and next-to-leading terms appearing in eqs.~(\ref{eq:chiralformulae}) expanded in powers of $m_{ud}/m_s$ plus a lattice artifact term, i.e.
\bea
\left[\frac{\delta F_K}{\Delta m_{ud}}\right](m_{ud},a)=
\left[\frac{\delta F_K}{\Delta m_{ud}}\right]^{QCD}
+C_F a^2+B_1 \left(m_{ud}-m_{ud}^{QCD}\right) + B_2\, m_{ud}\log\left(\frac{m_{ud}}{m_{ud}^{QCD}}\right) \, .
\label{eq:fitfk}
\eea
The systematics associated to this extrapolation has been estimated by replacing the (chiral) logarithmic term appearing into the previous expression with a quadratic term and it has been found of the order of $5\%$.

The value $m_{ud}^{QCD}=m_{ud}^{QCD}(\overline{MS},2GeV)=3.6(2)$ MeV has been taken from refs.~\cite{Colangelo:2010et,Blossier:2010cr}. The fitted values
\bea
\left[\frac{\Delta M_K^2}{\Delta m_{ud}}\right]^{QCD}(\overline{MS},2GeV)
&=& 2.57(8)\times 10^{3}\ \mbox{MeV}\, ,
\nonumber \\
\nonumber \\
\left[\frac{\delta F_K}{\Delta m_{ud}}\right]^{QCD}(\overline{MS},2GeV)
&=& 3.3(3)\times 10^{-3}\ \mbox{MeV}^{-1}\, ,
\label{eq:fitresults}
\eea
will be used in the next section to obtain our physical results for 
$\Delta m_{ud}^{QCD}=\Delta m_{ud}^{QCD}(\overline{MS},2GeV)$ and $F_{K^0}-F_{K^+}$.

\section{Electromagnetic corrections and $\mathbf{m_d-m_u}$}
\label{sec:electromagnetic}
In this section, by using as input the experimental determination of $M_{K^0}^2-M_{K^+}^2$, we shall determine $[m_d-m_u]^{QCD}$. 

When comparing the theoretical predictions with the experimental numbers we cannot neglect the isospin breaking corrections induced by electromagnetic interactions. In the literature, it has become popular to separate  electromagnetic and QIB effects and to give separately the theoretical value of these two contributions. Different calculations of the electromagnetic corrections, performed on the lattice or within other non-perturbative approaches,  are also  often compared~\cite{Colangelo:2010et}. However the separation of the electromagnetic and strong QIB effects is ambiguous~\cite{Gasser:2003hk,Bijnens:1993ae}, i.e. the different contributions   depend on the definition by which they are separated, whereas they  do not correspond to any physical observable because of ultraviolet divergences.

If we work at first order in the QED coupling constant and $\Delta m_{ud}$ and neglect terms of ${\cal O}(\alpha_{em}\Delta m_{ud})$, the relevant diagrams entering the difference  of kaon two point functions are
\bea
\Delta C_{K K}(t) &=& \gdsi  -\frac{e_d^2 -e_u^2}{2}\gdslselfl -e_s \frac{e_d -e_u}{2}\gdslexch
+ {\cal O}(\alpha_{em}\Delta m_{ud}) \, ,
\nonumber \\
\label{eq:kemcorr}
\eea
The electromagnetic corrections to $C_{KK}(t)$ are logarithmically divergent, corresponding to the renormalization of the quark masses\footnote{For simplicity we neglect in the discussion the renormalization of the meson sources/sinks, which in any case are finite if we use  vector or axial currents.}, and the divergent part has two components: one proportional to $\bar q q $ and the other proportional to $\bar q \tau_3 q$.  Alternatively one can treat electromagnetism on the lattice to all orders by exploiting the QED non compact formulation~\cite{Duncan:1996xy} but, in any case, two independent renormalization conditions have to be imposed in order to fix the counter-terms and separately renormalize the up and down quark masses. This can be achieved by matching  the physical masses of the charged and neutral kaons of the present example (the mass of the strange quark could be eventually fixed by the mass of the $\Omega^-$ baryon). Having extracted the light quark masses, i.e. $m_{ud}$ and $\Delta m_{ud}$,
one can predict $F_{K^0}-F_{K^+}$, the proton and neutron masses and all the other observables by including ``physical" isospin breaking effects. 

In this paper we only consider the QCD corrections because we want to show that our method works  for this part (we shall present a proposal for the ${\cal O}(\alpha_{em})$ corrections in a separate paper).  This is equivalent to say that we follow the common procedure of separating the two isospin breaking contributions by switching off electromagnetism.  Obviously the attempt to use physical quantities to fix $\Delta m_{ud}^{QCD}$ fails, since there are no data with electromagnetic interactions switched off and, for this reason,  we shall use the definition and determination of the electromagnetic corrections worked out by other groups.

According to Dashen's theorem~\cite{Dashen:1969eg}, electromagnetic corrections are the same in the chiral limit for $M_{K^0}^2-M_{K^+}^2$ and $M_{\pi^0}^2-M_{\pi^+}^2$ while, as discussed in the previous sections, pion masses are not affected by first order QCD corrections.
Beyond the chiral limit it is customary~\cite{Colangelo:2010et} to parametrize violations to the Dashen's theorem in terms of small parameters and, concerning $M_{K^0}^2-M_{K^+}^2$, we have
\bea
\left[M_{K^0}^2-M_{K^+}^2\right]^{QCD}=
\left[M_{K^0}^2-M_{K^+}^2\right]^{exp}-(1+\varepsilon_{\gamma})\left[M_{\pi^0}^2-M_{\pi^+}^2\right]^{exp}\, ,
\eea
where we have neglected QCD contributions of the second order $O(\Delta m_{ud}^2)$ in the pion mass difference. By using chiral perturbation theory and results from lattice QCD calculations of the electromagnetic corrections~\cite{Duncan:1996xy,Basak:2008na,Blum:2010ym,Portelli:2010yn}, ref.~\cite{Colangelo:2010et} estimates
\bea
&&\varepsilon_{\gamma}=0.7(5) \, ,
\nonumber \\
\nonumber \\
&&\left[M_{K^0}^2-M_{K^+}^2\right]^{QCD}=6.05(63)\times 10^3\ \mbox{MeV}^2\, .
\label{eq:input}
\eea 
On the one hand,  it is important to realize that to fix a value for  the electromagnetic term  $\varepsilon_{\gamma}$ is equivalent to define  \emph{a prescription} to separate (first order) QCD and QED isospin breaking effects and that for this reason we might ignore the error on this quantity. On the other hand,  one may wonder whether the different results combined in ref.~\cite{Colangelo:2010et} have really been obtained by using the same prescription. For this reason, in the following,  we shall quote our results by considering the error on $\varepsilon_{\gamma}$ as a way of  taking largely into account this ``scheme uncertainty" (see ref.~\cite{Gasser:2003hk} for a detailed discussion of this point).

By using eqs.~(\ref{eq:input}) and the results for $[\Delta M_K^2/\Delta m_{ud}]^{QCD}$ and $[\delta F_K/\Delta m_{ud}]^{QCD}$ given in eqs.~(\ref{eq:fitresults}), we get the following results
\bea
&&\left[m_d-m_u\right]^{QCD}(\overline{MS},2GeV)=2\Delta m_{ud}^{QCD} = 2.35(8)(24)\ \mbox{MeV} 
\quad \times \quad 
\frac{\left[M_{K^0}^2-M_{K^+}^2\right]^{QCD}}{6.05\times 10^3\ \mbox{MeV}^2} \, ,
\nonumber \\
\nonumber \\
\nonumber \\
&&\left[\frac{F_{K^0}-F_{K^+}}{F_{K}}\right]^{QCD} = 0.0078(7)(4)
\quad \times \quad 
\frac{\left[M_{K^0}^2-M_{K^+}^2\right]^{QCD}}{6.05\times 10^3\ \mbox{MeV}^2}\, ,
\label{eq:kaonphysres}
\eea 
where the first error comes from our calculation and combines in quadrature statistics and systematics while the second comes from the uncertainty on $\varepsilon_\gamma$.
 
At first order in $\Delta m_{ud}$, thanks to the fact that pions don't get corrections and that $K^+$ and $K^0$ get opposite corrections, we have
\bea
\left[\frac{F_{K^+}/F_{\pi^+}}{F_{K}/F_{\pi}}-1\right]^{QCD} &=& -0.0039(3)(2)
\quad \times \quad 
\frac{\left[M_{K^0}^2-M_{K^+}^2\right]^{QCD}}{6.05\times 10^3\ \mbox{MeV}^2} \, ,
\eea
a value that is higher than the estimate obtained in ref.~\cite{Cirigliano:2011tm} by using chiral perturbation theory, namely
\bea
\left[\frac{F_{K^+}/F_{\pi^+}}{F_{K}/F_{\pi}}-1\right]^{\chi pt} &=& -0.0022(6) \, .
\eea

\section{Nucleon masses}
\label{sec:nucleons}
Having determined $\Delta m_{ud}^{QCD}$,  we can now predict the QCD isospin breaking corrections on other observables, as already done in the previous section for the kaon decay constants. In this section we calculate the difference between the masses of the neutron and of the proton. We consider the correlation functions $C_{nn}^\pm(t)$ and $C_{pp}^\pm(t)$ at zero momentum (see  eqs.~(\ref{eq:twopoint})) and, in order to decrease statistical errors, we extract nucleon masses from the combinations
\bea
C_{nn}(t)&=& C_{nn}^+(t)-C_{nn}^-(T-t) \, , 
\nonumber \\
\nonumber \\
C_{pp}(t)&=& C_{pp}^+(t)-C_{pp}^-(T-t) \, . 
\eea
The quark fields entering the sink interpolating operators have been ``Gaussian smeared'' according to
\bea
\ell^{smeared} &=& (1+\alpha_g H)^{n_g} \ell \, , 
\nonumber \\
\nonumber \\
H(\vec x,\vec y) &=& \sum_{i=1}^3\left[
U_i(\vec x,t)\delta_{\vec x, \vec y-i}+
U_i^\dagger(\vec x-i,t)\delta_{\vec x, \vec y+i}
\right] \, , 
\eea
with the parameters $\alpha_g$ and $n_g$ fixed at the values optimized in ref.~\cite{Alexandrou:2008tn} where the same gauge configurations of this study have been used.

The extraction of physical informations from nucleon euclidean two point functions proceeds along the same lines described in detail in the case of the kaons. By using the diagrammatic analysis of eqs.~(\ref{eq:nucleondia}) we have
\bea
C_{NN}(t)&=& -\protondlll+\protonclll= W_{N} e^{-M_{N}t}+\cdots\, ,
\eea
and
\bea
\delta C_{NN}(t) 
&=&
-\frac{\protondill-\protondlil-\protondlli}{-\protondlll+\protonclll}
+\frac{\protoncill-\protonclil-\protonclli}{-\protondlll+\protonclll}
\nonumber \\
\nonumber \\
\nonumber \\
&=& \delta W_{N}-t \Delta M_{N}+\cdots \, ,
\nonumber \\
\label{eq:nucleonscorrs}
\eea
where the dots represent sub-leading exponentials contributing to the correlation functions.
\begin{figure}[!t]
\begin{center}
\includegraphics[width=0.49\textwidth]{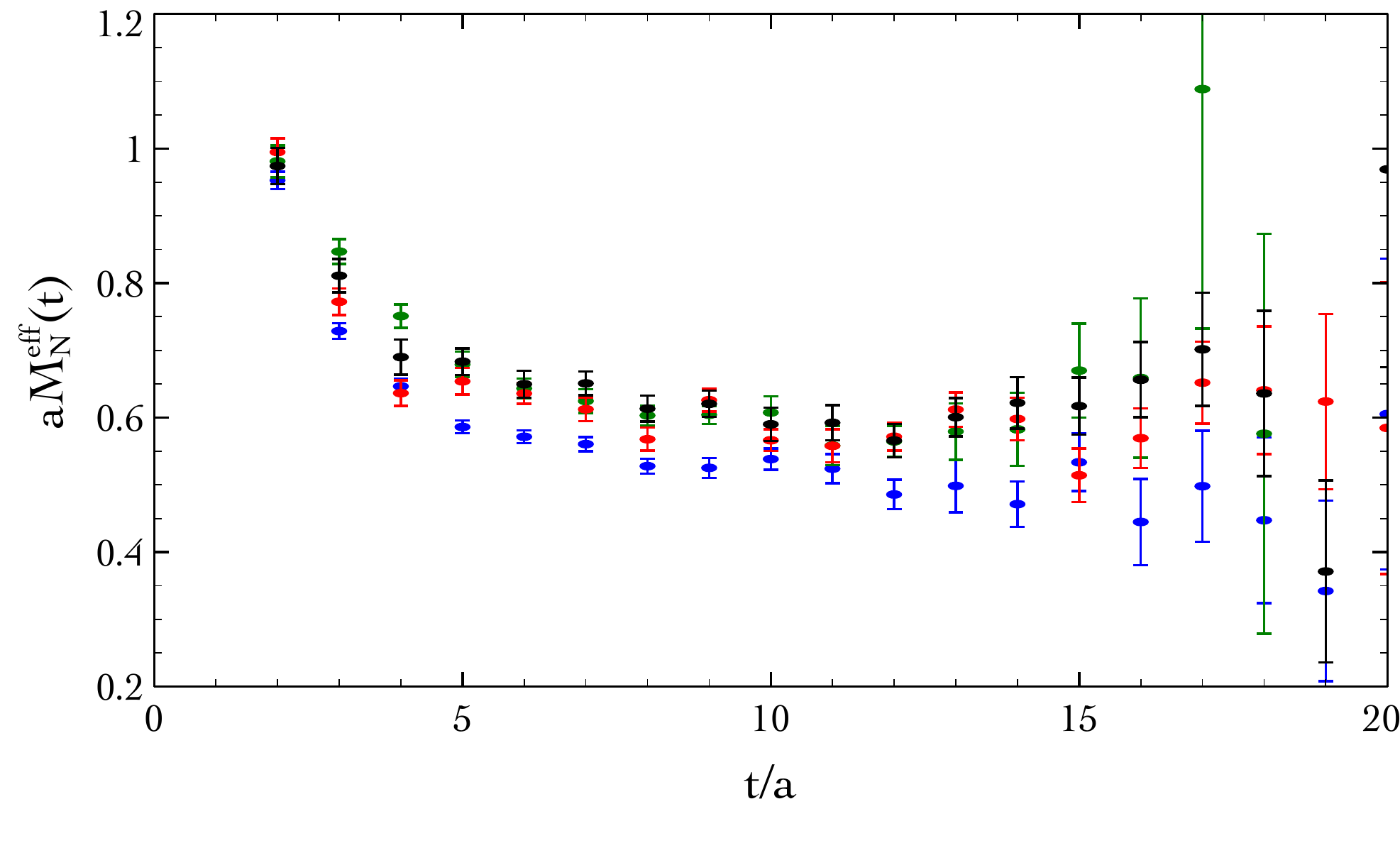}\hfill
\includegraphics[width=0.49\textwidth]{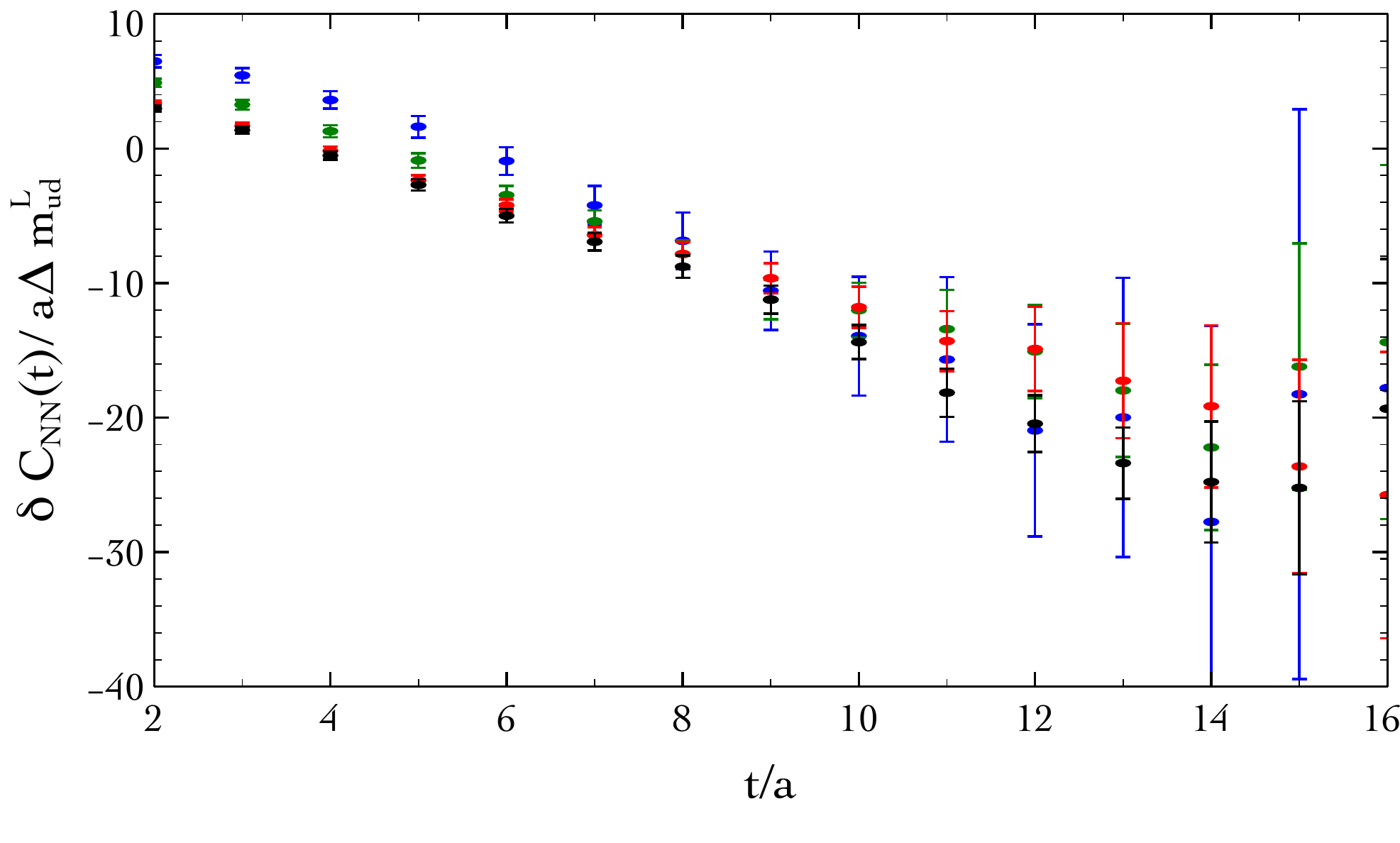}
\caption{\label{fig:nucleons} \footnotesize
{\it Left panel}: effective mass plots of nucleon correlation functions $C_{NN}(t)$.
{\it Right panel}: Correlation functions $\delta C_{NN}(t)/a\Delta m_{ud}^L$.
The data are at fixed lattice spacing $a=0.085$~fm for different values of $m_{ud}$ (see Appendix~\ref{sec:tm}).
}
\end{center}
\end{figure}
\begin{figure}[!t]
\begin{center}
\includegraphics[width=0.49\textwidth]{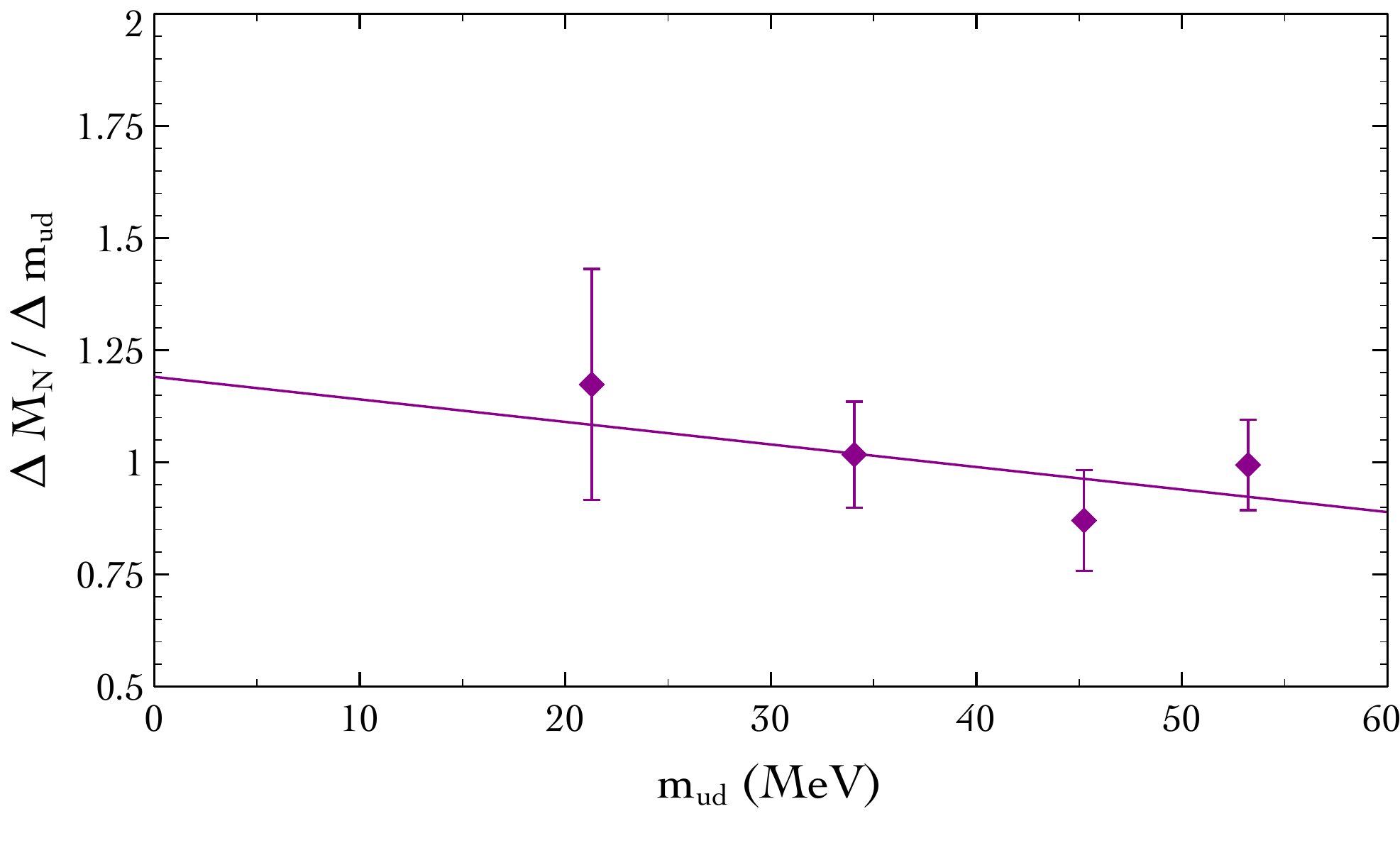}
\caption{\label{fig:nucleonschiral} \footnotesize
Chiral extrapolation of $\Delta M_N/\Delta m_{ud}$ at fixed lattice spacing $a=0.085$~fm.
}
\end{center}
\end{figure}
By extracting the slope in  $t$ of $\delta C_{NN}(t)$,  we can determine $\Delta M_{N}=(M_n-M_p)/2$. In the left panel of Figure~\ref{fig:nucleons} we show the effective masses $M_N^{eff}(t)$ as extracted from the correlation functions $C_{NN}(t)$ at fixed lattice spacing $a=0.085$~fm for different values of $m_{ud}$. In the right panel of the same figure we show the corresponding correlation functions $\delta C_{NN}(t)/a\Delta m_{ud}^L$ that we have fitted to straight lines, according to eq.~(\ref{eq:nucleonscorrs}), i.e. without taking into account the finite time extent of the lattice because it affects the correlation functions only at large times where no signal can anyway be extracted within our errors.

In Figure~\ref{fig:nucleonschiral} we show the chiral extrapolation of $\Delta M_N/\Delta m_{ud}$ performed by using the following fitting function~\cite{Gasser:1982ap}
\bea
\left[\frac{\Delta M_N}{\Delta m_{ud}}\right](m_{ud})=
\left[\frac{\Delta M_N}{\Delta m_{ud}}\right]^{QCD}+B_N (m_{ud}-m_{ud}^{QCD})\, .
\eea
By using the results of the fit and the value of $\Delta m_{ud}^{QCD}$ given in eqs.~(\ref{eq:kaonphysres}), we get
\bea
\left[M_n-M_p\right]^{QCD}= 2\Delta m_{ud}^{QCD} 
\left[\frac{\Delta M_N}{\Delta m_{ud}}\right]^{QCD}
=2.8(6)(3)\ \mbox{MeV}
\quad \times \quad 
\frac{\left[M_{K^0}^2-M_{K^+}^2\right]^{QCD}}{6.05\times 10^3\ \mbox{MeV}^2} \, ,
\eea 
where the first error takes into account the lattice uncertainties while the second comes from the uncertainty on $\varepsilon_\gamma$.
This number is our best estimate at present but it has been obtained at fixed lattice spacing and with limited statistics. We plan to refine the calculation in a separate publication.

\section{Semileptonic decays}
\label{sec:semileptonic}
In order to illustrate our calculation of the QCD isospin breaking corrections to the form factors parametrizing semileptonic decays,  we start by the following ``double ratio" of three point correlation functions~\cite{Becirevic:2004ya},
\bea
R^\mu_{D^+K^0}(t)= \frac{\gtcds\gtsdc}{\gtcdc\gtsds} \, ,
\eea 
where the correlators at numerator are given by
\bea
-\gtcds &=&
\langle\,
\bar d \gamma^5 s(\vec p_i, T/2)\
\bar s \gamma^\mu c(\vec p_f-\vec p_i, t)\
\bar c \gamma^5 d(0)     
\, \rangle
\nonumber \\
\nonumber \\
&=&\rho^\star_{K^0}\rho_{D^+}\ \bra{K^0} V^\mu_{cs} \ket{D^+}\ 
e^{-tE_{D^+}} e^{-(T/2-t)E_{K^0}}  +\cdots \, , 
\nonumber \\
\nonumber \\
\nonumber \\
-\gtsdc &=&
\langle\,
\bar d \gamma^5 c(\vec p_f, T/2)\
\bar c \gamma^\mu s(\vec p_i-\vec p_f, t)\
\bar s \gamma^5 d(0)
\, \rangle  
\nonumber \\
\nonumber \\
&=&\rho_{K^0}\rho^\star_{D^+}\ \bra{K^0} V^\mu_{cs} \ket{D^+}^\star\ 
e^{-tE_{K^0}} e^{-(T/2-t)E_{D^+}} +\cdots \,  , 
\label{eq:dk3corrs1}
\eea
while those at denominator are
\bea
-\gtcdc &=&
\langle\,
\bar d \gamma^5 c(\vec p_i, T/2)\
\bar c \gamma^0 c(\vec 0, t)\
\bar c \gamma^5 d(0)
\, \rangle
=\vert \rho_{D^+} \vert^2 \ \bra{D^+} V^0_{cc} \ket{D^+}\ 
e^{-TE_{D^+}/2}  +\cdots \, , 
\nonumber \\
\nonumber \\
\nonumber \\
-\gtsds &=&
\langle\,
\bar d \gamma^5 s(\vec p_f, T/2)\
\bar s \gamma^0 s(\vec 0, t)\
\bar s \gamma^5 d(0)
\, \rangle
= \vert \rho_{K^0} \vert^2 \ \bra{K^0} V^0_{ss} \ket{K^0}\ 
e^{-TE_{K^0}/2}  +\cdots \, . 
\label{eq:dk3corrs2}
\eea
By using the previous expressions and the conservation of flavour diagonal vector currents, $Z_V\bra{n}V^0\ket{n}=2E_n$, we obtain the well known and useful result 
\bea
R^\mu_{D^+K^0}(t) = \frac{1}{4E_{D^+}E_{K^0}}\ 
\left\vert \bra{K^0} V^\mu_{cs} \ket{D^+} \right\vert^2
+\cdots \, , 
\label{eq:rmudk}
\eea
i.e. the fact that, by neglecting sub leading exponentials, $R^\mu_{D^+K^0}(t)$ is a constant with respect to $t$ from which it is possible to extract the matrix elements with high statistical accuracy, thanks to the statistical correlation between the different correlation functions.   
The form factors can be extracted by using the  standard expressions of the matrix elements computed at   different initial and final meson momenta (with $\vec p_i\neq 0$ and/or $\vec p_f\neq 0$)
\bea
\bra{K^0} V^0_{cs} \ket{D^+} &=& (E_{D^+}+E_{K^0}) f_+^{D^+K^0}(q^2)+(E_{D^+}-E_{K^0})f_-^{D^+K^0}(q^2) \, , 
\nonumber \\
\nonumber \\
\bra{K^0} \vec{V}_{cs} \ket{D^+} &=& (\vec{p}_i+\vec{p}_f) f_+^{D^+K^0}(q^2)+(\vec{p}_i-\vec{p}_f)f_-^{D^+K^0}(q^2) \, ,  
\eea
where as usual $f_-^{D^+K^0}(q^2) $ can be expressed in terms of the scalar form factor
\bea  f_0^{D^+K^0}(q^2)  = f_+^{D^+K^0}(q^2)  +\frac{q^2}{M_{D^+}^2-M_{K^0}^2} \, f_-^{D^+K^0}(q^2) \, . \eea
Matrix elements with non vanishing momentum transfer between initial and final states have been computed by using flavour twisted boundary conditions for the valence quarks~\cite{Bedaque:2004kc,deDivitiis:2004kq}.

A very important observation concerning the calculation of isospin breaking corrections is that  $R^\mu_{D^+K^0}(t)$  in eq.~(\ref{eq:rmudk}) has the same form, when expressed in terms of diagrams, not only in the perturbed theory (the one discussed up to now), but also in the unperturbed isospin symmetric theory. This happens, as in the case of two point correlation functions, because ${\hat {\mathcal L}}$ does not generate decays of the $D$ or $K$ mesons. Thanks to this observation we simply have
\bea
R^\mu_{DK}(t) &=& \frac{\gtcls\gtslc}{\gtclc\gtsls}
=
\frac{1}{4E_{D}E_{K}}\ 
\left\vert \bra{K} V^\mu_{cs} \ket{D} \right\vert^2
+\cdots \, , 
\nonumber \\
\nonumber \\
\nonumber \\
\delta R^\mu_{DK}(t)&=&
-\frac{\gtcis}{\gtcls}
-\frac{\gtsic}{\gtslc}
+\frac{\gtcic}{\gtclc}
+\frac{\gtsis}{\gtsls}
\nonumber \\
\nonumber \\
&=& 2 \delta \left[ \bra{K} V^\mu_{cs} \ket{D} \right] -\delta E_D -\delta E_K+\cdots \, . 
\label{eq:corrfdk}
\eea
From the expressions of the form factors in terms of the matrix elements $\bra{K} V^\mu_{cs} \ket{D}$ and by using $\delta[\bra{K} V^\mu_{cs} \ket{D}]$ and the values of $\delta E_{D,K}$ obtained as explained in the previous sections it is possible to extract $\delta f_\pm^{DK}(q^2)$. A more detailed derivation of eqs.~(\ref{eq:corrfdk}) is provided in Appendix~\ref{sec:proof}.

The calculation of $\delta f_\pm^{K\pi}(q^2)$ proceeds along similar lines but there are some important differences that require a separate and detailed discussion. The starting point are the following diagrams  for $R^\mu_{K\pi}(t)$ and its variation, see eqs.~(\ref{eq:k0pim}) and (\ref{eq:kppi0}),
\bea
R^\mu_{K\pi}(t) &=& \frac{\gtsll\gtlls}{\gtsls\gtlll}
=
\frac{1}{4E_{K}E_{\pi}}\ 
\left\vert \bra{\pi} V^\mu_{su} \ket{K} \right\vert^2
+\cdots \, , 
\nonumber \\
\nonumber \\
\nonumber \\
\delta R^\mu_{K\pi}(t)&=&
-\frac{\gtsil-\discgtsli}{\gtsll}
-\frac{\gtlis-\discgtils}{\gtlls}
+\frac{\gtsis}{\gtsls}
\nonumber \\
\nonumber \\
&=& 2 \delta \left[ \bra{\pi} V^\mu_{su} \ket{K} \right] -\delta E_K+\cdots \, . 
\label{eq:corrfkpi0}
\eea
Note the differences of eqs.~(\ref{eq:corrfkpi0}) with respect to eqs.~(\ref{eq:corrfdk}), i.e. the presence of disconnected contributions and the absence of the correction to the pion correlation function (all black quark lines) in the denominator  of  $R^\mu_{K\pi}(t)$. The latter is a consequence of the vanishing of the QIB corrections at first order in $\Delta m_{ud}$ in the pion case.

In this work we have not calculated disconnected diagrams and we cannot show results for $\delta f_\pm^{K\pi}(q^2)$. These will be given in a separate publication but, for the time being and in order to show that our method works also in the case of three point functions and form factors, we have calculated the difference of $f_+^{K^0\pi^-}(q^2)$ with respect the isospin symmetric value $f_+^{K\pi}(q^2)$, i.e $\delta_f f_+^{K\pi}(q^2)$. 
This is a quantity that cannot be measured directly because the missing contribution, $\delta_b f_+^{K\pi}(q^2)$,  is neither  equal nor  related in a simple way to $\delta_f f_+^{K\pi}(q^2)$. The two different contributions are in fact associated to two  independent isospin channels and, according to ref.~\cite{Gasser:1984ux}, the $\pi^0$-$\eta$ mixing is expected to enhance considerably $\delta_b f_+^{K\pi}(q^2)$ with respect to $\delta_f f_+^{K\pi}(q^2)$. One may be tempted, to a first approximation, to neglect the disconnected diagrams but this cannot be done because they are needed in order to cancel non physical terms ($t\Delta E_P$) contributing to the slope of the connected diagrams,
\bea
-\frac{\gtsil}{\gtsll}
&=&
\mbox{const. } -t \Delta E_K +t \Delta E_P
+\cdots \, , 
\nonumber \\
\nonumber \\
-\frac{\gtsil-\discgtsli}{\gtsll}
&=&
\mbox{const. } -t \Delta E_K
+\cdots \, .
\eea
The $\Delta E_P$ contribution to the slope corresponds to the QIB correction to the energy of a meson, a copy of the physical pions, having as valence quarks a physical $u$ (or $d$) and an additional light quark, also of mass $m_{ud}$, but not contained into the isospin doublet. This term is of the same size of $\Delta E_K$, as we have explicitly checked numerically by using the slopes extracted from two point functions (eq.~(\ref{eq:ckkone})), whereas it cannot be present in physical kaon-to-pion three point correlation
functions because of isospin symmetry. Indeed the correction to the energy of the physical pions vanishes at first order in $\Delta m_{ud}$.

In order to calculate $\delta_f f_+^{K\pi}(q^2)$ we need an expression for $\delta_f R_{K\pi}^\mu(t)$. From eq.~(\ref{eq:k0pim}) we get
\bea
\delta_f R^\mu_{K\pi}(t)&=&
-\frac{\gtsil-\gtsli}{\gtsll}
-\frac{\gtlis-\gtils}{\gtlls}
+\frac{\gtsis}{\gtsls}
\nonumber \\
\nonumber \\
&=& 2 \delta_f \left[ \bra{\pi} V^\mu_{su} \ket{K} \right] -\delta E_K+\cdots \, . 
\label{eq:corrfkpi}
\eea

\begin{figure}[!t]
\begin{center}
\includegraphics[width=0.49\textwidth]{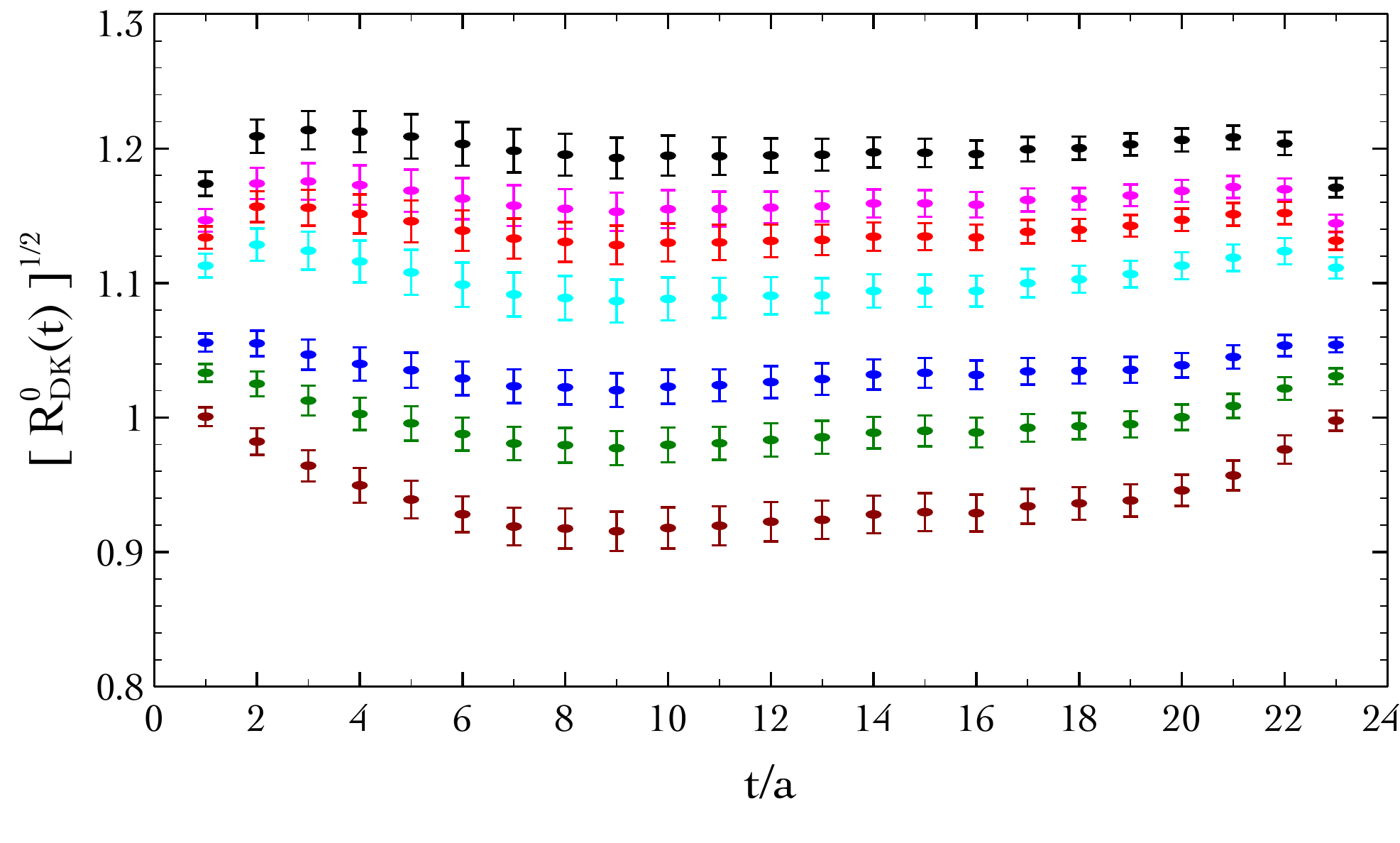}
\includegraphics[width=0.49\textwidth]{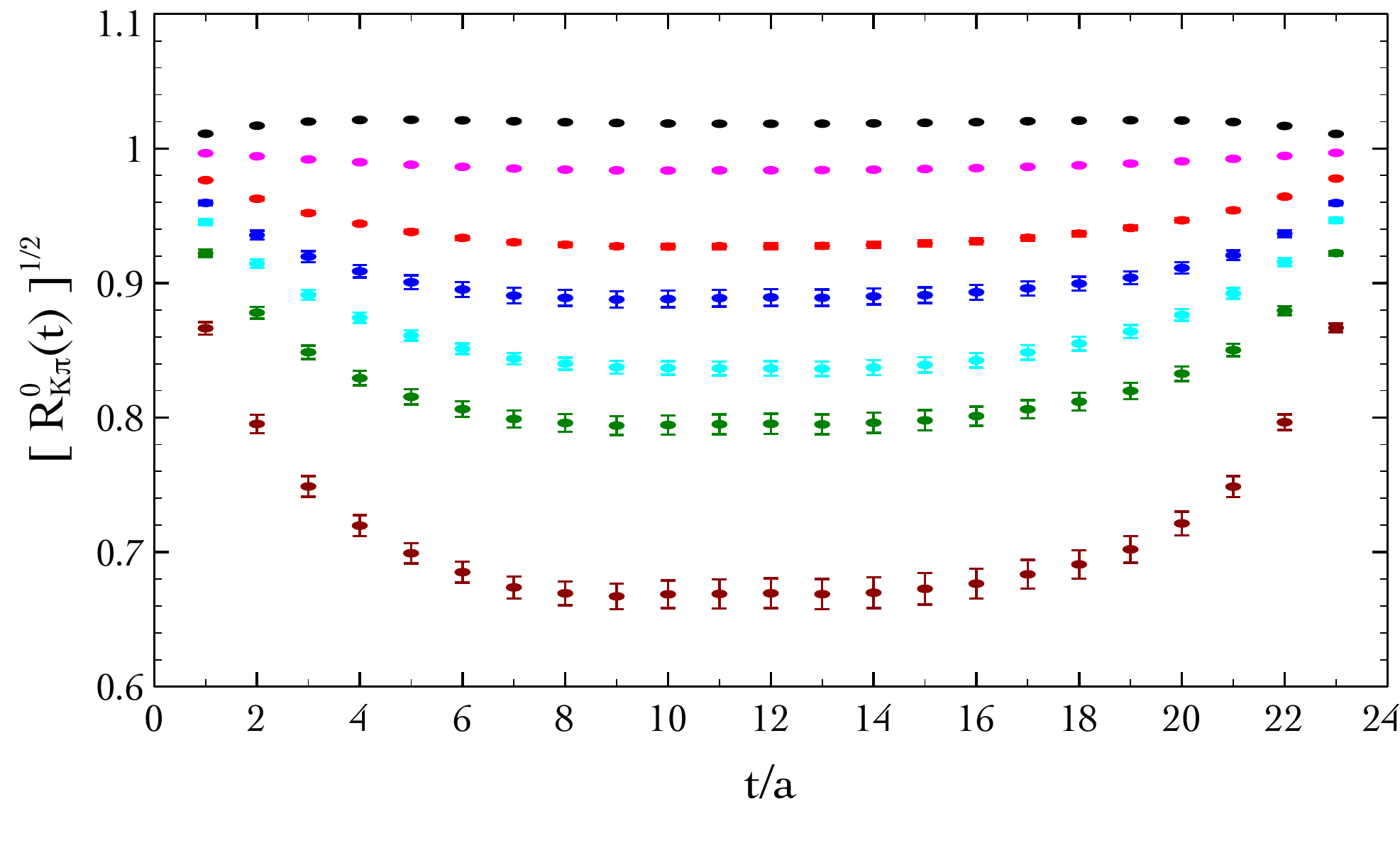}
\caption{\label{fig:mel} \footnotesize
{\it Left panel}: we show our results for $\sqrt{R^0_{DK}(t)}$ for several values of the momentum transfer. 
{\it Right panel}: we show our results for $\sqrt{R^0_{K\pi}(t)}$ for several values of the momentum transfer. 
The data are obtained at fixed lattice spacing $a=0.085$~fm  and at fixed $am_{ud}^L=0.0064$ (see Appendix~\ref{sec:tm}). 
}
\end{center}
\end{figure}
\begin{figure}[!t]
\begin{center}
\includegraphics[width=0.49\textwidth]{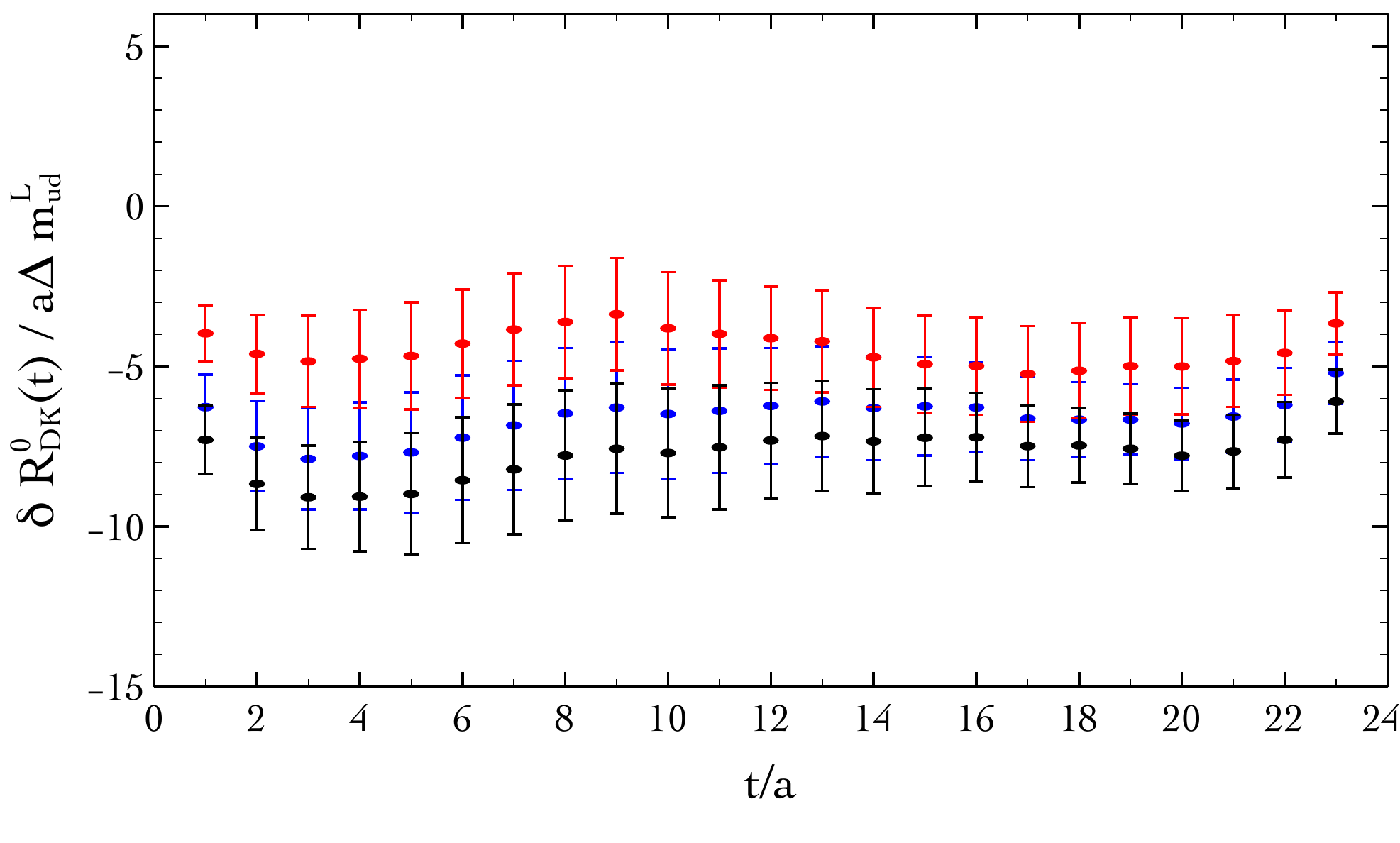}
\includegraphics[width=0.49\textwidth]{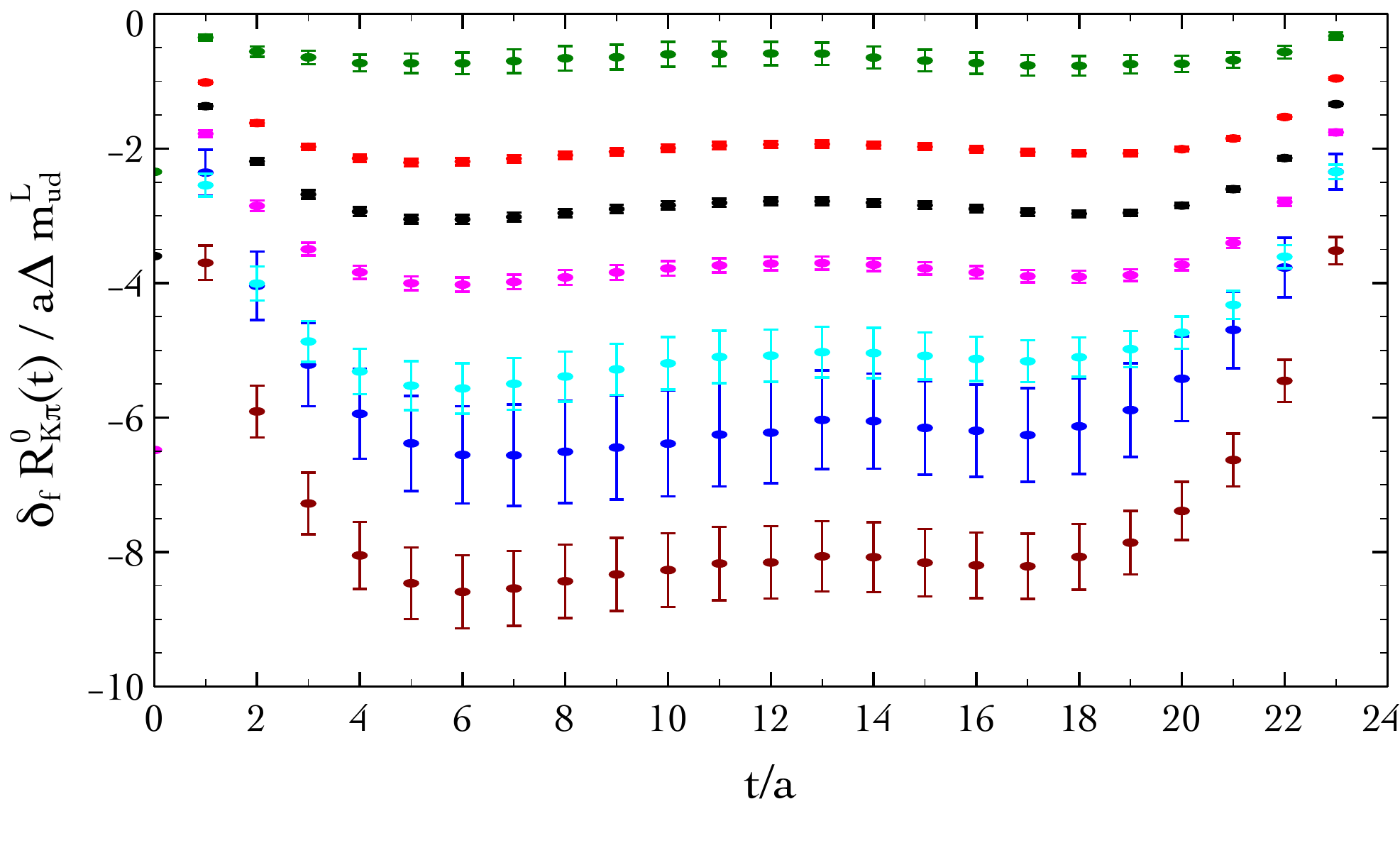}
\caption{\label{fig:deltamel} \footnotesize
{\it Left panel}: we show our results for $\delta R^0_{DK}(t)/ a\Delta m_{ud}^L$ for several values of the momentum transfer. 
{\it Right panel}: we show our results for $\delta_f R^0_{K\pi}(t)/ a\Delta m_{ud}^L$ for several values of the momentum transfer. 
The data are obtained at fixed lattice spacing $a=0.085$~fm  and at fixed $am_{ud}^L=0.0064$ (see Appendix~\ref{sec:tm}). 
}
\end{center}
\end{figure}
\begin{figure}[!t]
\begin{center}
\includegraphics[width=0.49\textwidth]{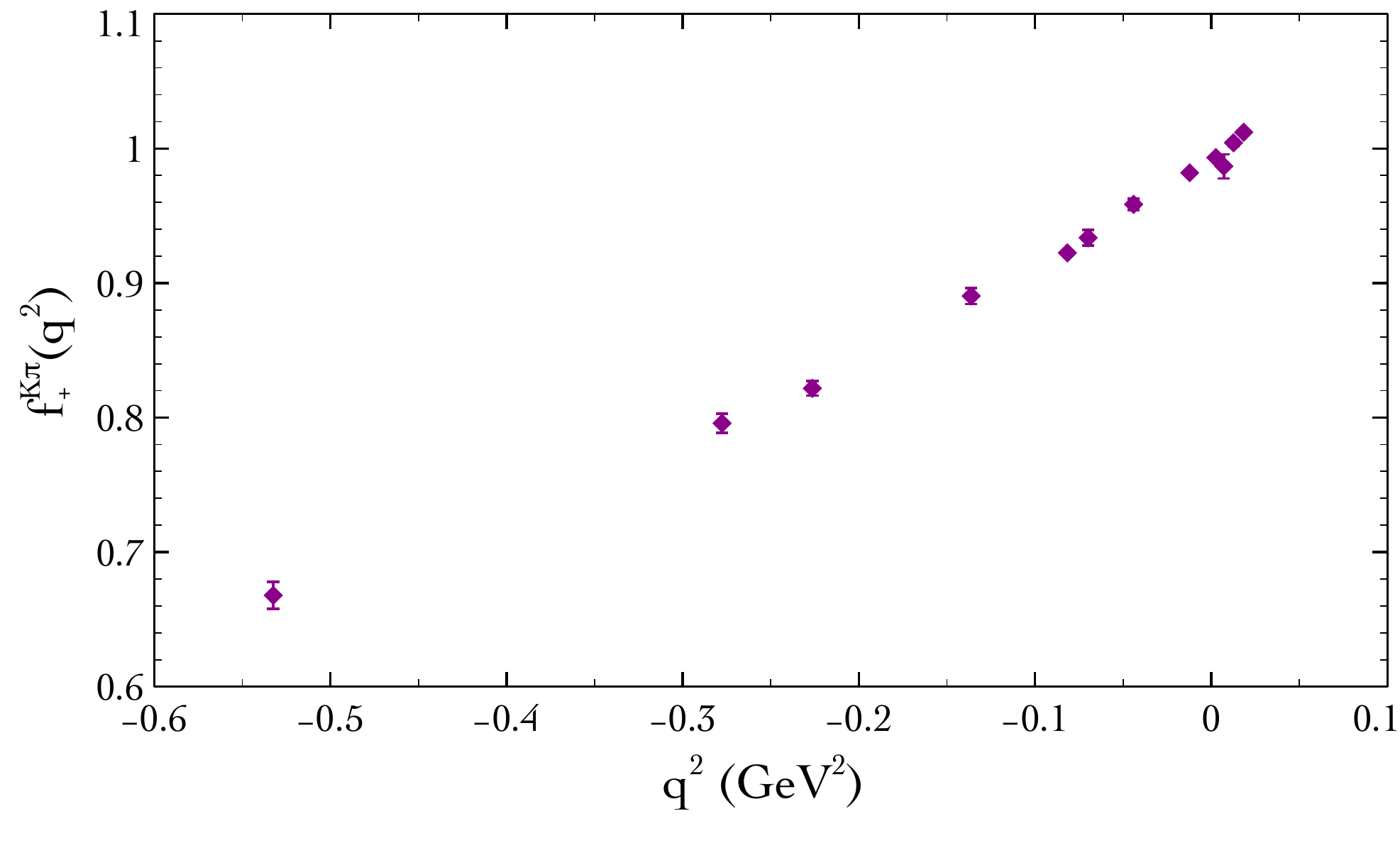}
\includegraphics[width=0.49\textwidth]{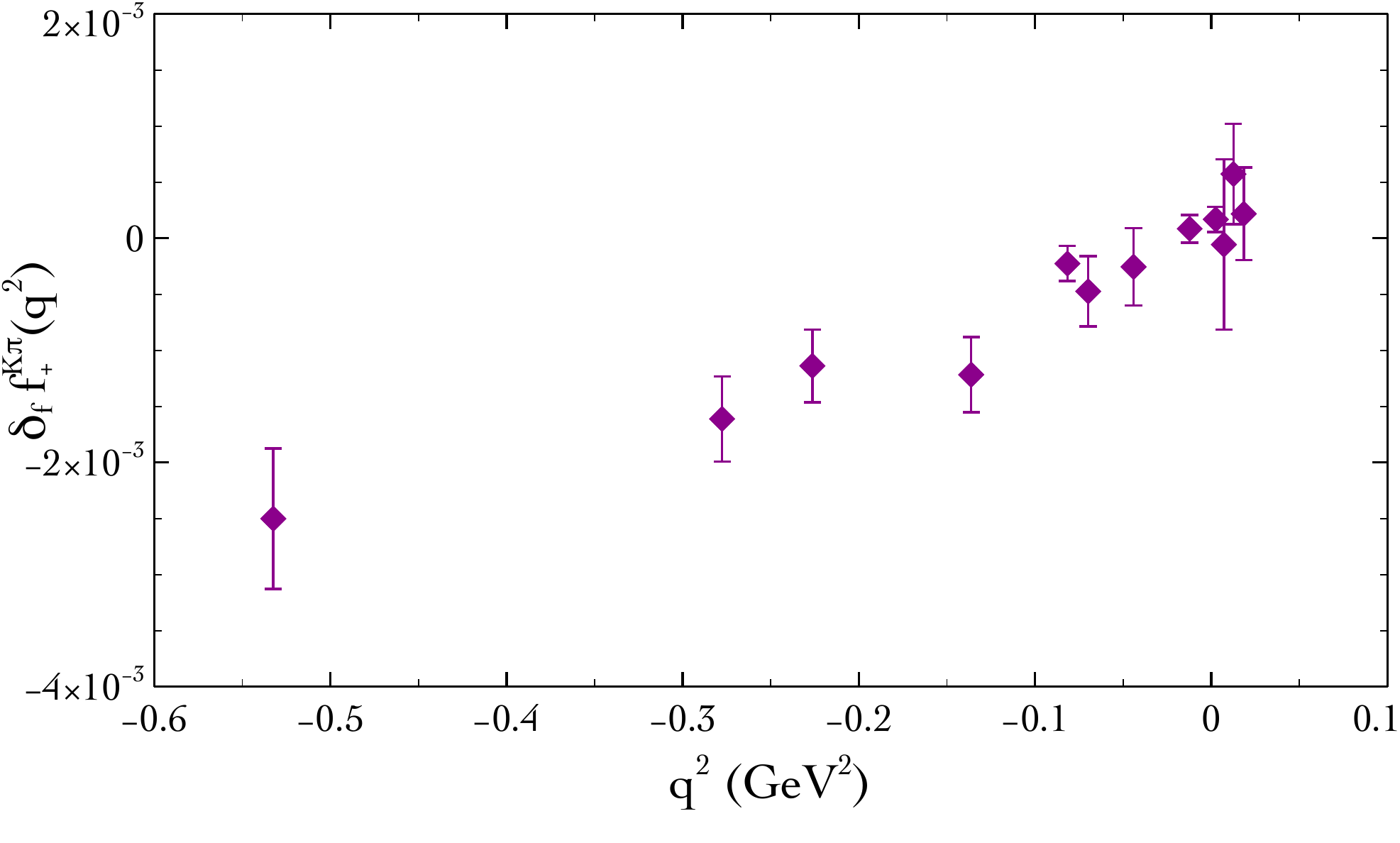}
\caption{\label{fig:deltaf} \footnotesize
{\it Left panel}: we show our results for $f_+^{K\pi}(q^2)$.
{\it Right panel}: we show our results for $\delta_f f_+^{K\pi}(q^2)$. 
The data are obtained at fixed lattice spacing $a=0.085$~fm  and at fixed $am_{ud}^L=0.0064$ (see Appendix~\ref{sec:tm}). 
}
\end{center}
\end{figure}

We now come to the numerical results. In Figure~\ref{fig:mel} we show the ratios $R^0_{DK}(t)$, left panel, and $R^0_{K\pi}(t)$, right panel, from which we extract  the form factors. In Figure~\ref{fig:deltamel} we show $\delta R^0_{DK}(t)/a\Delta m_{ud}^L$, left panel, and $\delta_f R^0_{K\pi}(t)/a\Delta m_{ud}^L$. As expected according to eqs.~(\ref{eq:corrfdk}) and~(\ref{eq:corrfkpi}) both $\delta R^0_{DK}(t)/a\Delta m_{ud}^L$ and $\delta_f R^0_{K\pi}(t)/a\Delta m_{ud}^L$ are constant with respect to $t$ in the middle of the lattice, within the  statistical errors that, in the case of $\delta R^0_{DK}(t)/a\Delta m_{ud}^L$,  are rather large. Finally, in Figure~\ref{fig:deltaf} we show our results for $f_+^{K\pi}(q^2)$, left panel, and for $\delta_f f_+^{K\pi}(q^2)$. These results, obtained only at fixed lattice spacing $a=0.085$~fm and fixed light quark mass $am_{ud}^L=0.0064$, 
show that our method works also in the case of  complicated observables, extracted from ratios of integrated three point correlation functions with non vanishing spatial momenta. 

In a separate paper we  shall calculate disconnected diagrams and the missing contribution $\delta_b f_+^{K\pi}(q^2)$ and refine our findings by improving the statistics, by repeating the calculation at several values of $m_{ud}$ and by performing chiral and continuum extrapolations. For the time being, we get
\bea
\left[ \frac{f_+^{K^0\pi^-}(0)-f_+^{K\pi}(0) }{f_+^{K\pi}(0)}\right]^{QCD}
= 0.85(18)(1)\times 10^{-4} \quad \times \quad 
\frac{\left[M_{K^0}^2-M_{K^+}^2\right]^{QCD}}{6.05\times 10^3\ \mbox{MeV}^2} \, .
\eea
As anticipated a few paragraph above, the fact that this result is two order of magnitude smaller than the one quoted in eq.~(\ref{eq:expdeltafkpi}) is not surprising in view  of the chiral perturbation theory analysis performed in ref.~\cite{Gasser:1984ux}. The enhancement of the missing contribution, $\delta_b f_+^{K\pi}(q^2)$, is indeed traced back to the $\pi^0$-$\eta$ mixing.

\section{Conclusions and Outlooks}
\label{sec:outlooks}
In this paper we have proposed a new method to compute with high precision leading QCD isospin breaking effects in relevant physical quantities at the lowest non trivial order in the up-down mass difference. The method can be easily extended with minor modifications to higher orders. 
We have computed the corrections to meson and nucleon masses, meson decay constants and weak form factors, showing that, in spite of the limited statistics, our approach is already competitive, or even better, than other non perturbative calculations based on effective chiral lagrangians. 

To obtain the complete physical results, our method has to be combined with  calculations of the electromagnetic corrections which will be the subject of a future investigation. In this paper, for a comparison with calculations in different theoretical frameworks, we have taken  the electromagnetic corrections to the meson masses  evaluated in ref.~\cite{Colangelo:2010et}.

As the method looks very promising, we are planning to extend this pioneering work to other physical observables.

\section*{Acknowledgements}
We are particularly grateful to M.~Testa for illuminating discussions on the renormalization of the electromagnetic corrections.
We thank N.~Christ and C.~Sachrajda for discussions on arguments close to the ones presented in this paper.
We thank the members of the ETMC collaboration for having generated and made publicly available the gauge configurations
used for this study. We thank the staff of the AURORA computing center for allocating computer resources to this project and for 
their assistance.
Work partially supported  by the Programme IDEAS, ERC-2010-AdG, DaMESyFla
Grant Agreement Number: 267985, and by the MIUR (Italy) under the contracts PRIN08 and PRIN09. 
V.L. acknowledges the support of CNRS and the Laboratoire de Physique Th\'eorique d'Orsay, 
Universit\'e Paris-Sud 11, where part of this work was completed.

\begin{appendices}

\section{}
\label{sec:tm}
%
\begin{table}[!h]
\begin{center}
\begin{tabular}{ccccccc}
$\quad$ $\beta$    $\quad$ & 
$\quad$ $am_{ud}^L$  $\quad$ & 
$\quad$ $am_s^L$     $\quad$ & 
$\quad$ $L/a$      $\quad$ & 
$\quad$ $N_{conf}$ $\quad$ & 
$\quad$ $a$~(fm)   $\quad$ & 
$\quad$ $Z_P(\overline{MS},2GeV)$      $\quad$\tabularnewline
\hline 
\tabularnewline
3.80 & 0.0080 & 0.0194 & 24 & 150 & 0.0977(31) & 0.411(12)\tabularnewline
        & 0.0110 &        & 24 & 150 &            &          \tabularnewline[4ex]

3.90 & 0.0030 & 0.0177 & 32 & 150 & 0.0847(23) & 0.437(07)\tabularnewline
     & 0.0040 &        & 32 & 150 &            &          \tabularnewline
     & 0.0040 &        & 24 & 150 &            &          \tabularnewline
     & 0.0064 &        & 24 & 150 &            &          \tabularnewline
     & 0.0085 &        & 24 & 150 &            &          \tabularnewline
     & 0.0100 &        & 24 & 150 &            &          \tabularnewline[4ex]

4.05 & 0.0030 & 0.0154 & 32 & 150 & 0.0671(16) & 0.477(06)\tabularnewline
     & 0.0060 &        & 32 & 150 &            &          \tabularnewline
     & 0.0080 &        & 32 & 150 &            &          \tabularnewline[4ex]

4.20 & 0.0020 & 0.0129 & 48 & 100 & 0.0536(12) & 0.501(20)\tabularnewline
     & 0.0065 &        & 32 & 150 &            &          \tabularnewline[4ex]

\hline 
\end{tabular}
\end{center}
\label{tab:gaugeconfigs}
\caption{{\footnotesize Guage ensambels used in this work. The gauge configurations have been generated with $N_f=2$ dynamical flavours of maximally twisted quarks of mass $am_{ud}^L$. The strange quark mass $am_s^L$ it has been used for valence propagators. At $\beta=3.90$, where we calculate quantities related to $D$ mesons and correlation functions at non-vanishing spatial momenta, we have set $am_c^L=0.2123$ and, by using flavour twisted boundary conditions, 
$\vec{p}L/2\pi=\{0.00,\pm 0.15, \pm 0.35 \}$.}}
\end{table}
In this work we have used the $N_f=2$ dynamical gauge ensambles generated and made publicly available by the European Twisted Mass Collaboration (see Table~\ref{tab:gaugeconfigs}). These gauge configurations have been generated by using the so called Twisted Mass lattice discretization of the QCD action~\cite{Frezzotti:2000nk}. The maximally twisted fermion action is given by
\begin{eqnarray}
{\mathcal L}_{TM}[U]
= \bar q\left(\ D[U]+m+i\gamma_5\tau^3 W[U]\ \right)q\, ,
\label{eq:tmaction3}
\end{eqnarray}
where $q^T=(\ell_+,\ell_-)$, $D[U]$ is the naive lattice action and $W[U]$ the critical Wilson term (with $a=1$),
\begin{eqnarray}
\nabla_\mu[U]\ q(x) &=& U_\mu(x)q(x+\mu) - q(x)\, ,
\nonumber \\
\nonumber \\
\nabla_\mu[U]^\dagger\ q(x) &=& q(x)-U_\mu^\dagger(x-\mu)q(x-\mu)\, ,
\nonumber \\
\nonumber \\
D[U]\ q(x) &=& \gamma^\mu \frac{\nabla_\mu[U]+\nabla_\mu^\dagger[U]}{2}q(x)\, ,
\nonumber \\
\nonumber \\
W[U]\ q(x) &=& \left[
\sum_\mu{\frac{\nabla_\mu[U]-\nabla_\mu^\dagger[U]}{2}}+m^{cr}\right]
q(x)\, .
\end{eqnarray}
The critical mass $m^{cr}(g_0^2)$ has been taken from ref.~\cite{Baron:2009wt}. 
Concerning our work, the choice of the maximally twisted Wilson lattice formulation has advantages and drawbacks.
The big advantage is automatic $O(a)$ improvement~\cite{Frezzotti:2003ni}. The drawback is the breaking of isospin symmetry at finite lattice spacing even with $\Delta m_{ud}=0$.
Indeed, by letting the physical $u$ and $d$ fields to coincide with the fields $\ell_+$ and $\ell_-$, by taking $m=m_{ud}$ and by identifying $\mathcal{L}_0$ with $\cal{L}_{TM}$, eq.~(\ref{eq:nodet}) ceases to be valid at fixed cutoff. This happens because of the interference between the $\tau_3$ matrix appearing within $\hat S$ and the $\tau_3$ appearing in the twisted critical Wilson term,
\bea
&&\left\langle 
\left[ \mbox{fermionic Wick contractions of }{\cal O} \right]
\times \mbox{tr}[\hat S]\ \right\rangle_0
\nonumber \\
\nonumber \\
&&\qquad\qquad
=
\left\langle 
\left[ \mbox{fermionic Wick contractions of }{\cal O} \right]\times 
\left\{\mbox{Tr}\left[G_{\ell_+}(x,x)\right]-\mbox{Tr}\left[G_{\ell_-}(x,x)\right]\right\}\ \right\rangle_0
\nonumber \\
\nonumber \\
&&\qquad\qquad= O(a^2) \neq 0\, .
\label{eq:tmlatticeartifact}
\eea  
Notice however that it remains true that $\langle \hat S \rangle_0=0$ owing to the invariance of the TM lattice action under parity times $\ell_+ \leftrightarrow \ell_-$ interchange.
Since eq.~(\ref{eq:tmlatticeartifact}) represent a mere $O(a^2)$ cutoff effect we have chosen to \emph{neglect} the corresponding contributions to the correlation functions considered in the text. This procedure actually corresponds to work within the mixed action approach of ref.~\cite{Frezzotti:2004wz} and, at the price of introducing $O(a^2)$ unitarity violations, preserves $O(a)$ improvement of physical quantities.

To clarify the point, let's first consider the discretized version of eqs.~(\ref{eq:kpcorr}), namely
\bea
C_{K^+K^-}(t) &=& -\overset{s_-} {\underset{u_+} {\gdsu}}=   
-\overset{s_-} {\underset{\ell_+} {\gdsl}} 
-\overset{s_-} {\underset{\ \ell_+ \rightarrow\ \ell_+} {\gdsi}} + {\cal O}(\Delta m_{ud})^2\, ,
\nonumber \\
\nonumber \\
C_{K^0K^0}(t) &=& -\overset{s_-} {\underset{d_+} {\gdsd}}= 
-\overset{s_-} {\underset{\ell_+} {\gdsl}} 
+\overset{s_-} {\underset{\ \ell_+ \rightarrow\ \ell_+} {\gdsi}} + {\cal O}(\Delta m_{ud})^2  \, .   
\label{eq:kpcorrlatt}
\eea
In the previous equations we have explicitly shown a label indicating the flavour of each propagator and the corresponding sign of the term $\pm\;\gamma_5W[U]$ appearing in its kinetic operator. The important point to note is that in the valence we have chosen the signs of the Wilson terms of the up and down quarks independently from the choice made in the sea, where the two quarks must necessarily have opposite signs in order to deal with a real positive fermionic action, and that this choice is legitimately \emph{observable dependent}. In this particular case the results are correlation functions with much smaller lattice artifacts and statistical errors with respect to the other possible choices, e.g. $\mbox{Tr}\left[\gamma_5 G_{s_+}(0,x) \gamma_5 G_{\ell_+}(x,0)\right]$. 

Analogously we can choose conveniently the sign in front of the Wilson term for the valence quarks entering three point correlation functions. As an example we discuss explicitly the case of eq.~(\ref{eq:k0pim}), namely
\bea
C^\mu_{K^0\pi^-}(t)&=&
-\underset{d_+}{ \sideset{^{s_-}}{^{u_-}}{\operatorname{\gtsdu}}}
\nonumber \\
\nonumber \\
&=&
-\underset{\ell_+}{ \sideset{^{s_-}}{^{\ell_-}}{\operatorname{\gtsll}}}
\ + \
\underset{\ell_+\rightarrow \ell_+}{ \sideset{^{s_-}}{^{\ell_-}}{\operatorname{\gtsil}}} 
\ -\
\underset{\ell_+}{ \sideset{^{s_-}}{^{\ell_-\rightarrow \ell_-}}{\operatorname{\gtsli}}}
\quad +\
{\cal O}(\Delta m_{ud})^2\, .
\label{eq:k0pimlatt}
\eea
As it can be seen, in order to have mesons interpolated by operators of the form $\bar\ell_+ \gamma_5 \ell_-$ (the ones entailing smaller discretization and statistical errors), here we have chosen the sign of the $u$ quark Wilson term opposite with respect to that in eqs.~(\ref{eq:kpcorrlatt}). 
The discussion of eq.~(\ref{eq:kppi0}) is considerably more involved because of the presence of a neutral pion in the physical correlation functions. A convenient lattice discretization of the correlator of eq.~(\ref{eq:kppi0}) can be also obtained within the mixed action approach but since the calculation of $\delta_b f^{K\pi}(q^2)$ will be the subject of a future work we don't discuss here this point.

For the different gauge ensembles used in this work the values of the lattice spacing $a$ (ref.~\cite{Blossier:2010cr}), of the strange valence quark mass (ref.~\cite{Blossier:2010cr}) and of the renormalization constant $Z_P$ (ref.~\cite{Constantinou:2010gr}) are given in Table~\ref{tab:gaugeconfigs}. The values of $Z_P$ are relevant because in the maximally TM formulation one has~\cite{Frezzotti:2000nk,Frezzotti:2004wz}
\bea
\Delta m_{ud}\ \bar q \tau_3 q \ = \ 
Z_{\Delta m} \Delta m_{ud}^L\ \ Z_P \left[\bar q \tau_3 q\right]^L
\qquad \longrightarrow \qquad
Z_{\Delta m } = \frac{1}{Z_P}\, .
\eea

\section{}
\label{sec:proof}
A more detailed derivation of eqs.~(\ref{eq:corrfdk}), i.e. of
\bea
\delta R^\mu_{DK}(t)&=&
-\frac{\gtcis}{\gtcls}
-\frac{\gtsic}{\gtslc}
+\frac{\gtcic}{\gtclc}
+\frac{\gtsis}{\gtsls}
\nonumber \\
\nonumber \\
&=& 2 \delta \left[ \bra{K} V^\mu_{cs} \ket{D} \right] -\delta E_D -\delta E_K+\cdots \, ,
\label{eq:corrfdkhere}
\eea
can be obtained by applying perturbation theory with respect to $\Delta m_{ud}$ (see refs.~\cite{Cabibbo:1983xa,Martinelli:1982cb,Maiani:1987by} for related works). Let us analyze in detail the case of
\bea
-\gtcds=
\rho^\star_{K^0}\rho_{D^+}\ \bra{K^0} V^\mu_{cs} \ket{D^+}\ 
e^{-tE_{D^+}} e^{-(T/2-t)E_{K^0}} \, , 
\label{eq:proof1}
\eea
where $\rho^\star_{K^0}=\bra{0}\bar d \gamma^5 s(0) \ket{K^0}/2E_{K^0}$ and $\rho_{D^+}=\bra{D^+}\bar c \gamma^5 d(0) \ket{0}/2E_{D^+}$ and where we have neglected sub leading exponentials. 
The perturbation ${\mathcal V}$,
\bea
{\mathcal V}=\sum_{\vec x}\hat {\cal L}(0,\vec x)=\sum_{\vec x}\left[
\bar u u -\bar d d
\right](0,\vec x)\, ,
\eea
is flavour diagonal and does not open any decay channel for the $K$ and $D$ mesons. Furthermore charged meson states do not mix with the corresponding neutral states.
By considering lattice states $\ket{n_L}$ normalized to one, $\braket{n_L}{n_L}=1$, we can use the well known formulae
\bea
&&\ket{K^0_L} \ =\ \ket{K_L} + \ket{\Delta K_L} 
\ =\ \ket{K_L} + 
\Delta m_{ud}\sum_{n\neq K} \frac{\ket{n_L}\bra{n_L}{\mathcal V}\ket{K_L}}{E_n-E_K} \, ,
\nonumber \\
\nonumber \\
&&E_{K^0}\ =\ E_K+\Delta E_K \ =\ E_K+\Delta m_{ud}\bra{K_L} {\mathcal V} \ket{K_L}  \, ,
\nonumber \\
\nonumber \\
\nonumber \\
&&\ket{D^+_L}\ =\ \ket{D_L} + \ket{\Delta D_L} 
\ =\ \ket{D_L} + 
\Delta m_{ud}\sum_{n\neq D} \frac{\ket{n_L}\bra{n_L}{\mathcal V}\ket{D_L}}{E_n-E_D} \, ,
\nonumber \\
\nonumber \\
&&E_{D^+}\ =\ E_D+\Delta E_D \ =\ E_D+\Delta m_{ud}\bra{D_L} {\mathcal V} \ket{D_L}  \, .
\label{eq:nondegpt}
\eea
connecting, at first order, the states and the eigenvalues of the unperturbed isospin symmetric theory with the corresponding quantities of the perturbed theory. First order corrections for the relativistically covariant states, $\braket{n}{n}=2E_n$, are then readily obtained by changing the normalization,
\bea
\ket{K^0}&=&\ket{K}+\ket{\Delta K}
=\sqrt{2E_{K^0}}\ \ket{K^0_L}
\nonumber \\ 
\nonumber \\
&=&\sqrt{2E_{K}}\left(1+\frac{\Delta E_K}{2E_K} \right)
\left(\ket{K_L}+\ket{\Delta K_L} \right)
\nonumber \\
\nonumber \\
\nonumber \\
&=& \ket{K} + \frac{\delta E_K}{2}\ket{K}
+\Delta m_{ud}
\sum_{n\neq K} \frac{\ket{n}\bra{n}{\mathcal V}\ket{K}}{2E_n(E_n-E_K)} \, .
\eea
In the case of $\ket{D^+}$, by proceeding along the same lines, we get
\bea
\ket{\Delta D}= \frac{\delta E_D}{2}\ket{D}
+\Delta m_{ud}
\sum_{n\neq D} \frac{\ket{n}\bra{n}{\mathcal V}\ket{D}}{2E_n(E_n-E_D)} \, .
\eea
The expansion of the matrix element $\bra{K^0} V^\mu_{cs} \ket{D^+}$ appearing into eq.~(\ref{eq:proof1}) is thus given by
\bea
&&\bra{K^0} V^\mu_{cs} \ket{D^+}
\nonumber \\
\nonumber \\
&&=
\bra{K} V^\mu_{cs} \ket{D}+\bra{\Delta K} V^\mu_{cs} \ket{D}+\bra{K} V^\mu_{cs} \ket{\Delta D}
\nonumber \\
\nonumber \\
\nonumber \\
&&=
\left(1+\frac{\delta E_K + \delta E_D}{2}\right)\bra{K} V^\mu_{cs} \ket{D}+
\Delta m_{ud}
\left\{
\sum_{n\neq K} \frac{\bra{K}{\mathcal V}\ket{n}\ \bra{n}V^\mu_{cs} \ket{D}}{2E_n(E_n-E_K)}
+
\sum_{n\neq D} \frac{\bra{K}V^\mu_{cs}\ket{n}\ \bra{n} {\mathcal V} \ket{D}}{2E_n(E_n-E_D)}
\right\}\, .
\nonumber \\
\nonumber \\
\label{eq:proof2}
\eea
In order to obtain the explicit expression of the QIB correction to the correlator of eq.~(\ref{eq:proof1}) we should also expand the exponentials and the matrix elements of the interpolating operators. The explicit expression of $\delta \rho_{D,K}$ is not needed, because these terms cancel in the final expression of $\delta R^\mu_{DK}$, while the expansion of the exponential factors is easily obtained,
\bea
e^{-tE_{D^+}} e^{-(T/2-t)E_{K^0}} 
&=&
e^{-tE_{D}} e^{-(T/2-t)E_{K}}\left[
1-t\Delta E_D -(T/2-t)\Delta E_K
\right]
\nonumber \\
\nonumber \\
&=&
e^{-tE_{D}} e^{-(T/2-t)E_{K}}\left[
1-t\Delta m_{ud}\frac{\bra{D} {\mathcal V} \ket{D}}{2E_K} 
-(T/2-t)\Delta m_{ud}\frac{\bra{K} {\mathcal V} \ket{K}}{{2E_K}}
\right] \, .
\label{eq:proof3}
\eea 
Eqs.~(\ref{eq:proof2}) and~(\ref{eq:proof3}), combined with the diagrammatic expansion of our correlation function derived in section~\ref{sec:observables}, i.e.
\bea
-\gtcds=-\gtcls+\gtcis  +{\cal O}(\Delta m_{ud})^2\, ,
\eea
give us an explicit expression for the first term entering the $\delta R^\mu_{DK}$ formula, namely
\bea
-\gtcls&=&
\rho^\star_{K}\rho_{D}\ \bra{K} V^\mu_{cs} \ket{D}\ 
e^{-tE_{D}} e^{-(T/2-t)E_{K}} \, ,
\nonumber \\
\nonumber \\
-
\frac{\gtcis}{\gtcls}&=&
\left\{1+\delta\rho^\star_{K}\right\} \
\left\{1 +\delta\rho_{D}\right\} \
\left\{1+\delta[\bra{K} V^\mu_{cs} \ket{D}]\right\}\ 
\left\{
1-t\Delta E_D-(T/2-t)\Delta E_{K}
\right\} -1\, .
\nonumber \\
\nonumber \\
\label{eq:proofterm1}
\eea
By repeating the same arguments for the other correlation functions of eqs.~(\ref{eq:dk3corrs1}) and~(\ref{eq:dk3corrs2}),  
it is straightforward to obtain explicit expressions for the remaining three terms appearing into the expression of $\delta R^\mu_{DK}$, i.e.
\bea
-
\frac{\gtsic}{\gtcls}&=&
\left\{1+\delta\rho_{K}\right\} \
\left\{1 +\delta\rho^\star_{D}\right\} \
\left\{1+\delta[\bra{K} V^\mu_{cs} \ket{D}]\right\}\ 
\left\{
1-t\Delta E_K-(T/2-t)\Delta E_{D}
\right\} -1\, ,
\nonumber \\
\nonumber \\
-
\frac{\gtcic}{\gtclc}&=&
\left\{1+\delta\rho_{D}\right\} \
\left\{1 +\delta\rho^\star_{D}\right\} \
\left\{1+\delta E_D\right\}\ 
\left\{
1-T\Delta E_{D}/2 
\right\} -1\, ,
\nonumber \\
\nonumber \\
-
\frac{\gtsis}{\gtsls}&=&
\left\{1+\delta\rho_{K}\right\} \
\left\{1 +\delta\rho^\star_{K}\right\} \
\left\{1+\delta E_K\right\}\ 
\left\{
1-T\Delta E_{K}/2 
\right\} -1\, .
\label{eq:proofterm2}
\eea
The proof of the $\delta R^\mu_{DK}$ formula, second of eqs.~(\ref{eq:corrfdk}) or eq.~(\ref{eq:corrfdkhere}), is finally obtained by substituting in that relation eqs.~(\ref{eq:proofterm1}) and eqs.~(\ref{eq:proofterm2}).

In deriving eqs.~(\ref{eq:proofterm1}) and eqs.~(\ref{eq:proofterm2}) we have not shown terms proportional to the correction of the ``vacuum energy" because in our case $\bra{0}{\mathcal V} \ket{0}=0$. If different from zero, such kind of contributions would appear at an intermediate stage of the calculation whereas they would cancel in the final expression as happens to the terms proportional to $\delta \rho_{D,K}$.

\end{appendices}


\end{document}